\documentclass[12pt]{iopart}
\include{epsf}
\usepackage{amsfonts}
\usepackage{iopams}
\usepackage{color}
\usepackage{graphics}
\newtheorem{THEOREM}{Theorem}[section]
\renewcommand{\theequation}{\thesection.\arabic{equation}}
\newcounter{labelflag} 

\newcommand{\Label}[1]{
                       \ifnum\thelabelflag=1
                          \ifmmode
                             \makebox[0in][l]{\qquad\fbox{\rm#1}}
                          \else
                             \marginpar{\vspace{0.7\baselineskip}
                                        \hspace{-1.1\textwidth}
                                        \fbox{\rm#1}}
                          \fi
                       \fi
                       \label{#1}
                      }
\newcommand{\be}{\begin{equation}}
\newcommand{\ee}{\end{equation}}
\newcommand{\dx}{\partial_x}
\newcommand{\dxx}{\partial_{xx}}
\newcommand{\R}{\mathbf{R}}

\newcommand{\N}{\mathbf{N}}
\newcommand{\eps}{\varepsilon}
\newcommand{\I}{\mathcal{I}}

\newcommand{\Lfpu}{\mathcal{T}^+}
\newcommand{\Nf}{\mathcal{N}}

\newcommand{\DLf}{\mathcal{DT}}
\newcommand{\DLfpu}{\mathcal{DT}^+}
\newcommand{\Lo}{\mathcal{L}_0}
\newcommand{\LK}{\mathcal{L}_K}

\newcommand{\sL}{\sigma_0}
\newcommand{\sR}{\sigma_1}
\newcommand{\sst}{\sigma_*}
\newcommand{\ex}{\mathrm{e}}
\newcommand{\V}{\mathrm{v}}
\newcommand{\Ai}{{\rm Ai}}
\newcommand{\Bi}{{\rm Bi}}
\newcommand{\bmp}[3]{\Gamma_{#3}\left(#1 \,; #2\right)}
\newcommand{\G}[2]{{\mathcal G}\left(#1 \,; #2\right)}
\newcommand{\xo}{x_0}
\newcommand{\xs}{{x_*}}
\newcommand{\xss}{{x^{**}}}
\newcommand{\Fo}{\bar{F}}
\newcommand{\Wslave}{G}

\newcommand{\dWeq}{{{\rm d}\Omega}}

\newcommand{\dPsi}{{{\rm d}\Psi}}
\newcommand{\w}{\omega}
\newcommand{\wpu}{\omega^+}
\newcommand{\wmu}{\omega^-}
\newcommand{\ypu}{\psi^+}
\newcommand{\ymu}{\psi^-}
\newcommand{\upu}{u^+}

\newcommand{\vmu}{v^-}

\newcommand{\hwpu}{{\hat{\omega}^+}}
\newcommand{\y}{\psi}

\newcommand{\n}{\eta}
\newcommand{\hn}{{\hat{\eta}}}
\newcommand{\s}{\zeta}
\newcommand{\abs}[1]{\left\vert#1\right\vert}
\newcommand{\norm}[1]{\left\vert\left\vert#1\right\vert\right\vert}
\newcommand{\hide}[1]{}
\newcommand{\im}{{\rm i}}
\begin{document}

\title[Emergence of localized structures in a  phytoplankton--nutrient model]{Emergence
of steady and oscillatory localized structures in a phytoplankton--nutrient model}

\author{A.~Zagaris$^1$ and A.~Doelman$^{2,3}$}

\address{$^1$ University of Twente,
Department of Applied Mathematics,
P.O. Box 217, 7500 AE Enschede,
the Netherlands}
\address{$^2$ University of Leiden,
Mathematisch Instituut,
P.O. Box 9512, 2300 RA Leiden,
the Netherlands}
\address{$^3$ CWI, P.O. Box 94079,
1090 GB Amsterdam,
the Netherlands}

\ead{a.zagaris@ewi.utwente.nl}

\maketitle

\begin{abstract}
Co-limitation of marine phytoplankton growth
by light and nutrient,
both of which are essential
for phytoplankton,
leads to complex dynamic behavior
and
a wide array of coherent patterns.
The building blocks
of this array
can be considered to be
deep chlorophyll maxima, or DCMs,
which are structures
localized in a finite depth
interior to the water column.
From an ecological point of view,
DCMs are evocative of a balance
between the inflow of light
from the water surface
and
of nutrients from the sediment.
From a (linear) bifurcational point of view,
they appear through a transcritical bifurcation
in which
the trivial, no-plankton steady state
is destabilized.
This article is devoted to
the analytic investigation
of the weakly nonlinear dynamics
of these DCM patterns,
and
it has two overarching themes.
The first of these
concerns the fate
of the destabilizing stationary DCM mode
beyond the center manifold regime.
Exploiting the natural
singularly perturbed nature
of the model,
we derive an explicit reduced model
of asymptotically high dimension
which fully captures these dynamics.
Our subsequent and fully detailed study
of this model---which involves
a subtle asymptotic analysis
necessarily transgressing the boundaries
of a local center manifold reduction---establishes that
a stable DCM pattern indeed appears
from a transcritical bifurcation.
However,
we also deduce
that asymptotically close
to the original destabilization,
the DCM looses its stability
in a secondary bifurcation
of Hopf type.
This is in agreement with indications
from numerical simulations
available in the literature.
Employing the same methods,
we also identify
a much larger DCM pattern.
The development of the method
underpinning this work---which,
we expect,
shall prove useful
for a larger class of models---forms
the second theme
of this article.
\end{abstract}


\submitto{Nonlinearity}

\section{Introduction \label{s-intro}}
\setcounter{equation}{0}
Phytoplanktonic photosynthesis
provides the major biological component
of the transport mechanism
carrying atmospheric carbon dioxide
into the deep ocean.
Concurrently,
plankton forms the basis
of the aquatic food chain.
As a consequence,
phytoplankton growth and decay
plays a crucial role
in understanding climate dynamics \cite{FBS-1998}
and
forms an integral part
of oceanographic research.
Conversely,
climate changes---such as
global temperature variations---have a direct impact
on the aquatic ecosystem
and thus also on phytoplankton \cite{BLW-2010,SHSM-1998}:
there is a subtle
and
under-explored interplay
between the dynamics of phytoplankton concentrations
and climate variability.
At the same time,
phytoplankton concentrations exhibit
surprisingly rich spatio-temporal dynamics.
The character of those dynamics
is determined in an intricate fashion
by (changes in) the external conditions,
see \cite{HPKS-2006}
and the references therein.
The building blocks
for the observed complex patterns
are \emph{deep chlorophyll maxima} (DCMs)
or \emph{phytoplankton blooms},
in which
the phytoplankton concentration
exhibits a maximum
at a certain,
well-defined depth
of the basin.
These patterns are the manifestation
of a fundamental balance
between the supply
of light from the surface
and of nutrients from the depths of the basin.
For the simplest models,
in which spatiotemporal fluctuations
in the nutrient concentration
are omitted
(\emph{eutrophic} environment),
it has been shown that
there can only be a stationary global attractor
\cite{IT-1982}.
In particular,
if the trivial state (no phytoplankton) is unstable,
then there can only be
a stationary globally attracting phytoplankton bloom
with its maximum either
at the surface (a surface layer),
at the bottom (a benthic layer, BL),
or in between (a DCM)
\cite{EATSH-2001,GM-2003,HOW-1999a,IT-1982}.
This is no longer the case
in coupled phytoplankton--nutrient systems
(\emph{oligotrophic} environment),
although DCMs do tend to appear
in those systems, also,
for certain parameter combinations
\cite{DH-2008a,DH-2008b,FB-2003,HOW-1999a,HS-2002,KL-2001}.
The detailed numerical studies
reported in \cite{HPKS-2006},
however,
show that the appearance of a DCM
only triggers a complex sequence of bifurcations:
as parameters vary,
a DCM may be time-periodic,
undergo a sequence
of period doubling bifurcations,
and eventually behave chaotically.

In this paper,
we focus on the effect
that varying environmental conditions,
and in particular
nutrient levels at the ocean bed,
have on the dynamics
generated by the one-dimensional model
for phytoplankton ($W$)--nutrient ($N$) interactions
originally introduced in \cite{HPKS-2006},
\be
\left\{
\begin{array}{l}
 W_t
=
 D \, W_{zz}
-
 V \, W_z
+
 \left[ \mu \, P(L,N) - l \right] W ,
\\
 N_t
=
 D \, N_{zz}
-
 Y^{-1} \, \mu \, P(L,N) \, W .
\end{array}
\right.
\Label{PDE-orig}
\ee
In this model,
the vertical coordinate $z$
measures the depth
in a water column
spanned by $[0,z_B]$,
while
$W(z,t)$ and $N(z,t)$
are the phytoplankton and nutrient concentrations,
respectively,
at depth $z$ and time $t$.
As in \cite{HPKS-2006,ZDPS-2009},
the system is assumed
to be in
the turbulent mixing regime
\cite{EATSH-2001,HOW-1999a},
so that
the diffusion coefficient $D$
is identically the same
for phytoplankton and nutrient.
The phytoplankton is characterized by
its sinking speed $V$,
its (species-specific) loss rate $l$,
its maximum specific production rate $\mu$,
and
its yield $Y$ on light and nutrient.
The model is equipped
with natural no-flux boundary conditions
at the surface
for both phytoplankton and nutrients;
the bottom is a source of nutrients
but impenetrable for phytoplankton,
\be
 D \, W_z - V \, W \vert_{z=0,z_B} = 0 ,
\quad
 N_z \vert_{z=0} = 0 ,
\quad\mbox{and}\quad
 N \vert_{z=z_B} = N_B .
\Label{BC-dim}
\ee
The constant nutrient concentration $N_B$
will act as the primary bifurcation parameter
in this work.
The nonlinear expression $P(L,N)$
models phytoplankton growth
due to light and nutrient,
\be
 P(L,N)
=
 \frac{L N}{(L + L_H)(N + N_H)},
\Label{P}
\ee
in which
$L_H$ and $N_H$
are the half-saturation constants
of light and nutrient,
respectively.
(See \cite{ZDPS-2009}
for a short discussion
on the nature and specificity
of $P(L,N)$.)
The light intensity $L$
at depth $z$ and time $t$
is determined by
the total amount of
planktonic and non-planktonic components
in the column $[0,z]$,
\be
 L(z,t)
=
 L_I
\,
 \ex^{-K_{bg} z - R \int_0^z W(s,t) ds}.
\Label{L-expl}
\ee
Hence,
the system is non-local---a typical feature
of most realistic phytoplankton models.
The light intensity term
introduces an extra three parameters:
$L_I$,
the intensity of
the incident light
at the water surface;
$K_{bg}$,
the light absorption coefficient
due to non-planktonic, background components
and hence
a measure of \emph{turbidity};
and $R$,
the light absorption coefficient
due to plankton
(\emph{self-shading}).
The first two of these parameters,
together with
$z_B$, $D$, $Y$, and $N_B$
quantify the effect
that the environment has
on the planktonic population.
It is by varying these parameters
that we examine the effect
of changing environmental conditions
on plankton.

It is shown in \cite{ZDPS-2009}
that the system (\ref{PDE-orig})
has a natural
singularly perturbed nature.
This can be seen
by rescaling time and space
via $\tau = \mu \, t$ and $x = z/z_B$
and the phytoplankton concentration $W$,
nutrient concentration $N$,
and
light intensity $L$ via
\[
\fl
 \wpu(x,\tau)
=
 \frac{l z_B^2}{D Y N_B} W(z,t) ,
\quad
 \n(x,\tau)
=
 1
-
 \frac{N(z,t)}{N_B} ,
\quad\mbox{and}\quad
 j(x,\tau)
=
 \frac{L(z,t)}{L_I} .
\]
Substitution into (\ref{PDE-orig}) then yields,
\be
\begin{array}{l}
 \wpu_\tau
=
 \eps \wpu_{xx}
-
 2 \sqrt{\eps \V}
\,
 \wpu_x
+
 ( p(\wpu , \n , x) - \ell)
\,
 \wpu ,
\\
 \n_\tau
=
 \eps
 \left(
 \n_{xx}
+
 \ell^{-1}
 p(\wpu , \n , x)
\,
 \wpu
 \right) ,
\end{array}
\Label{PDE}
\ee
with boundary conditions,
\be
\fl
 ( \wpu_x
-
 2 \sqrt{\V/\eps} \, \wpu )(0)
=
 ( \wpu_x
-
 2 \sqrt{\V/\eps} \, \wpu )(1)
=
 0
\quad\mbox{and}\quad
 \n_x(0)
=
 \n(1)
=
 0 .
\Label{BC}
\ee
For realistic choices
of the original parameters
of (\ref{PDE-orig}),
\[
 \eps
=
 \frac{D}{\mu z_B^2}
\approx
 10^{-5} ,
\]
cf. \cite{HPKS-2006,ZDPS-2009}.
Effectively,
$\eps^{1/4}$ characterizes
the extent of the zone
where DCMs appear
relative to the depth
of the ocean.
In this paper,
we follow \cite{ZDPS-2009}
and treat the parameter $\eps$
as an asymptotically small parameter,
\emph{i.e.},
we assume that $0 < \eps \ll 1$
so that (\ref{PDE}) has, indeed,
a singularly perturbed character.
The nonlinearity $p$ in (\ref{PDE})
is given by
\be
 p(\wpu , \n , x)
=
 \frac{1 - \n}{(\n_H + 1 - \n) \, (1 + j_H/j(\wpu , x))} ,
\Label{p}
\ee
with rescaled light intensity
\be
 j(\wpu , x)
=
 \exp\left(-\kappa x - r \int_0^x \wpu(s,\tau) ds\right) .
\Label{j}
\ee
The remaining six rescaled parameters of (\ref{PDE}),
\be
\fl
 \V
=
 \frac{V^2}{4 \mu D} ,
\quad
 \ell
=
 \frac{l}{\mu} ,
\quad
j_H = \frac{L_H}{L_I} ,
\quad
\n_H = \frac{N_H}{N_B} ,
\quad
\kappa
=
 K_{bg} z_B ,
\quad\mbox{and}\quad
 r
=
 \frac{R D Y N_B}{l z_B} ,
\Label{pardefs}
\ee
are all considered
to be $\Or(1)$
with respect to $\eps$
in the forthcoming analysis
(cf. \cite{ZDPS-2009}).

\begin{figure}[t]
\hspace*{-2cm}
\scalebox{0.45}[0.4]{
\includegraphics{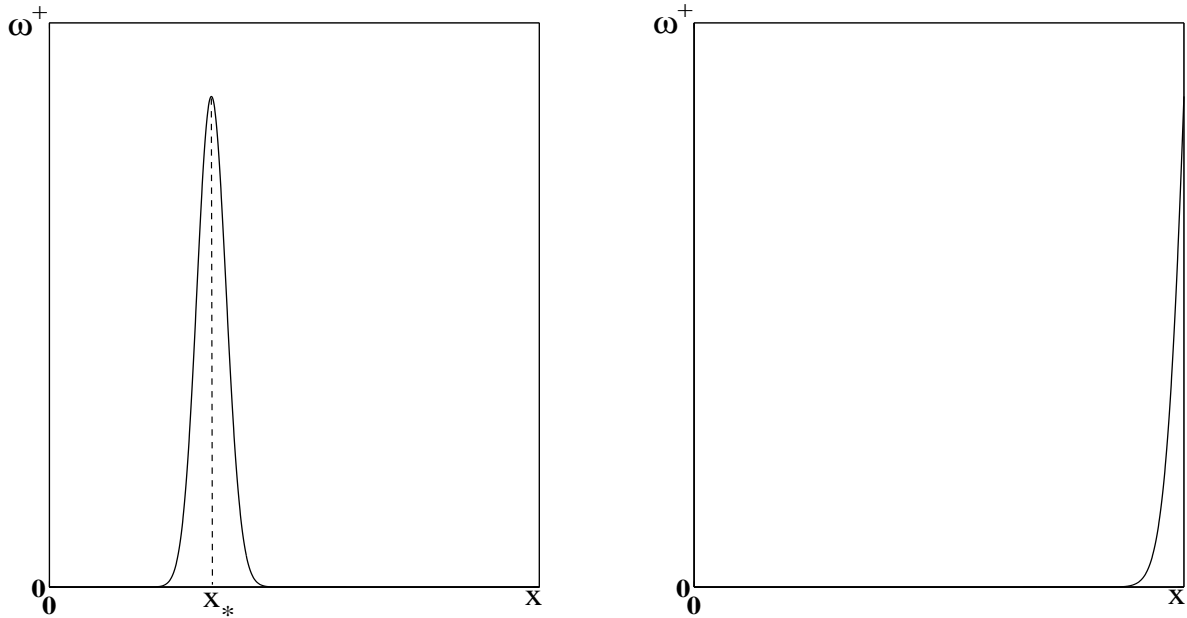}
}
\caption{
\label{f-profiles}
\emph{Left panel}:
a DCM profile
for the planktonic component
of (\ref{PDE})--(\ref{BC}).
Essentially all plankton
is concentrated in
an $\Or(\eps^{1/4})$ region
around a finite depth $\xs$.
\emph{Right panel}:
a BL profile
for the planktonic component
of (\ref{PDE})--(\ref{BC}).
Here,
essentially all plankton
is concentrated in
an $\Or(\eps^{1/2})$ region
over the depth of the basin.
}
\end{figure}

Our attempt to comprehend
the mechanism underpinning
the appearance of
phytoplankton patterns,
as well as
the character of such patterns,
begins with the determination
of the spectral stability
of the trivial steady state
$\upu = (0,0)^{\rm T}$.
At that state,
and in terms of
the original system (\ref{PDE-orig}),
there is no phytoplankton---$W(z,t) \equiv 0$---and
the nutrient concentration
remains constant
throughout the column---$N(z,t) \equiv N_B$,
the value at
the bottom of the basin (\ref{BC-dim}).
The system (\ref{PDE})
may be written compactly
in the form
\be
 \upu_\tau
=
 \Lfpu(\upu)
=
\left(
 \begin{array}{c}
 \eps \, \wpu_{xx}
-
 2 \sqrt{\eps \V} \, \wpu_x
+
 ( p(\wpu,\n,x) - \ell ) \, \wpu
\\
 \eps \, \n_{xx} + \eps \, \ell^{-1} \, p(\wpu,\n,x) \, \wpu
 \end{array}
\right) ,
\Label{L-def}
\ee
where
\[
 \upu
=
\left(
 \begin{array}{c}
 \omega^+ \\ \eta
 \end{array}
\right) .
\]
Here,
the nonlinear operator $\Lfpu$
is densely defined in
${\rm L}^2(0,1) \times {\rm L}^2(0,1)$.
The associated spectral problem
has been investigated
in full asymptotic detail
in \cite{ZDPS-2009},
where
we worked with
the linearization of (\ref{L-def})
around $\upu = (0,0)^{\rm T}$,
\be
 \DLfpu
=
\left(
 \begin{array}{cc}
 \eps \, \dxx - 2 \sqrt{\eps \V} \, \dx + f - \ell & 0
\\
 \eps \, \ell^{-1} \, f & \eps \, \dxx
 \end{array}
\right) ,
\Label{DL-def}
\ee
in which
\be
 f(x)
=
 \frac{\nu}{1 + j_H \ex^{\kappa x}}
\quad\mbox{and}\quad
\nu = \frac{1}{1+\n_H} \in (0,1) .
\Label{f}
\ee
The spectrum
$\sigma(\DLfpu)
=
\{ \nu_n \}_{n \ge 0} \cup \{ \lambda_n \}_{n \ge 0}$
of the operator $\DLfpu$
consists of two distinct, real parts
associated with the two diagonal blocks of $\DLfpu$,
cf.~(\ref{DL-def}).
Here,
the eigenvalues
$\nu_n = -\eps \, (n + 1/2)^2 \, \pi^2$
are negative,
independent of all parameters,
and associated with
the lower block.
These eigenvalues,
together with
the corresponding sinusoidal eigenfunctions
$(0,\cos((n+1/2) \pi x))^{\rm T}$,
describe nutrient diffusion
in the complete absence of phytoplankton.
It follows that
the spectral stability
of the trivial state
is governed solely
by $\{ \lambda_n \}_{n \ge 0}$,
the set of eigenvalues
associated with the upper block.
In \cite{ZDPS-2009},
we identified
two different linear destabilization mechanisms.
In the regime $\V < f(0) - f(1)$,
corresponding to reduced oceanic diffusivity
or increased turbidity
(cf. (\ref{pardefs}) and (\ref{f})),
the planktonic component $\omega^+_0$
of the eigenfunction $w_0^+$
associated with
the critical eigenvalue $\lambda_0$
has the character of a DCM:
$\omega^+_0$ is localized
in an $\Or(\eps^{1/4})$ region
centered around
a certain depth $\xs$
at which it attains its maximal value,
see Figure~\ref{f-profiles}.
This depth can be determined explicitly:
to leading order,
$f(\xs) = f(0) - \V$ \cite{ZDPS-2009}.
Hence,
$\xs$ increases monotonically
from $\xs = 0$ to $\xs = 1$
as $\V$ increases from $\V = 0$
to the transitional value $\V = f(0) - f(1)$.
In the complementary case $\V > f(0) - f(1)$,
corresponding to increased oceanic diffusivity
or decreased turbidity,
the planktonic component
of the critical eigenfunction
destabilizing the trivial state
has the character
of a BL:
that is,
it increases monotonically with depth
and
essentially all phytoplankton is concentrated
in an $\Or(\eps^{1/2})$ region
over the bottom,
see again Figure~\ref{f-profiles}.

In this article,
we focus exclusively
on the regime in which
DCMs may appear,
\emph{i.e.},
we assume throughout the article
that $\V < f(0) - f(1)$.
In that regime,
we investigate the nature
of the bifurcation
associated with
the destabilization mechanism
of DCM type.
We know from \cite{ZDPS-2009} that,
in this case,
\be
 \lambda_n
=
 \lambda_*
-
 \eps^{1/3} \sL^{2/3} \abs{A_{n+1}}
+
 \Or(\eps^{1/2}) ,
\Label{lambdan}
\ee
with
\be
 \lambda_*
=
 f(0) - \ell - \V
=
 \frac{\nu}{1 + j_H} - \ell - \V
\Label{lambda*}
\ee
and where
\be
 \sL
=
 F'(0)
=
 -f'(0)
=
 \frac{\kappa \, \nu \, j_H}{(1 + j_H)^2} ,
\quad\mbox{with}\quad
 F(x) = f(0) - f(x) .
\Label{defsigF}
\ee
Here,
$A_n < 0$ is
the $n-$th root
of $\Ai$,
the Airy function
of the first kind.
The bifurcation occurs
as $\lambda_0$ crosses zero,
yielding the bifurcation diagram
in the left panel of Figure~\ref{f-fullbif}.
More specifically,
we focus on
the (weakly nonlinear) dynamics
generated by (\ref{L-def})
for parameter choices such that
\be
 \lambda_0
=
 \frac{\nu}{1 + j_H}
-
 \ell
-
 \V
-
 \eps^{1/3} \sL^{2/3} \abs{A_1}
+
 \Or(\eps^{1/2})
=
 \eps^\rho \Lambda_0 ,
\Label{defalphaLamb}
\ee
where
$\rho > 0$ is fixed
and
$\Lambda_0$ is allowed to be
at most logarithmically large
with respect to $\eps$.
Note that
one can \emph{tune} the appearance
of a destabilization
of DCM type
(\emph{i.e.},
of the simplest phytoplankton pattern)
by choosing appropriately
the parameters in (\ref{L-def});
also,
that $\lambda_0$ depends
on all parameters
with the exception of $r$,
the rescaled self-shading coefficient,
see the definitions
of $f$ and $\sL$
in (\ref{f}) and (\ref{defsigF}).
We remark, further,
that the parameter $\V$
depends on
the diffusion coefficient $D$
(cf. (\ref{pardefs})),
the main parameter
varied in \cite{HPKS-2006}
and
the one that most strongly depends
on varying external conditions
such as global temperature \cite{SHSM-1998}.
Finally,
$\Lambda_0$ is an increasing function
of our bifurcation parameter $N_B$
through its dependence on $\nu$,
see (\ref{lambdan})--(\ref{lambda*})
together with the definitions
of $\nu$ in (\ref{f})
and
of $\eta_H$ in (\ref{pardefs}).
Based on this final observation,
we will often treat $\Lambda_0$
as our bifurcation parameter.

The first step
in analyzing the dynamics
generated by a linear destabilization mechanism
is to perform
a center manifold analysis
to determine
the local character
of the bifurcation
associated with the destabilization
(see, for instance, \cite{BJ-1989,C-1981}).
This is a well-established procedure.
In the setting of (\ref{defalphaLamb}),
this amounts to assuming
that $\lambda_0$ is (asymptotically) smaller
than all other eigenvalues,
and
it corresponds to
the case $\rho > 1$
and
$\Lambda_0 = \Or(1)$.
In this regime,
the remaining eigenvalues
$\{\nu_n\}_{n\ge0} \cup \{\lambda_n\}_{n\ge1}$
are negative and asymptotically larger
than $\lambda_0$,
so that
the local flow near
the trivial pattern $(0,0)^{\rm T}$
is determined by the flow
on the one-dimensional center manifold.
The tangent space
of this manifold
at the trivial steady state
is spanned by
the critical eigenfunction $w_0^+$
associated with $\lambda_0$.
Hence,
this flow can be determined
by expanding $\upu$ as
$\upu(x,\tau)
=
\eps^{\rho-1/6} \, \Omega_0(\tau) \, w_0^+(x) + R(x,t)$,
with $\Omega_0$ being
an unknown,
time-dependent amplitude
and
the higher order remainder $R$
encapsulating the component of $\upu$
along directions associated with
the stable eigenvalues---the additional $1/6$
in the exponent of $\eps$
follows from the projection analysis
by which the equation for $\Omega_0$ is determined
(see below and Section~\ref{s-a000}).
An ODE for the unknown amplitude $\Omega_0$
is obtained through a projection procedure
which is straightforward
but can nevertheless be highly technical,
especially in a PDE setting.
In the case at hand,
this equation reads
\be
 \dot{\Omega}_0
=
 \Lambda_0 \, \Omega_0 - a_{000}(0) \, \Omega_0^2 ,
\Label{cmred}
\ee
to leading order.
Thus,
the procedure reveals
the existence of
a \emph{nontrivial} fixed point
which is stabilized
through a standard,
co-dimension one
transcritical bifurcation.
This fixed point corresponds to
an asymptotically small DCM pattern,
the amplitude of which grows linearly
with $\Lambda_0$,
\be
 \wpu(x)
\sim
 \eps^{\rho-1/6} \, \Omega^*_0 \, \wpu_0(x) ,
\quad\mbox{with}\
 \Omega^*_0
=
 \frac{\Lambda_0}{a_{000}(0)} .
\Label{cmred-fp}
\ee
In general,
one cannot expect
to be able to compute
the coefficient $a_{000}(0)$
explicitly.
Here,
we exploit
the singularly perturbed nature of (\ref{L-def})
and
the localized character
of the eigenfunction $w_0^+$
to do exactly that;
in particular,
it follows
from the analysis
to be presented
in this article
that
\be
 a_{000}(0)
=
 (1 -\nu) (1-\xs)
\,
\frac{
 \sL^{1/3} \, f(0) \, \exp(\abs{A_1}^{3/2})
}{
\left(
 \abs{f'(\xs)} \int_{A_1}^\infty \Ai^2(s) \, ds
\right)^{1/2}
}
>
 0 ,
\Label{a000(0)}
\ee
see Section~\ref{s-a000}.
In addition to yielding an explicit,
leading order formula
for the amplitude
of the emerging (stable) DCM,
this first result also implies
that this DCM
is ecologically relevant
since the planktonic component
of the primary eigenfunction is positive,
$\wpu_0 > 0$,
and hence also $\wpu > 0$
by (\ref{cmred-fp})--(\ref{a000(0)}).

The main aim of this paper
is to develop an analytic approach
through which one can go \emph{beyond}
the direct, finite-dimensional
center manifold reduction
outlined above.
The original ideas
underlying this approach---namely,
the method of
\emph{weakly nonlinear stability analysis}---qualify
as classical \cite{WMZ-1994}.
However,
this particular method
does not always provide more insight
than the rigorously established
center manifold reduction method:
for instance,
it also reduces the flow
to a one-dimensional ODE
of the form (\ref{cmred}).
The situation is strikingly different here,
as we can exploit
the singularly perturbed nature of (\ref{L-def}),
in conjunction with
the asymptotic information
on the eigenfunctions of $\DLfpu$
obtained in \cite{ZDPS-2009},
to study in full analytic detail
the case $\lambda_0 = {\cal O}(\eps)$---see
Section~\ref{s-emerge}---and even
extend our analysis
to the regime
$\lambda_0 = {\cal O}(\eps\log^2\eps)$---see
Section~\ref{ss-destroy}.
This way,
we can analytically trace the fate
of the bifurcating DCM pattern
well into the regime
where the pattern undergoes
secondary and, possibly, even tertiary bifurcations.

For clarity of presentation,
we divide the rest of the material
in this Introduction
into two parts.
The first one
focuses on the bifurcations
undergone by the DCM patterns
and on
the ecological interpretation
of our findings,
while the second one
focuses on the specifics
of the asymptotic method
developed in this work.
%

\subsection{The bifurcations of the DCM patterns}
\begin{figure}[t]
\hspace*{-1.6cm}
\scalebox{0.43}[0.4]{
\includegraphics{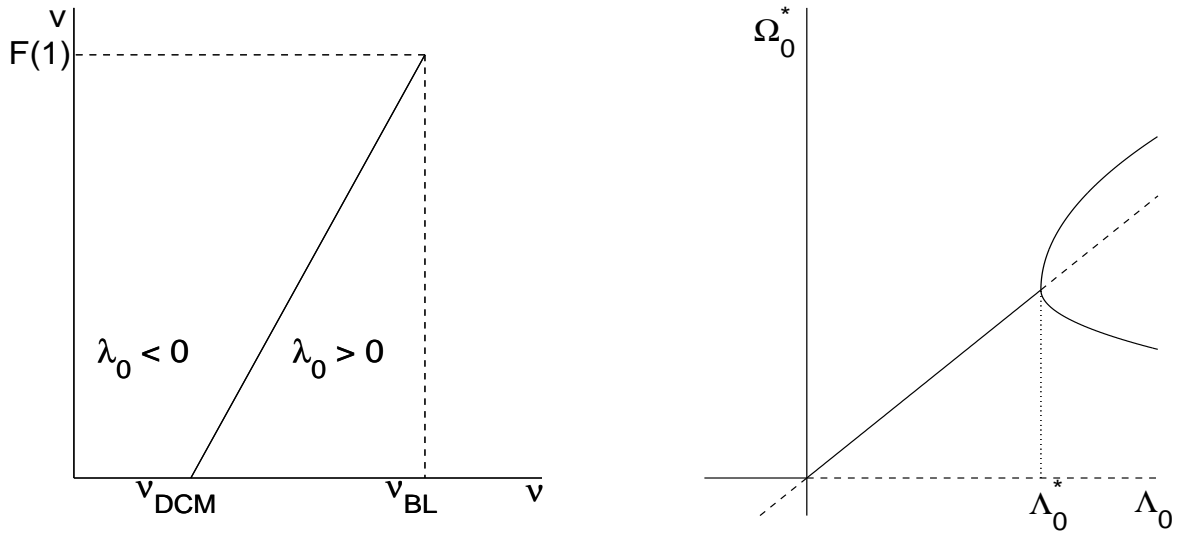}
}
\caption{
\label{f-fullbif}
\emph{Left panel}:
the bifurcation diagram
for the trivial steady state
of (\ref{PDE})
in the regime $\V < f(0) - f(1) = F(1)$.
The trivial steady state is stable
in the region $\lambda_0 < 0$
and
unstable in the region $\lambda_0 > 0$.
Here,
$\nu_{DCM} = \ell (1 + j_H)$
and
$\nu_{BL} = \ell (1 + {\rm e}^\kappa j_H)$.
\emph{Right panel}:
the bifurcation diagram
for the small-amplitude DCM
reported in (\ref{cmred}) and (\ref{a000(0)}).
The origin marks
the transcritical bifurcation
through which
the trivial steady state is destabilized
and
the small-amplitude DCM pattern emerges.
The value $\Lambda_0^*$ marks
the (first) Hopf bifurcation
where
this small-amplitude DCM
is destabilized
and
a time-periodic DCM pattern is generated.
}
\end{figure}

The outcome
of our asymptotic analysis
is summarized in
the right panel of Figure~\ref{f-fullbif}.
The localized DCM
that  bifurcates
as $\lambda_0$ crosses zero
is a stable attractor
of the flow
generated by (\ref{PDE-orig}),
for all $\rho > 1$
and
$\Lambda_0 = {\cal O}(1)$
with respect to $\eps$,
cf. (\ref{defalphaLamb}).
As we remarked above,
the amplitude $\Omega_0^*$ of this localized DCM,
and thus also
the biomass associated with it,
grows linearly with $\Lambda_0$
in that regime,
cf. (\ref{cmred-fp})--(\ref{a000(0)}).
Quite remarkably,
from the point of view
of our weakly nonlinear stability analysis,
$\Omega_0^*$ continues growing
linearly with $\Lambda_0$
also beyond the region
where the center manifold reduction is valid.
In particular,
(\ref{cmred-fp})--(\ref{a000(0)}) remain valid
in the regime
$\rho = 1$ and $\Lambda_0 = \Or(1)$,
see (\ref{fp-Labda0=O(1)}).
The corresponding biomass
turns out to be
\be
 \int_0^1
 \wpu(x)
\,
 dx
=
 \eps
\frac{(1 + j_H)}{(1 - \nu) \, \nu \, (1-\xs)}
\,
 \Lambda_0
=
\frac{
 (1 + j_H) \, \nu - \ell - \V
}{
 (1 - \nu) \, (1-\xs) \, (\ell + \V)
} ,
\Label{biomass}
\ee
to leading order.
This second result establishes that,
in the $\lambda_0 = {\cal O}(\eps)$ regime,
the DCM pattern grows with $\nu$
and hence also with $N_B$,
the primary parameter
measuring nutrient availability
in the water column
(see (\ref{pardefs}) and (\ref{f})).
This fact certainly reinforces
our ecological intuition.

The stability properties
of the DCM mode
corresponding to $\Omega_0^*$,
on the other hand,
undergo a drastic change
in that same regime.
Our rather involved stability analysis
of this emergent nontrivial steady state
reveals that it becomes \emph{unstable},
in this same
$\lambda_0 = {\cal O}(\eps)$ regime already,
as $\Lambda_0$ continues to grow
and through a standard \emph{Hopf} bifurcation;
this is our third result.
The appearance of this secondary bifurcation
can be determined explicitly
by our methods
and,
as we demonstrate,
its onset occurs for values of $\Lambda_0$
which increase unboundedly as $\xs \downarrow 0$
(equivalently, as $\V \downarrow 0$).
It is natural, then,
to attempt an extension
of our analysis
into a region where $\Lambda_0 \gg 1$.
In that regime,
we establish the existence
of a \emph{second} localized DCM-type pattern:
the associated reduced system
has two critical points.
Using our methods,
we trace this second localized structure
back to $\Or(1)$ values of $\Lambda_0$
and find that
it corresponds to
an $\Or(\eps^{1/2})$ biomass
depending \emph{nonlinearly} on $\Lambda_0$.
This is our fourth result.
The stability type of this pattern
can be also determined explicitly,
although we do not undertake this task
in the present work.

Hence,
our analysis yields
that the stationary, stable, localized DCM pattern
emerging at the transcritical bifurcation
through which
the trivial state becomes unstable
only persists
in an asymptotically small,
${\cal O}(\eps)$ region
in parameter space
before it yields to an oscillatory pattern
emerging through a Hopf bifurcation.
This fact reinforces our mathematical intuition
that the appearance
of this stationary DCM
is the first step
in a cascade of bifurcations
leading to the chaotic dynamics
reported in \cite{HPKS-2006}---see also
our discussion in \cite{ZDPS-2009}.
In light of this,
our analytical findings
seem to agree qualitatively
with these numerical results.
In the same vein,
our findings here suggest
that the chaotic dynamics
can be traced back
to the small amplitude patterns
emerging from the destabilization
of the trivial steady state.
(Of course,
one must always exercise caution
in interpreting numerical observations
from an asymptotic point of view,
especially when these simulations
concern an \emph{unscaled} system
as is the case here:
the authors of \cite{HPKS-2006}
have simulated the original system (\ref{PDE-orig})
and \emph{not}
the scaled system (\ref{PDE}).)
Additionally,
the fact that the onset
of the Hopf bifurcation
for $\V \downarrow 0$
occurs in the regime $\Lambda_0 \gg 1$---where
certain higher order terms
in our analysis
become leading order
and hence
the analysis must be necessarily extended---possibly
explains the absence
of oscillatory and chaotic dynamics
for small values of $\V$,
see \cite[Figure~3.3]{ZDPS-2009}.

Naturally,
the questions on the fate
of the oscillatory pattern
generated through the Hopf bifurcation
and
on the nature of
the larger DCM pattern
are intriguing.
At present,
this is the subject
of ongoing research.
We do not pursue
these questions further
in this article,
apart from a short discussion
in its concluding section.

\subsection{The asymptotic method}
Parallel to understanding the character and fate
of the linear destabilization mechanism
established in \cite{ZDPS-2009},
this article has a second---and
from a mathematical point of view
at least equally important---theme.
Here,
we have developed
a powerful approach
by which we can study
the weakly nonlinear dynamics
generated by (\ref{PDE})
in full asymptotic detail
and
far from the region
covered by more standard techniques
(such as
the center manifold
reduction method).
Indeed,
one cannot hope in general
to extend the analysis
beyond the one-dimensional
center manifold reduction discussed above
and into the regime
where $\lambda_0$ is \emph{not}
asymptotically closer to zero
than all other eigenvalues.
In other words,
the sole analytical insight
into the dynamics
of the flow
near the destabilization
that one can generically obtain
is the confirmation that
DCMs indeed appear
through a transcritical bifurcation.
Let us look into this last point
in more detail
and for
our specific model (\ref{PDE})--(\ref{BC}).
For $\lambda_0 = {\cal O}(\eps)$---equivalently,
for $\rho = 1$ in (\ref{defalphaLamb})---one
can no longer `project away'
the directions corresponding to the eigenvalues
$\nu_n = -\eps \, (n+1/2)^2 \, \pi^2$
associated with the operator $\DLfpu$.
Indeed,
these are ${\cal O}(\eps)$
for ${\cal O}(1)$ values of $n$,
and hence
of the same asymptotic magnitude as $\lambda_0$.
As a result,
the center manifold reduction approach
yields a leading order system
in at least
asymptotically many dimensions.
In general,
such a system
cannot be studied analytically,
and
one has to abandon
the idea of performing
an asymptotically accurate analysis.

The crucial ingredient
in our approach
is our ability to explicitly determine,
to leading order,
all relevant coefficients
in the reduced,
asymptotically high-dimensional system
that extends
the leading order,
one-dimensional
center manifold reduction.
All of these coefficients
are defined in a relatively standard manner
in terms of projections
based on the linear spectral analysis,
see (\ref{ab-coeffs}) in Section~\ref{s-WNLA}.
We report the outcome
of this part of our work
in (\ref{coeffs-ae}).
These leading order formulas
clearly reveal a certain structure
in these coefficients,
which in turn reflects on
the system of ODEs
for the Fourier modes.
It is this structure
that allows us to extend
our stability and bifurcation analysis.
The sometimes remarkably subtle
and laborious analysis
by which these coefficients are computed
provides the foundation
for the strength and success
of our program.
Therefore,
this analysis is
a central component
of our approach
and
lies at the core
of the forthcoming presentation,
see especially Sections~\ref{s-a000}
and \ref{s-bm00}--\ref{s-bm0k}.

An understanding of the conditions
under which similar structure
may be expected to appear
is apposite to deciphering
the fundamental mechanisms
underpinning the success of our method
and
to determining a more general setting
where this method is applicable.
Naturally,
what enables us
to compute these coefficients,
and thus also determine
how they are related,
is the accurate asymptotic control
over the eigenfunctions
that we establish.
It is neither clear,
\emph{a priori},
that the structure
present in the reduced system
is a necessary consequence
of that control,
nor how much of that control
is necessary to establish
the presence of sufficient structure.
These issues are the subject
of current research
undertaken by the authors.
Below we offer a brief sketch
of the ideas behind
this work in progress,
as it also encapsulates
the essentials of the method
developed in the present work.

To avoid the computational complexities
associated with the weakly nonlinear analysis,
we consider a much simpler,
autonomous,
coupled,
reaction--diffusion system,
\be
\begin{array}{lcrll}
 U_t
&=&
 U_{xx}
&+
 \mu \, U
&+
 F(U, V; \eps) ,
\\
 V_t
&=&
\eps
\left(
 V_{xx}
\right.
&+
 \nu \, V
&+
\left.
 G(U; V,\eps)
\right) .
\end{array}
\Label{model}
\ee
Here,
$U$ and $V$ are defined in
$[0,1] \times \mathbb{R}^+$
and obey certain boundary conditions,
\emph{e.g.},
of homogeneous Neumann or Dirichlet type.
The nonlinearities $F(U,V)$ and $G(U,V)$
are assumed to be smooth
and
at least quadratic in $U$ and $V$;
finally,
$0 < \eps \ll 1$ is
an asymptotically small parameter.
The spectral problem
associated with the trivial state
$(U,V) \equiv (0,0)$
decouples into two scalar problems
of harmonic oscillator type.
It immediately follows that,
for $\nu$ below a certain critical value $\nu^*$,
this trivial state loses stability
when $\mu$ crosses
a threshold $\mu^*$.
Moreover,
the eigenvalues $\{\lambda^U_n\}_{n \ge 0}$
associated with the $U-$component
(and hence also with $\mu$)
are ${\cal O}(1)-$apart,
while the eigenvalues $\{\lambda^V_n\}_{n \ge 0}$
associated with the $V-$component
(and also with $\nu$)
are ${\cal O}(\eps)-$apart.
Both sets of eigenvalues
are naturally paired with
simple trigonometric eigenfunctions.
A straightforward center manifold reduction
suffices to determine
the nature of the bifurcation
as $\mu$ crosses $\mu^*$
and
{\it in the regime $\mu -\mu^* \ll \eps$}.
This situation corresponds directly
to our---technically more involved---center
manifold problem (\ref{cmred})--(\ref{a000(0)})
briefly discussed earlier.
Note that, here,
the leading order analog
of the DCM pattern
identified in that discussion
is a sinusoidal function.

As long as $\mu -\mu^* \ll \eps$,
the modes associated with
the eigenvalues $\{\lambda^V_n\}_n$
remain slaved to
the critical $\lambda^U_0$-mode,
exactly as in
our phytoplankton--nutrient model.
However,
this is no longer the case
when $\mu -\mu^* = {\cal O}(\eps)$;
in that regime,
asymptotically many $\lambda^v_n$-modes
are nonlinearly triggered
by that critical mode.
Nevertheless,
the remaining $\lambda^U_n$-modes
stay slaved,
so that
one obtains a reduced system
of asymptotically high dimension.
Here also,
the coefficients
of the leading nonlinearities
can all be expressed
in terms of projections
along the eigenmodes,
albeit they correspond to
much simpler integrals.
This process should enable us
to study the conditions
under which
one is able to infer relations
between these coefficients
similar to those reported in (\ref{coeffs-ae}).
This, in turn,
should lead to conditions
under which
the reduced system
has sufficient structure
to allow a secondary bifurcation analysis---and
perhaps even the identification of
a cascade of subsequent bifurcations---of
the nontrivial state bifurcating at $\mu = \mu^*$.
An additional benefit
of working in
a simple setting
of this sort
is its amenability
to \emph{rigorous} analysis,
which is beyond the scope
of this article.

A natural question to ask
at this point
is whether
the model problem (\ref{model}) shares
enough structure with (\ref{PDE})
to enjoy similarly complex
yet tractable dynamics.
Note, in particular,
the absence of
nonlocal and non-autonomous terms
from (\ref{model}).
Mathematically speaking,
we expect these aspects
to be insignificant
for the type of dynamics
that the model exhibits
close to bifurcation.
(The situation is very different
from the ecological point of view,
naturally.)
In the setting of (\ref{PDE}),
the nonlocality only complicates our analysis
and thus clouds our understanding
of the secondary and subsequent bifurcations
beyond the center manifold reduction.
Indeed, one expects the self-shading effect
that a \emph{small} DCM pattern has on itself
to be much smaller
than the shading
due to the water column above it.
This is most evident
in Sections~\ref{s-a000} and \ref{s-a00k},
where self-shading
(quantified by the parameter $r$)
is finally shown to contribute
higher order terms only.
Similarly,
the sole role of
the non-autonomous features of (\ref{PDE-orig})
is seemingly to introduce two spectra,
$\{\nu_n\}_n$ and $\{\lambda_n\}_n$,
with different asymptotic properties.
In our model problem (\ref{model}),
this is achieved instead
by choosing disparate diffusivities
for the two model components.

Finally, it should be noted
that our work resembles,
but is certainly not identical to,
Lange's work in
\cite{L-1981a,LK-1985}.
Lange has devised
a powerful asymptotic method
applicable to problems with
closely spaced branch points
which allows one to track
the evolution of solution branches
well into the regime where
center manifold reduction breaks down.
In our work also,
the spectrum is asymptotically closely spaced,
as are also then
the branch points.
Nevertheless,
the differences
between our work
and
the work in \cite{L-1981a}
are substantial.
Most prominently,
Lange essentially defines branch points
as points in parameter space
where the linearization
around the steady state
admits a zero eigenvalue,
see the derivation of \cite[(3.10)]{L-1981a} in particular.
In our work, instead,
the secondary bifurcation
is induced by
the \emph{parameter-independent} negative spectrum
related to pure diffusion
and
occur before any eigenvalues
other than $\lambda_0$
have bifurcated.
As such,
these branch points
are \emph{not} captured
by Lange's method.
In fact,
this secondary bifurcation---and,
we expect,
part of the cascade
toward chaotic dynamics---occurs
in a region of parameter space
which is asymptotically small
compared to the magnitude of
the next critical eigenvalue
$\lambda_1$.
Viewed from this perspective,
then,
the existence of
the rich dynamics
reported here for the regime
$\lambda_0 = {\cal O}(\eps)$
acts as a paradigmatic manifestation
of nonlinear interactions.
The \emph{linearly stable} modes
manage to have
a decisive impact
on the dynamics
solely through nonlinear couplings
and although
a \emph{strictly linear} point of view
dictates that these modes
should be utterly irrelevant.

\section{Evolution of the Fourier coefficients
\label{s-WNLA}}
\setcounter{equation}{0}
Our aim in this section
is to write the PDE system (\ref{L-def})
as an infinite-dimensional system
of nonlinear ODEs
and
subsequently reduce it
by relaxing
the fast stable directions.
To achieve this,
we need explicit formulas
for the (point) spectrum $\sigma(\DLfpu)$,
as well as
for the corresponding eigenbasis
and its dual.
The spectrum and the eigenbasis
have been determined
in \cite{ZDPS-2009};
we summarize
the relevant formulas
in Section~\ref{ss-eigen} below.
We then obtain
the dual basis
in Section~\ref{ss-dual}
by solving the eigenproblem
for the adjoint operator $(\DLfpu)^*$.
Finally,
in Section~\ref{ss-evol},
we derive the desired ODEs
for the Fourier coefficients
close to the bifurcation point.

\subsection{The spectrum and the corresponding eigenbasis of $\DLfpu$
\label{ss-eigen}}
For completeness,
we let
$\mathcal{H}_{\wpu}$
and
$\mathcal{H}_\n$
be the subspaces of $L^2(0,1)$
associated with
the boundary conditions (\ref{BC}),
$\mathcal{H}_\w$
be associated with
the boundary conditions
\be
 ( \dx \w
-
 \sqrt{\V/\eps} \, \w )(0)
=
 ( \dx \w
-
 \sqrt{\V/\eps} \, \w )(1)
=
 0 ,
\Label{BCt}
\ee
and we write
$\mathcal{H}_+
=
\mathcal{H}_{\wpu} \times \mathcal{H}_\n$
and
$\mathcal{H}
=
\mathcal{H}_\w \times \mathcal{H}_\n$.
Both product spaces
can be equipped
with the inner product
\[
 \langle \upu_1 , \upu_2 \rangle
=
 \left\langle
 \left(\begin{array}{c}\wpu_1 \\ \n_1\end{array}\right)
,
 \left(\begin{array}{c}\wpu_2 \\ \n_2\end{array}\right)
 \right\rangle
=
 \int_0^1
 \Big(\wpu_1(x) \, \wpu_2(x) + \n_1(x) \, \n_2(x)\Big)
 dx .
\]
Subsequently,
we define
the function
$E(x)
=
\exp(\sqrt{\V/\eps} \, x)$
and the operator
$\mathcal{E}: \mathcal{H} \to \mathcal{H}_+$
corresponding to an application
of the Liouville transform,
\be
\fl
 \mathcal{E} u
=
\left(
\begin{array}{c}
 E \, \w
\\
 \n
\end{array}
\right)
=
\left(
\begin{array}{c}
 \wpu
\\
 \n
\end{array}
\right)
=
 \upu
\ \mbox{with inverse} \
 u
=
\left(
\begin{array}{c}
 \w
\\
 \n
\end{array}
\right)
=
\left(
\begin{array}{c}
 \wpu/E
\\
 \n
\end{array}
\right)
=
 \mathcal{E}^{-1} \upu .
\Label{Liouville}
\ee
(It is straightforward to check
that the boundary conditions (\ref{BC}) for $u$
yield the boundary conditions (\ref{BCt}) for $\upu$.)
Both $\mathcal{E}$ and $\mathcal{E}^{-1}$
are self-adjoint and bounded
and
\be
 \DLf
=
 \mathcal{E}^{-1} \DLfpu \mathcal{E}
=
\left(
 \begin{array}{cc}
 \eps \dxx + f - \ell - \V & 0
\\
 \eps \ell^{-1} f E & \eps \dxx
 \end{array}
\right) ,
\Label{DLt-def}
\ee
with $\DLf$ densely defined
and
having self-adjoint diagonal blocks.

As mentioned in the Introduction,
the eigenvalues $\nu_n$
associated with $\DLfpu$
correspond
to the pure diffusion problem
for the nutrient
in the absence of plankton.
In particular,
they are solutions to
the eigenvalue problem
\[
 \eps \dxx\s_n
=
 \nu_n \s_n ,
\quad\mbox{with}\quad
 \dx\s_n(0)
=
 \s_n(1)
=
 0 ,
\]
and
may be calculated explicitly,
\be
 \nu_n
=
 -\eps N_n ,
\quad\mbox{where}\quad
 N_n
=
 (\pi/2 + n\pi)^2
\quad\mbox{for} \
 n \ge 0 .
\Label{Nn-def}
\ee
The corresponding eigenfunctions
have a zero $\wpu-$component,
and
they are
\be
 v_n
=
\left(
\begin{array}{c}
 0 \\ \s_n
\end{array}
\right),
\quad \mbox{where} \
 \s_n(x)
=
 \sqrt{2} \cos(\sqrt{N_n} \, x) .
\Label{vs}
\ee
These are normalized so that
$\norm{\s_n}_2 = 1$.

The eigenvalues $\lambda_n$,
on the other hand,
correspond to the eigenvectors
\[
 w^+_n
=
\left(
\begin{array}{c}
 \wpu_n \\ \n_n
\end{array}
\right) .
\]
Here,
the functions
$\wpu_n$ and $\n_n$
are solutions to
\[
\begin{array}{rcl}
 \eps \, \dxx\wpu_n
-
 2 \sqrt{\eps \V} \, \dx\wpu_n
+
(f(x) - \ell - \lambda_n) \, \wpu_n
&=&
 0 ,
\\
 ( \dx \wpu_n
-
 2 \sqrt{\V/\eps} \, \wpu_n )(0)
=
 ( \dx \wpu_n
-
 2 \sqrt{\V/\eps} \, \wpu_n )(1)
&=&
 0 ,
\end{array}
\]
cf. (\ref{DL-def}),
together with
the self-adjoint, inhomogeneous,
boundary value problem
for the component $\n_n$,
\be
 \eps \, \dxx\n_n
-
 \lambda_n \, \n_n
=
 -\eps \ell^{-1} f \, \wpu_n ,
\quad\mbox{where} \
 \dx \n_n(0)
=
 \n_n(1)
=
 0 .
\Label{ODE-eta}
\ee
Equivalently,
they are solutions to
the self-adjoint,
Sturm--Liouville problem
\be
\begin{array}{rcl}
 \eps \, \dxx\w_n
+
 (f(x) - \ell - \V - \lambda_n) \, \w_n
&=&
 0 ,
\\
 ( \dx \w_n
-
 \sqrt{\V/\eps} \, \w_n )(0)
&=&
 ( \dx \w_n
-
 \sqrt{\V/\eps} \, \w_n )(1)
=
 0 ,
\end{array}
\Label{ODEt-omega}
\ee
cf. (\ref{Liouville})--(\ref{DLt-def}).
As already stated,
in \cite{ZDPS-2009}
we derived
the asymptotic expressions
\[
 \lambda_n
=
 \lambda_*
-
 \eps^{1/3} \sL^{2/3} \abs{A_{n+1}}
+
 \Or(\eps^{1/2}) ,
\quad\mbox{with}\
 n \ge 0 ,
\]
cf. (\ref{lambdan}).
Here,
$\lambda_* = f(0) - \ell - \V$,
$\sL = F'(0) = -f'(0)$,
and
$A_n < 0$
is the $n-$th root
of the Airy function $\Ai$,
cf. (\ref{defsigF}).
A formula for the $n-$th eigenfunction $\w_n$
can also be derived
using the WKB method,
cf. \cite{ZDPS-2009}.
The corresponding eigenfunctions for $\DLfpu$ are
$w_n^+ = (\wpu_n , \n_n)^{\rm T}$,
where
$\wpu_n = E \, \w_n$---cf. (\ref{Liouville}).
As we will see
in the next section,
it is natural to impose
the normalization condition
$\norm{\w_n}_2 = 1$.

\subsection{The dual eigenbasis of $\DLfpu$
\label{ss-dual}}
To carry out
the weakly nonlinear stability analysis
of the bifurcating DCM profile,
we also need to obtain
the dual eigenbasis
$\{\hat{w}^+_n\}_{n\ge0} \cup \{\hat{v}_n\}_{n\ge0}$
uniquely determined
by the conditions
\[
 \langle w^+_n , \hat{w}^+_m \rangle
=
 \langle v_n , \hat{v}_m \rangle
=
 \delta_{nm}
\quad\mbox{and}\quad
 \langle w^+_n , \hat{v}_m \rangle
=
 \langle v_n , \hat{w}^+_m \rangle
=
 0 ,
\]
for all
$n, m \ge 0$.
In this section,
we show that
\be
 \hat{w}^+_n
=
\left(
\begin{array}{c}
 \wmu_n \\ 0
\end{array}
\right)
\quad\mbox{and}\quad
 \hat{v}_n
=
\left(
\begin{array}{c}
 \ymu_n \\ \s_n
\end{array}
\right) .
\Label{dual}
\ee
Here,
$\wmu_n \equiv \w_n/E$,
where
$\w_n$ solves the eigenvalue problem (\ref{ODEt-omega})
and
satisfies the normalization condition
$\norm{\w_n}_2 = 1$.
Further,
expressions for the functions
$\{\s_n\}_n$
were reported in (\ref{vs}),
while
the functions
$\{\ymu_n\}_n$
may be found
by solving
the inhomogeneous problem
\be
\begin{array}{rcl}
 \eps \, \dxx\ymu_n
+
 2 \sqrt{\eps \V} \, \dx\ymu_n
+
 (f(x) - \ell - \nu_n) \, \ymu_n
&=&
 -\eps \ell^{-1} f \, \s_n ,
\\
 \dx \ymu_n(0)
=
 \dx \ymu_n(1)
&=&
 0 .
\end{array}
\Label{ODE-psi}
\ee
Alternatively,
$\ymu_n = \y_n/E$,
where
$\y_n$ solves
the self-adjoint inhomogeneous problem
\be
\begin{array}{rcl}
 \eps \, \dxx\y_n
+
 (f(x) - \ell - \V - \nu_n) \, \y_n
&=&
 -\eps \ell^{-1} f E \, \s_n ,
\\
 ( \dx \y_n
-
 \sqrt{\V/\eps} \, \y_n )(0)
&=&
 ( \dx \y_n
-
 \sqrt{\V/\eps} \, \y_n )(1)
=
 0 .
\end{array}
\Label{ODEt-psi}
\ee

To verify the above,
we start from the observation
that the dual basis
may be obtained
by solving
the corresponding eigenvalue problem
for $(\DLfpu)^*$,
the adjoint of
the operator $\DLfpu$.
To calculate $(\DLfpu)^*$,
we write $\vmu = \mathcal{E}^{-1} v$,
recall (\ref{DLt-def}),
and note that
\[
\fl
 \langle \DLfpu \upu , \vmu \rangle
=
 \langle \DLfpu \mathcal{E} u , \vmu \rangle
=
 \langle \mathcal{E} \DLf u , \vmu \rangle
=
 \langle \DLf u , \mathcal{E} \vmu \rangle
=
 \langle \DLf u , v \rangle
=
 \langle u , \DLf^* v \rangle .
\]
This implies, further, that
\[
 \langle \DLfpu \upu , \vmu \rangle
=
 \langle u , \DLf^* v \rangle
=
 \langle \mathcal{E}^{-1} \, \upu , \DLf^* \mathcal{E} \vmu \rangle
=
 \langle \upu , \mathcal{E}^{-1} \DLf^* \mathcal{E} \vmu \rangle ,
\]
whence
$(\DLfpu)^* = \mathcal{E}^{-1} \DLf^* \mathcal{E}$.
Here,
$\upu$ satisfies
the boundary conditions
(\ref{BC}),
whereas
the boundary conditions
for $\vmu$
are determined from
$\vmu = \mathcal{E}^{-1} v$
and
the boundary conditions (\ref{BCt})
for $v$---in
particular,
\be
\fl
 \dx \ymu(0)
=
 \dx \ymu(1)
=
 0
\quad\mbox{and}\quad
 \dx\s(0)
=
 \s(1)
=
 0 ,
\quad\mbox{where}\
 v
=
\left(
\begin{array}{c}
 \ymu \\ \s
\end{array}
\right) .
\Label{BCt*}
\ee
It is straightforward to show that
\[
 \DLf^*
=
\left(
 \begin{array}{cc}
 \eps \dxx + f - \ell - \V
&
 \eps \ell^{-1} f E
\\
 0 & \eps \dxx
 \end{array}
\right) ,
\]
and,
since also
$(\DLfpu)^* = \mathcal{E}^{-1} \DLf^* \mathcal{E}$,
\be
 (\DLfpu)^*
=
\left(
 \begin{array}{cc}
 \eps \, \dxx + 2 \sqrt{\eps \V} \, \dx + f - \ell
&
 \eps \ell^{-1} f
\\
 0 & \eps \, \dxx
 \end{array}
\right) .
\Label{DL*}
\ee

In view of (\ref{DL*}),
the eigenvalue problem
$(\DLfpu)^* \hat{w}^+_n = \lambda_n \hat{w}^+_n$
for $\hat{w}^+_n = (\hwpu_n , \hn_n)^{\rm T}$
reads
\begin{eqnarray*}
 \eps \, \dxx\hwpu_n
+
 2 \sqrt{\eps \V} \, \dx\hwpu_n
+
 (f - \ell - \lambda_n) \, \hwpu_n
=
 -\eps \ell^{-1} f \, \hn_n ,
\\
 \eps\dxx\hn_n
=
 \lambda_n \, \hn_n ,
\end{eqnarray*}
subject to
the boundary conditions
(\ref{BCt*}).
The latter equation yields immediately $\hn_n \equiv 0$,
so that
the former equation becomes homogeneous.
It is now trivial to check that
$\hwpu_n = \wmu_n \equiv \w_n/E$,
where
$\w_n$ solves
the eigenvalue problem (\ref{ODEt-omega}).
This establishes
the first part of (\ref{dual}).

Similarly,
(\ref{DL*}) shows that
the eigenvalue problem
$(\DLfpu)^* \hat{v}_n = \nu_n \hat{v}_n$
has solutions
\[
 \hat{v}_n
=
\left(
\begin{array}{c}
 \ymu_n \\ \s_n
\end{array}
\right) ,
\]
where
the functions
$\{\ymu_n\}_n$
satisfy the boundary value problem (\ref{ODE-psi}).
An application
of the Liouville transform
$\y_n = E \ymu_n$
leads directly to
the self-adjoint problem (\ref{ODEt-psi}).

\subsection{Evolution of the Fourier coefficients
\label{ss-evol}}
Our aim in this section
is to write the PDE system (\ref{L-def})
as an infinite-dimensional system
of nonlinear ODEs.
We start by expanding
the solution of $\partial_\tau \upu = \Lfpu(\upu)$
in terms of the eigenbasis
associated with the
linear stability problem,
\be
 \upu(x,\tau)
=
 \eps^{c-1/6} \, \delta
 \sum_{n\ge 0}
 \Omega_n(\tau) \, w^+_n(x)
+
 \eps^c
 \sum_{n\ge 0}
 \Psi_n(\tau) \, v_n(x) ,
\Label{e-decomp}
\ee
where
$c > 0$
is yet undetermined
and
the coefficients
$\Omega_n$ and $\Psi_n$
are determined by
\be
 \Omega_n
=
 \eps^{-c} \, \delta^{-1}
 \langle \upu , \hat{w}^+_n \rangle
\quad\mbox{and}\quad
 \Psi_n
=
 \eps^{-c}
 \langle \upu , \hat{v}_n \rangle .
\Label{h,h'}
\ee
The exponent of $1/6$
in the first sum of (\ref{e-decomp})
is related to the localized nature of $\wpu_0$,
the planktonic component of $w^+_0$.
In particular,
$\wpu_0$ is shaped
as a DCM
with an $\Or(\eps^{1/6})$ biomass $\norm{\wpu_0}_1$
(recall from our discussion
following (\ref{dual}) that,
in contrast,
$\norm{\w_0}_2 = 1$).
More details on this issue
will be presented
in Section~\ref{sss-eco}.
Moreover,
we have introduced
the exponentially small parameter
\be
 \delta
=
\exp
\left(
 \frac{-J_-(\xs)}{\sqrt{\eps}}
\right)
\ll
 1 ,
 \Label{delta}
\ee
the role of which is to counterbalance
the exponentially large amplitudes
of the eigenfunctions $w^+_n$ and $v_n$.
In particular,
\be
 J_{\pm}(x)
=
 \sqrt{\V} \, x
\pm I(x)
\quad\mbox{and}\quad
I(x) =
 \int_{\xo}^x
 \sqrt{F(s) - F(\xo)} \, ds .
\Label{defIJpm}
\ee
Here,
the $\Or(\eps^{1/3})-$parameter $\xo$
corresponds to the turning point of (\ref{ODEt-omega}),
\be
 \xo
=
 F^{-1}(\lambda_* - \lambda_0)
=
 \eps^{1/3} \sL^{-1/3} \abs{A_1} + \Or(\eps^{1/2}) ,
\Label{barx0}
\ee
while $\xs$
is the location of the DCM, the unique point
where $J_-(\cdot)$ attains its (positive) maximum
(\cite{ZDPS-2009}---see also \ref{s-w0-ae}), \emph{i.e.},
\be
 \xs
=
 F^{-1}(\V + F(\xo))
=
 F^{-1}(\V) + \Or(\eps^{1/3}) .
\Label{xstar}
\ee
Thus, $\delta^{-1}$ is a measure
for the amplitude of the $\omega$-component
of the (linear) mode
associated with a bifurcating DCM.
The introduction of $\delta$
in the decomposition (\ref{e-decomp})
allows us to identify \emph{small patterns}
($\upu \ll 1$)
and
is motivated
by the observation
that this decomposition yields
\be
\begin{array}{rcl}
 \wpu(x,\tau)
&=&
 \eps^{c-1/6} \, \delta
 \sum_{n\ge 0}
 \Omega_n(\tau) \, \wpu_n(x) ,
\\
 \n(x,\tau)
&=&
 \eps^{c-1/6} \, \delta
 \sum_{n\ge 0}
 \Omega_n(\tau) \, \n_n(x)
+
 \eps^c
 \sum_{n\ge 0}
 \Psi_n(\tau) \, \s_n(x) .
\end{array}
\Label{e-decomps}
\ee
The principal part of $\wpu_0$
is derived
in \ref{s-w0-ae},
while
asymptotic formulas for $\wpu_n$,
with $n \ge 1$,
can be derived
in a similar manner.
For $\Or(1)$ values of $n$,
it follows
that $\wpu_n$ is
exponentially small everywhere
apart from
an asymptotically small neighborhood of $\xs$
where it attains
its maximum value
of asymptotic magnitude
at most $\Or(\eps^{-1/12} \delta^{-1})$.
Similarly,
the principal part of $\n_0$
is given in \ref{s-n0-ae},
together with
an ${\rm L}^\infty-$estimate
which shows that
$\n_0$ is at most
$\Or(\eps^{1/6} \delta^{-1})$ in $[0,1]$.
As a result,
the coefficients
of the eigenmodes
$\Omega_n$ ($n \ge 0$)
in (\ref{e-decomp})
are bounded uniformly in $L^\infty(0,1)$
by an $\Or(\eps^{c-1/12})$ constant,
while those of $\Psi_n$ ($n \ge 0$)
are $\Or(1)$.
In what follows,
we derive the ODEs
governing the evolution
of these eigenmodes.

\subsubsection{Eigenbasis decomposition of $\Lfpu(\upu)$
\label{sss-Lfpu-decomp}}
To derive the ODEs for the eigenmodes,
we need to express $\Lfpu(\upu)$
in the eigenbasis
$\{w^+_n\}_{n\ge0} \cup \{v_n\}_{n\ge0}$.
In particular,
we show that
\begin{eqnarray}
\fl
 \Lfpu(\upu)
&=&
 \eps^{c-1/6} \, \delta
 \sum_{k\ge0}
\left[
 \lambda_k \, \Omega_k
-
 \eps^c
\,
 \sum_{m\ge 0}
 \sum_{n\ge 0}
\left(
 a_{mnk} \, \Omega_m \Omega_n
+
 b_{mnk} \, \Psi_m \Omega_n
\right)
\right]
 w^+_k
\nonumber\\
\fl
&{}&
+
 \eps^c
 \sum_{k\ge0}
\left[
 \nu_k
 \Psi_k
-
 \eps^c
\,
 \sum_{m\ge 0}
 \sum_{n\ge 0}
\left(
 a'_{mnk} \, \Omega_m \Omega_n
+
 b'_{mnk} \, \Psi_m \Omega_n
\right)
\right]
 v_k ,
\hspace*{1cm}
\Label{Lu}
\end{eqnarray}
where
we have omitted
an $\Or(\eps^{3c-1/2})$ remainder.
The coefficients appearing
in this equation
are given
by the formulas
\be
\fl
\begin{array}{rcrclcl}
 a_{mnk}
&=&
 \eps^{-1/6}
\left\langle
\left(\begin{array}{c}
 1 \\ \eps \ell^{-1}
\end{array}\right)
 a_m \, \wpu_n , \hat{w}^+_k
\right\rangle
&=&
 \eps^{-1/6}
 \langle a_m \w_n , \w_k \rangle ,
\vspace*{1mm}\\
 a'_{mnk}
&=&
 \eps^{-1/3}
\left\langle
\left(\begin{array}{c}
 1 \\ \eps \ell^{-1}
\end{array}\right)
 \delta \, a_m \, \wpu_n , \hat{v}_k
\right\rangle
&=&
 \eps^{-1/3} \delta
\Big[
 \langle a_m \w_n , \y_k \rangle
+
 \eps \ell^{-1}
 \langle a_m \wpu_n , \s_k \rangle
\Big] ,
\vspace*{1mm}\\
 b_{mnk}
&=&
\left\langle
\left(\begin{array}{c}
 1 \\ \eps \ell^{-1}
\end{array}\right)
 b_m \, \wpu_n , \hat{w}^+_k
\right\rangle
&=&
 \langle b_m \w_n , \w_k \rangle ,
\vspace*{1mm}\\
 b'_{mnk}
&=&
 \eps^{-1/6}
\left\langle
\left(\begin{array}{c}
 1 \\ \eps \ell^{-1}
\end{array}\right)
 \delta \, b_m \, \wpu_n , \hat{v}_k
\right\rangle
&=&
 \eps^{-1/6} \delta
\Big[
 \langle b_m \w_n , \y_k \rangle
+
 \eps \ell^{-1}
 \langle b_m \wpu_n , \s_k \rangle
\Big] .
\end{array}
\Label{ab-coeffs}
\ee
Here,
we have defined the functions
\be
\fl
\begin{array}{rcl}
 a_m
&=&
 \delta
\,
\left[
 (1 - \nu)
\,
 \n_m
+
 (1 - \nu^{-1} f)
\,
 r
\,
 s_m
\right]
 f ,
\quad\mbox{with}\
 s_n(x)
=
 \int_0^x \wpu_n(s) ds ,
\\
 b_m
&=&
 (1 - \nu)
\,
 f
\,
 \s_m .
\end{array}
\Label{am-vs-bm}
\ee
Note that we use
$\langle \cdot , \cdot \rangle$
to denote all inner products---in
$\mathcal{H}$,
$\mathcal{H}_{\wpu}$,
and
${\mathcal H}_\n$---as
there is no danger of confusion.

We start by decomposing $\Lfpu(\upu)$
into linear and nonlinear terms
by means of
\be
\fl
 \Lfpu(\upu)
=
 \DLfpu \upu
+
 \Nf(\upu) ,
\quad\mbox{where}\
 \Nf(\upu)
=
\left(\begin{array}{c}
 1 \\ \eps \ell^{-1}
\end{array}\right)
 \left(p - f\right)
 \wpu .
\Label{N-def}
\ee
Substitution of the decomposition (\ref{e-decomp})
into the linear term
yields the eigendecomposition
of that linear term,
\begin{eqnarray}
 \DLfpu \upu
&=&
 \eps^{c-1/6} \, \delta
 \sum_{k\ge 0}
 \Omega_k \, \DLfpu \, w^+_k
+
 \eps^c
 \sum_{k\ge 0}
 \Psi_k \, \DLfpu \, v_k
\nonumber\\
&=&
 \eps^{c-1/6} \, \delta
 \sum_{k\ge0}
 \lambda_k \, \Omega_k
 w^+_k
+
 \eps^c
 \sum_{k\ge0}
 \nu_k
 \Psi_k
 v_k ,
\Label{DLu-decomp}
\end{eqnarray}
where we have also used
that $w^+_n$ and $v_n$
are eigenvectors of $\DLfpu$
(see Section~\ref{ss-eigen}).
It remains to express
the nonlinearity $\Nf(\upu)$
with respect to
that same eigenbasis.
First,
since $p - f$ contains
the nonlocal term $\int_0^x \wpu(s) ds$,
see (\ref{p})--(\ref{j}),
we write (cf. (\ref{e-decomps}))
\be
 S(x,\tau)
:=
 \eps^{-c+1/6}
 \int_0^x \wpu(s,\tau) ds
=
 \delta
 \sum_{n\ge 0}
 \Omega_n(\tau) \, s_n(x) ,
\Label{S-def}
\ee
where $s_n$
was introduced in (\ref{am-vs-bm}).
We subsequently obtain,
by (\ref{p}) and (\ref{f}),
\begin{eqnarray*}
 p
&=&
 \frac{1 - \n}{\nu^{-1} - \n}
\,
 \frac{1}{1 + j_H \exp(\kappa x) \exp(\eps^{c-1/6} \, r \, S)}
\\
&=&
 f
\,
 \frac{1 - \n}{1 - \nu\n}
\,
 \frac{1}{1 + (1 - \nu^{-1} f) ( \exp(\eps^{c-1/6} \, r \, S) - 1 )} .
\end{eqnarray*}
Substituting from (\ref{e-decomps})
for $\wpu$ and $\n$
into this formula
and
expanding asymptotically,
we find further
\begin{eqnarray}
 p(\wpu,\n,x)
&=&
 f
-
 \eps^{c-1/6}
 \sum_{m\ge 0}
 a_m \Omega_m
-
 \eps^c
 \sum_{m\ge 0}
 b_m \Psi_m
+
 \Or\left(\eps^{2c-1/3}\right) ,
\Label{p-ae}
\end{eqnarray}
with $a_m$ and $b_m$
as defined in (\ref{am-vs-bm}).
We remark
for later use
that this asymptotic expansion remains valid
for $o(\eps^{1/4-c})$ values
of $\Omega_n$ ($n \ge 0$)
and
$o(\eps^{-c})$ values
of $\Psi_n$ ($n \ge 0$)
(see our discussion following (\ref{e-decomps})).
Next,
(\ref{e-decomps}) and (\ref{p-ae}) yield
\[
\fl
 (p-f) \, \wpu
=
-
 \eps^{2c-1/3} \, \delta
 \sum_{m\ge 0}
 \sum_{n\ge 0}
 a_m \wpu_n \, \Omega_m \Omega_n
-
 \eps^{2c-1/6} \, \delta
 \sum_{m\ge 0}
 \sum_{n\ge 0}
 b_m \wpu_n \, \Psi_m \Omega_n ,
\]
where we have again omitted
an $\Or(\eps^{3c-1/2})$ remainder.
By virtue of (\ref{N-def}),
then,
\begin{eqnarray*}
 \Nf(\upu)
&=&
-
 \eps^{2c-1/3} \, \delta
 \sum_{m\ge 0}
 \sum_{n\ge 0}
\left(\begin{array}{c}
 1 \\ \eps \ell^{-1}
\end{array}\right)
 a_m \wpu_n \, \Omega_m \Omega_n
\\
&{}&
-
 \eps^{2c-1/6} \, \delta
 \sum_{m\ge 0}
 \sum_{n\ge 0}
\left(\begin{array}{c}
 1 \\ \eps \ell^{-1}
\end{array}\right)
 b_m \wpu_n \, \Psi_m \Omega_n
+
 \Or\left(\eps^{3c-1/2}\right) .
\end{eqnarray*}
We may now decompose
the spatial components
in these sums
with respect to the eigenbasis,
\begin{eqnarray*}
\left(\begin{array}{c}
 1 \\ \eps \ell^{-1}
\end{array}\right)
 \delta \, a_m \, \wpu_n
=
 \sum_{k\ge0}
\left(
 \eps^{1/6} \delta \, a_{mnk} \, w^+_k
+
 \eps^{1/3} \, a'_{mnk} \, v_k
\right) ,
\\
\left(\begin{array}{c}
 1 \\ \eps \ell^{-1}
\end{array}\right)
 \delta \, b_m \, \wpu_n
=
 \sum_{k\ge0}
\left(
 \delta \, b_{mnk} w^+_k
+
 \eps^{1/6} \, b'_{mnk} v_k
\right) ,
\end{eqnarray*}
where
the coefficients
$a_{mnk}$, $a'_{mnk}$, $b_{mnk}$, and $b'_{mnk}$
are found
by means of (\ref{ab-coeffs}).
Using this decomposition,
we finally write
(omitting throughout an $\Or(\eps^{3c-1/2})$ term)
\begin{eqnarray}
 \Nf(\upu)
&=&
-
 \eps^{2c}
 \sum_{m\ge 0}
 \sum_{n\ge 0}
 \sum_{k\ge0}
\left(
 \eps^{-1/6} \delta \, a_{mnk} \, w^+_k
+
 a'_{mnk} \, v_k
\right)
 \Omega_m \Omega_n
\nonumber\\
\fl
&{}&
-
 \eps^{2c}
 \sum_{m\ge 0}
 \sum_{n\ge 0}
 \sum_{k\ge0}
\left(
 \eps^{-1/6} \delta \, b_{mnk} \, w^+_k
+
 b'_{mnk} \, v_k
\right)
 \Psi_m \Omega_n
\nonumber\\
&=&
-
 \eps^{2c-1/6} \delta
 \sum_{m\ge 0}
 \sum_{n\ge 0}
 \sum_{k\ge0}
\left(
 a_{mnk} \Omega_m \Omega_n
+
 b_{mnk} \Psi_m \Omega_n
\right)
 w^+_k
\nonumber\\
\fl
&{}&
-
 \eps^{2c}
 \sum_{m\ge 0}
 \sum_{n\ge 0}
 \sum_{k\ge0}
\left(
 a'_{mnk} \Omega_m \Omega_n
+
 b'_{mnk} \Psi_m \Omega_n
\right)
 v_k .
\Label{N-decomp}
\end{eqnarray}
Combining (\ref{DLu-decomp}) and (\ref{N-decomp}),
then,
we arrive at
the desired result (\ref{Lu}).

\subsubsection{ODEs near the bifurcation point
\label{sss-bifODEs}}
We are now in a position
to derive the ODEs
for the amplitudes
$\{\Omega_n\}_{n\ge0}$
and
$\{\Psi_n\}_{n\ge0}$.
Differentiating
both members
of (\ref{e-decomp})
with respect to time,
we find
\be
 \partial_\tau \upu
=
 \eps^{c-1/6} \, \delta
 \sum_{k\ge 0}
 \dot{\Omega}_k \, w^+_k
+
 \eps^c
 \sum_{k\ge 0}
 \dot{\Psi}_k \, v_k ,
\Label{dotu}
\ee
where
the overdot denotes differentiation
with respect to $\tau$.
Next,
$\partial_\tau \upu = \Lfpu(\upu)$
and hence,
combining ({\ref{Lu}}) with (\ref{dotu}),
we obtain the ODEs
for the amplitudes,
\begin{eqnarray*}
 \dot{\Omega}_k
&=&
 \lambda_k
 \Omega_k
-
 \eps^c
\,
 \sum_{m\ge 0}
 \sum_{n\ge 0}
\left(
 a_{mnk} \, \Omega_m \Omega_n
+
 b_{mnk} \, \Psi_m \Omega_n
\right)
+
 \Or\left(\eps^{2c}\right) ,
\nonumber\\
 \dot{\Psi}_k
&=&
 \nu_k
 \Psi_k
-
 \eps^c
\,
 \sum_{m\ge 0}
 \sum_{n\ge 0}
\left(
 a'_{mnk} \, \Omega_m \Omega_n
+
 b'_{mnk} \, \Psi_m \Omega_n
\right)
+
 \Or\left(\eps^{2c}\right) .
\end{eqnarray*}

We now tune
the bifurcation parameter $\lambda_*$
so that
the largest eigenvalue,
$\lambda_0$,
is the only positive eigenvalue
while
the eigenvalues $\lambda_1, \lambda_2, \ldots$
are negative.
In particular,
we write (cf.~(\ref{defalphaLamb}))
\begin{eqnarray*}
 \lambda_0
&=&
 \eps \Lambda_0 ,
\ \mbox{where} \
0
<
 \Lambda_0
<<
 \eps^{-2/3} ,
\\
 \nu_k
&=&
 -\eps N_k ,
\ \mbox{where} \
 N_k
>
 0
\ \mbox{is} \
\Or(1)
\ \mbox{for} \
 k
=
 0, 1, \ldots ,
\\
 \lambda_k
&=&
 -\eps^{1/3} \Lambda_k ,
\ \mbox{where} \
 \Lambda_k
>
 0
\ \mbox{for} \
 k
=
 1, 2, \ldots .
\end{eqnarray*}
As we will see shortly,
the cases of particular interest
will turn out to be those where
$\Lambda_0$ is either $\Or(1)$
or logarithmically large.
Note also that,
since the distance
between $\lambda_0$ and $\lambda_k$
is $\Or(\eps^{1/3})$ by (\ref{lambdan}),
it follows that
$\lambda_1, \lambda_2, \ldots \ll \nu_1$.
Then,
the evolution equations
for the amplitudes
become
\begin{eqnarray}
\fl
 \dot{\Omega}_0
&=
 \eps^\rho \Lambda_0
 \Omega_0
-
 \eps^c
 \sum_{m\ge 0}
 \sum_{n\ge 0}
 a_{mn0} \Omega_m \Omega_n
-
 \eps^c
 \sum_{m\ge 0}
 \sum_{n\ge 0}
 b_{mn0} \Psi_m \Omega_n ,
\Label{Fourier-ODE-Omega0}\\
\fl
 \dot{\Psi}_k
&=
 -\eps
 N_k
 \Psi_k
-
 \eps^c
 \sum_{m\ge 0}
 \sum_{n\ge 0}
 a'_{mnk} \Omega_m \Omega_n
-
 \eps^c
 \sum_{m\ge 0}
 \sum_{n\ge 0}
 b'_{mnk} \Psi_m \Omega_n ,
\quad
 k \ge 0 ,
\Label{Fourier-ODE-Psi}
\\
\fl
 \dot{\Omega}_k
&=
 -\eps^{1/3}
 \Lambda_k
 \Omega_k
-
 \eps^c
 \sum_{m\ge 0}
 \sum_{n\ge 0}
 a_{mnk} \Omega_m \Omega_n
-
 \eps^c
 \sum_{m\ge 0}
 \sum_{n\ge 0}
 b_{mnk} \Psi_m \Omega_n ,
\quad
 k \ge 1 ,
\Label{Fourier-ODE-Omegak}
\end{eqnarray}
where
we have omitted
all higher order terms.

\section{Application of Laplace's method on $a_{000}$
\label{s-a000}}
\setcounter{equation}{0}
Explicit asymptotic expressions
for the coefficients
in the ODEs
(\ref{Fourier-ODE-Omega0})--(\ref{Fourier-ODE-Omegak})
obtained in the previous section
can be derived
by applying Laplace's method
and
the principle of stationary phase
to the integrals in (\ref{ab-coeffs}).
In this section,
we demonstrate the use of the former
in deriving an asymptotic formula for $a_{000}$.
Asymptotic expressions for the remaining coefficients
will be derived independently
in Sections~\ref{s-bm00}--\ref{s-bm0k},
after we have thoroughly analyzed
the bifurcations that
our system undergoes.
Although the analysis
in those sections
is substantially more involved,
our approach there
is very similar to that
in the present section. 

The main result of this section
is the leading order approximation
\be
 a_{000}
=
 A(\Lambda_0) 
=
 \alpha \, a(\Lambda_0) ,
\Label{a000-ae}
\ee
where
we have defined
the $\Or(1)$, positive, $\Lambda_0-$independent
constant $\alpha$
and
the function $a$
by means of
\be
\fl
 \alpha
=
 (1 - \nu) \, f(0) \, C_1 \, C_2 \, \sL^{1/3} \, \sst^{-1/2}
>
 0
\quad\mbox{and}\quad
 a(\Lambda_0)
=
\frac{
 \sinh\left(\sqrt{\Lambda_0}(1-\xs)\right)
}{
 \sqrt{\Lambda_0}
\,
 \cosh\sqrt{\Lambda_0}
} .
\Label{defbarCa}
\ee
Here,
$\sL$ is defined in (\ref{defsigF}),
while
\be
\fl
 C_1
=
 \left(\int_{A_1}^\infty \Ai^2(s) \, ds\right)^{-1/2} ,
\quad
 C_2
=
 \exp(\abs{A_1}^{3/2}) ,
\quad\mbox{and}\quad
\sst = F'(\xs) = -f'(\xs) ,
\Label{defC12sstar}
\ee
see \cite{ZDPS-2009} and \ref{s-w0-ae}.
We start by recalling that
the coefficient $ a_{000}$
is given by
\be
 a_{000}
=
 \eps^{-1/6}
\,
 \int_0^1
 a_0(x) \, \w^2_0(x) \, dx ,
\Label{a000-aux1}
\ee
cf. (\ref{ab-coeffs}),
where
\[
 a_0(x)
=
 \delta
\left[
 r (1 - \nu^{-1} f(x)) s_0(x)
+
 (1 - \nu) \n_0(x)
\right]
\,
 f(x) .
\]
Employing
(\ref{am-vs-bm}), (\ref{S-def}),
using the explicit approximation (\ref{eta0})
for $\eta_0$ from \ref{s-n0-ae},
and
defining the functions
\begin{eqnarray}
 h_1(x,y)
&=&
 f(x)
\,
\left[
 r
\,
 \left(1 - \frac{f(x)}{\nu}\right)
-
 \frac{1 - \nu}{\ell \sqrt{\Lambda_0}}
\,
 \sinh\left(\sqrt{\Lambda_0} (x-y)\right)
\,
 f(y)
\right] ,
\Label{h1-def}\\
 h_2(x,y)
&=&
\frac{
 (1 - \nu)
\,
 f(x)
\,
 \cosh\left(\sqrt{\Lambda_0} \, x\right)
}{
 \ell
\,
 \sqrt{\Lambda_0}
\,
 \cosh\sqrt{\Lambda_0}
}
\,
 f(y)
\,
 \sinh\left(\sqrt{\Lambda_0}(1-y)\right) ,
\Label{h2-def}
\end{eqnarray}
we find further
\[
 a_0(x)
=
 \eps^{-1/6}
 \delta
 \int_0^x
 h_1(x,y)
\,
 \wpu_0(y)
\,
 dy
+
 \eps^{-1/6}
 \delta
 \int_0^1
 h_2(x,y)
\,
 \wpu_0(y)
\,
 dy .
\]
Thus,
\begin{eqnarray}
\fl
 a_{000}
&=
 \eps^{-1/6}
 \delta
 \int_0^1
 \int_0^x
 h_1(x,y)
\,
 \w^2_0(x)
\,
 \wpu_0(y)
\,
 dy dx
+
 \eps^{-1/6}
 \delta
 \int_0^1
 \int_0^1
 h_2(x,y)
\,
 \w^2_0(x)
\,
 \wpu_0(y)
\,
 dy dx
\nonumber\\
\fl
&=
 \eps^{-1/6} \delta (\I_1 + \I_2) ,
\Label{a000-aux2}
\end{eqnarray}
where
$\I_1$ and $\I_2$
are the two double integrals
appearing in this expression.

We can obtain
the principal parts
of $\I_1$ and $\I_2$
using Theorem~\ref{t-Laplace1},
based on \cite{W-2001},
in \ref{s-Laplace}.
We start with
the latter integral which,
as we will see,
fully determines
the leading order behavior
of $a_{000}$.
First,
the normalization condition
$\| \w_0 \|_2 = 1$
yields
$\int_0^1
h_2(x,y) \, \w^2_0(x) \, dx
=
h_2(0,y)$
to leading order.
Since, also,
$\wpu_0$ has
a unique maximum
at the interior critical point $\xs$,
Theorem~\ref{t-Laplace1}.I
(with
$\lambda = \eps^{-1/2}$,
$\Pi = -J_-$,
and
$\Xi = h_2(0,\cdot)$)
yields
\be
\fl
 \I_2
=
 \int_0^1
 h_2(0,y)
\,
 \wpu_0(y)
\,
 dy
=
 \frac{1}{(\eps^{-1/2})^{1/2}}
\,
\frac{
 \sqrt{2 \pi} \, h_2(0,\xs)
}{
 \sqrt{-J_-''(\xs)}
}
\,
 \wpu_0(\xs)
=
 \eps^{1/6}
\,
 \delta^{-1}
\,
 C_3
\Label{a000-Int2-ae}
\ee
to leading order,
where
we have used the explicit leading order approximation (\ref{wo+})
of $\wpu_0$ from \cite{ZDPS-2009}
(see also \ref{s-w0-ae}),
recalled the definition (\ref{delta}) of $\delta$,
defined
\be
 C_3
=
\frac{
 \sqrt{2 \pi} \, h_2(0,\xs)
}{
 \sqrt{-J_-''(\xs)}
}
\,
\frac{
 C_1 \, C_2 \, \sL^{1/3}
}{
 2 \sqrt{\pi} \, F^{1/4}(\xs)
}
=
 C_1
\,
 C_2
\,
 \sL^{1/3}
\,
 \sst^{-1/2}
\,
 h_2(0,\xs) ,
\Label{a000-C2}
\ee
and
employed the identity
$J''_-
=
-2^{-1} F^{-1/2} F'$.

Next,
we show $\I_1$
to be exponentially smaller
than $\I_2$.
First,
we rewrite it as
\begin{eqnarray}
\fl
 \I_1
=
 \eps^{-1/4}
\,
\frac{
 C_1^3 \, C_2^3 \, \sL
}{
 8 \pi^{3/2}
}
\,
\sum_{j=1}^6
\theta_j
 \int\int_D
\frac{
 h_1(x,y)
}{
 \sqrt{F(x)} \, F^{1/4}(y)
}
\,
 \exp\left(\frac{\Pi_j(x,y)}{\sqrt{\eps}}\right)
\,
 dA_{xy} ,
\hspace*{0.75cm}
\Label{a000-aux3}
\end{eqnarray}
where
we have used
(\ref{wo+})
and
(\ref{wo}).
Here,
$D
=
\{ (x,y) \vert 0 \le y \le x , \, 0\le x \le 1 \}$
and
\be
\begin{array}{rclcrcl}
 \Pi_1(x,y)
&=&
 J_-(y)
-
 2 I(x)
&\quad\mbox{and}\quad&
 \theta_1
&=&
 1 ,
\\
 \Pi_2(x,y)
&=&
 J_-(y)
-
 2 I(1)
&\quad\mbox{and}\quad&
 \theta_2
&=&
 2 \theta ,
\\
 \Pi_3(x,y)
&=&
 J_-(y)
+
 2 I(x)
-
 4 I(1)
&\quad\mbox{and}\quad&
 \theta_3
&=&
 \theta^2 ,
\\
 \Pi_4(x,y)
&=&
 J_+(y)
-
 2 I(x)
-
 2 I(1)
&\quad\mbox{and}\quad&
 \theta_4
&=&
 \theta ,
\\
 \Pi_5(x,y)
&=&
 J_+(y)
-
 4 I(1)
&\quad\mbox{and}\quad&
 \theta_5
&=&
 2 \theta^2 ,
\\
 \Pi_6(x,y)
&=&
 J_+(y)
+
 2 I(x)
-
 6 I(1)
&\quad\mbox{and}\quad&
 \theta_6
&=&
 \theta^3 ,
\end{array}
\ee
where $I(x)$ and $J_{\pm 1}(y)$ have been defined in (\ref{defIJpm}), and
\be
 \theta
=
 \frac{\sqrt{\sR} + \sqrt{\V}}{\sqrt{\sR} - \sqrt{\V}}
\quad\mbox{with}\quad 
 \sR
=
 F(1) .
\Label{defgammas1}
\ee
Theorem~\ref{t-Laplace} yields,
for each integral,
a result
proportional to
$\exp(\max_{(x,y) \in D}\Pi_j(x,y) / \sqrt{\eps})$.
We first identify
$\max\Pi_1$
and then
show that
$\max\Pi_1 > \max\Pi_j$,
for $j = 2 , \ldots , 6$;
it follows that
the dominant term
in (\ref{a000-aux3})
corresponds to $\Pi_1$
and
the rest are
exponentially smaller than it.
Now,
$\Pi_1$ has no critical points in $D$,
and thus
its global maximum
lies on
\begin{eqnarray*}
\fl
 \partial D
=
\bigcup_{i=1}^3
 (\partial D)_i
=
 \{ (1,y) \vert 0\le y \le 1 \}
\cup
 \{ (x,x) \vert 0\le x \le 1 \}
\cup
 \{ (x,0) \vert 0\le x \le 1 \} .
\end{eqnarray*}
First,
the global maximum
cannot be on $(\partial D)_1$;
indeed,
$\mathring{D}$ lies
to the left of $(\partial D)_1$
and
$\partial_x \Pi_1(x,y) = -2 \sqrt{\Fo(x)} \le 0$,
where we have introduced $\Fo(x) = F(x) - F(\xo)$,
so that $\Pi_1$ assumes
higher values in $\mathring{D}$
than on $(\partial D)_1$.
Next,
$\Pi_1(x,x) = \sqrt{\V} x - 3 I(x)$
on $(\partial D)_2$,
and thus
$\max\Pi_1(x,x) = \Pi_1(\xss,\xss)$
with
$0 < \xss = \Fo^{-1}(\V/9) < \xs$
(recall (\ref{xstar}) and note that
$\Fo > 0$ is increasing).
Finally,
$\Pi_1(x,0) = -2 I(x) \le 0$
on $(\partial D)_3$,
and thus
$\max_{(\partial D)_3}\Pi_1 \le 0 < \Pi_1(\xss,\xss)$.
In total,
then,
we find that
$\max\Pi_1 = \Pi_1(\xss,\xss) > 0$.
Next,
$\Pi_2(x,y)
\le
\Pi_1(x,y)
\le
\Pi_1(\xss,\xss)$.
Since
the leftmost equality
holds only in
an $\Or(\eps^{1/2})$-neighborhood
of $x = 1$,
we find that
$\max\Pi_2
<
\Pi_1(\xss,\xss)$,
as desired.
Additionally,
$\Pi_3 \le \Pi_2$
on $D$,
and thus also
$\max_D\Pi_3 < \max_D\Pi_1$.
Next,
$\Pi_4$ has no critical points
in $\mathring{D}$,
and hence
we need to examine
its behavior
on $\partial D$.
First,
the maximum
cannot be on $(\partial D)_1$
by the same argument
we used for $\Pi_1$.
Next,
$\Pi_4(x,x)= J_-(x) - 2 I(1)$
on $(\partial D)_2$,
and thus
$\max_{(\partial D)_2}\Pi_4
=
\Pi_4(\xs,\xs)
=
J_-(\xs) - 2 I(1)$.
Finally,
$\Pi_4
\le
-2 I(1)
<
\Pi_4(\xs,\xs)$
on $(\partial D)_3$,
and hence
$\max\Pi_4
=
J_-(\xs) - 2 I(1)
=
\max\Pi_2
<
\max\Pi_1$,
as desired.
Finally,
$\Pi_5 \le \Pi_4$
and
$\Pi_6 \le \Pi_4$,
and
the desired result now follows.

These estimates
show, then, that
$\max\Pi_1 = \Pi(\xss,\xss) > \max\Pi_j$,
for $j = 2 , \ldots , 6$.
Since $(\xss,\xss) \in \partial D$
and
its Jacobian satisfies
$D\Pi_1(\xss,\xss) \ne 0$,
Theorem~\ref{t-Laplace} yields
for (\ref{a000-aux3})
the asymptotic formula
\[
\fl
 \I_1
=
 \eps^{3/4}
\,
 C'_1
\left(
 \eps^{-1/4}
\,
 \frac{C_1^3 \, C_2^3}{8 \pi^{3/2}}
\,
 \exp\left(\frac{\Pi_1(\xss,\xss)}{\sqrt{\eps}}\right)
\right)
=
 \eps^{1/2}
\,
 C''_1
\,
 \exp\left(\frac{\Pi_1(\xss,\xss)}{\sqrt{\eps}}\right) ,
\]
for some $\Or(1)$ constants
$C'_1 , C''_1 > 0$.
Since $\I_2 = {\cal O}(\eps^{1/6} \delta^{-1})$ (\ref{a000-Int2-ae}) and, by (\ref{delta}),
\[
\frac{\I_1}{\I_2} = \eps^{1/3} \frac{C''_1}{C_3} 
\exp\left(\frac{\Pi_1(\xss,\xss)- J_-(\xs)}{\sqrt{\eps}}\right)
\]
with
\[
\Pi_1(\xss,\xss)- J_-(\xs) = 
[J_-(\xss) - J_-(\xs)] - 2I(\xss) < 0
\]
(recall that
$\xs$ is defined in (\ref{xstar})
as the location
of the maximum of $J_-$),
it indeed follows that
$\I_1$ is exponentially small
compared to $\I_2$. 

We conclude that
$a_{000}$ is given by $\delta \, \I_2$
at leading order.
Combining the expressions
(\ref{a000-Int2-ae})--(\ref{a000-C2})
with the definition of $h_2$ in (\ref{h2-def}),
we obtain
the leading order result (\ref{a000-ae})
by using the fact that
$f(\xs) = \ell$,
also at leading order.
To derive this last identity,
observe that---in the regime
$\lambda_0 \ll 1$---it holds that
$\lambda_* = 0$ 
at $\Or(1)$,
see (\ref{lambda*}), (\ref{defalphaLamb}), or 
equivalently that
$\V = f(0) - \ell$;
further,
and also
to leading order,
$F(\xs) = \V$ by (\ref{xstar}),
so that
the desired identity follows
from the definition $F(x) = f(0) - f(x)$
applied at $x = \xs$.
Finally, we note that higher order terms
in formula (\ref{a000-ae})
may be obtained
solely by considering $\I_2$,
as $\I_1$ is exponentially smaller
than $\I_2$.

\section{Emergence of a stable DCM
\label{s-emerge}}
\setcounter{equation}{0}
The trivial (zero) state
is, by construction, a fixed point
of the ODEs
(\ref{Fourier-ODE-Omega0})--(\ref{Fourier-ODE-Omegak})
for the Fourier coefficients.
In this and the next section,
we identify the remaining
fixed points of that system
and
determine their stability.
In this entire section,
we work exclusively
in the regime
$\rho = 1$ and $\Lambda_0 = \Or(1)$.

\subsection{Asymptotic expressions for
$b_{m00}$, $a'_{00k}$, and $b'_{m0k}$
\label{ss-coeff-ae}}
As stated in the previous section,
where we derived
an asymptotic expression for $a_{000}$,
asymptotic expressions
for the coefficients
$b_{m00}$,
$a'_{00k}$,
and
$b'_{m0k}$
appearing in
(\ref{Fourier-ODE-Omega0})--(\ref{Fourier-ODE-Omegak})
are derived independently
in Sections~\ref{s-bm00}--\ref{s-bm0k} below.
Here,
we summarize the leading order behavior
of these coefficients,
including also (\ref{a000-ae}) for completeness:
\be
\begin{array}{cclr}
 a_{000}
&=&
 A(\Lambda_0) ,
&
{}
\\
 b_{m00}
&=&
 B ,
&
\ \mbox{for}\
 m \ll \eps^{-1/3} ,
\\
 a'_{00k}
&=&
-
 A'_k(\Lambda_0)
\,
 A(\Lambda_0) ,
&
\ \mbox{for}\
 0 \ne k \ll \eps^{-1/3} ,
\\
 b'_{m0k}
&=&
-
 A'_k(\Lambda_0)
\,
 B ,
&
\ \mbox{for}\
 0 \ne k , m \ll \eps^{-1/3} .
\end{array}
\Label{coeffs-ae}
\ee
The function $A$
was introduced in (\ref{a000-ae})--(\ref{defbarCa}),
whereas
$B = \sqrt{2} \, (1 - \nu) \, f(0)$
is a positive $\Or(1)$ constant.
Further,
we have introduced
the function $A'_k$ via
\begin{eqnarray}
\fl
 A'_k(\Lambda_0)
=
 \alpha'
\,
 a'_k(\Lambda_0) ,
\quad\mbox{where}\quad
 \alpha'
=
 \frac{\sqrt{2} \, C_2 \, \sL^{1/3}}{C_1 \, C_3 \, \sst^{1/2}}
\quad\mbox{and}\quad
 a'_k(\Lambda_0)
=
 \frac{\cos(\sqrt{N_k} \, \xs)}{N_k + \Lambda_0} .
\Label{5const}
\end{eqnarray}
Here, $C_3 = (\Ai'(A_1))^2$.
Note that,
similarly to $\alpha$ (cf.~(\ref{defbarCa})),
$\alpha'$ is an $\Or(1)$ constant
independent of $\Lambda_0$;
the constants
$\sL$, $\sst$, $C_1$ and $C_2$
have been defined in (\ref{defsigF}) and (\ref{defC12sstar}).
We also note
the following identity
concerning Airy functions
(see \cite[Section~9.11(iv), identity~(9.11.5)]{NIST})
\[
 \int_{A_1}^\infty \Ai^2(s) \, ds
=
 (\Ai'(A_1))^2 ,
\quad\mbox{or equivalently}\quad
 C_1^2 \, C_3
=
 1 ,
\]
which, in turn,
yields an identity
that will prove to be
of exceeding importance
in the rest of this section---namely,
\be
 2 \alpha
=
 \alpha' B .
\Label{2a=a'B}
\ee
Asymptotic formulas for
$b_{m00}$,
$a'_{00k}$,
and
$b'_{m0k}$
and
for higher values
of $m$ and $k$
can be derived similarly.
However,
seeing as such formulas
only contribute higher order terms
in our analysis below,
we refrain from presenting the details.
In what follows, instead,
we treat (\ref{coeffs-ae}) as being valid
for all values of $k$ and $m$.

\subsection{The reduced system
\label{ss-Lambda=O(1)-slaving}}
The system
(\ref{Fourier-ODE-Omega0})--(\ref{Fourier-ODE-Omegak})
exhibits asymptotically disparate timescales
depending on the value of $\rho$
and
associated with
the asymptotic magnitudes
of the eigenvalues.
In this section,
we investigate the case
$\rho = 1$,
in which regime
$\Omega_0$
and
$\Psi_0$,
$\Psi_1$,
$\ldots$
evolve on
a slow timescale
and
the higher-order modes
$\Omega_1$,
$\Omega_2$,
$\ldots$
become slaved
to them.
Setting, then, $\rho = 1$
and
rescaling time
(with a slight abuse of notation)
as $t = \eps \tau$,
the evolution equations become
\begin{eqnarray}
\fl
 \dot{\Omega}_0
&=
 \Lambda_0
 \Omega_0
-
 \eps^{c-1}
 \sum_{m\ge 0}
 \sum_{n\ge 0}
 a_{mn0} \Omega_m \Omega_n
-
 \eps^{c-1}
 \sum_{m\ge 0}
 \sum_{n\ge 0}
 b_{mn0} \Psi_m \Omega_n ,
\Label{Omega0'}
\\
\fl
 \dot{\Psi}_k
&=
 -N_k
 \Psi_k
-
 \eps^{c-1}
 \sum_{m\ge 0}
 \sum_{n\ge 0}
 a'_{mnk} \Omega_m \Omega_n
-
 \eps^{c-1}
 \sum_{m\ge 0}
 \sum_{n\ge 0}
 b'_{mnk} \Psi_m \Omega_n ,
\quad
 k \ge 0 ,
\Label{Psik'}
\\
\fl
 \eps^{2/3}
 \dot{\Omega}_k
&=
 -\Lambda_k
 \Omega_k
-
 \eps^{c-1/3}
 \sum_{m\ge 0}
 \sum_{n\ge 0}
 a_{mnk} \Omega_m \Omega_n
-
 \eps^{c-1/3}
 \sum_{m\ge 0}
 \sum_{n\ge 0}
 b_{mnk} \Psi_m \Omega_n ,
\quad
 k \ge 1 .
\Label{Omegak'}
\end{eqnarray}
(Here also,
the overdot denotes differentiation
with respect to $t$.)
It is natural to introduce
slaving relations
for the latter modes
in this system,
\be
 \Omega_k
=
 \eps^{c_k}
\,
 \Wslave_k(\Omega_0 , \Psi_1 , \Psi_2 , \ldots) ,
\quad\mbox{for all}\quad
 k \ge 1 ,
\Label{equil-a=1}
\ee
where
the positive constants
$c_1, c_2 , \ldots$
and
the $\Or(1)$ functions
(with $\Or(1)$ partial derivatives)
$\Wslave_1$,
$\Wslave_2$,
$\ldots$
are to be determined.
To do so,
we first write
the evolution equations
for $\Omega_0$
and
$\Psi_1$,
$\Psi_2$,
$\ldots$
under these slaving relations;
we find
\begin{eqnarray*}
 \dot{\Omega}_0
&=&
 \Lambda_0
 \Omega_0
-
 \eps^{c-1}
 a_{000} \Omega_0^2
-
 \eps^{c-1}
 \Omega_0
 \sum_{m\ge 0}
 b_{m00} \Psi_m ,
\\
 \dot{\Psi}_k
&=&
 -N_k
 \Psi_k
-
 \eps^{c-1}
 a'_{00k} \Omega_0^2
-
 \eps^{c-1}
 \Omega_0
 \sum_{m\ge 0}
 b'_{m0k} \Psi_m ,
\quad
 k \ge 0 ,
\end{eqnarray*}
where
we have retained
only the leading order terms
from each sum.
Dominant balance yields,
then,
$c = 1$.
Next,
the invariance equation
for $\Omega_k$
yields that
the right member of (\ref{Omegak'})
must vanish to leading order.
Dominant balance yields $c_k = 2/3$
and
\[
 \Wslave_k(\Omega_0 , \Psi_1 , \Psi_2 , \ldots)
=
-
 \frac{a_{00k}}{\Lambda_k}
 \Omega_0^2
-
 \frac{\Omega_0}{\Lambda_k}
 \sum_{m\ge 0}
 b_{m0k} \Psi_m .
\]
Recalling, also, (\ref{coeffs-ae}),
we arrive at the evolution equations
\be
\begin{array}{rcrl}
 \dot{\Omega}_0
&=&
 \Lambda_0
 \Omega_0
-
 A \, \Omega_0^2
-
 B \, \Omega_0
 \sum_{m\ge 0}
 \Psi_m
\ \ ,
&
 {}
\\
 \dot{\Psi}_k
&=&
 -N_k
 \Psi_k
+
 A'_k
\left[
 A \, \Omega_0^2
+
 B \, \Omega_0
 \sum_{m\ge 0}
 \Psi_m
\right] ,
&
 k \ge 0 .
\end{array}
\Label{redux-ODE-a=1}
\ee
Here also,
we have retained
only the leading order terms
from each sum.

\paragraph{Remark~4.1.}
The ODE (\ref{cmred})---describing the flow
on the one-dimensional center manifold
in the regime where
$\lambda_0 = \eps^\rho \Lambda_0 \ll \eps$---can be
obtained from the system (\ref{redux-ODE-a=1}) above
as its $\Lambda_0 \to 0$ limit.
Indeed,
the $\Psi$-modes
become slaved to
the mode $\Omega_0$
in this limit,
and
(\ref{redux-ODE-a=1}) reduces to (\ref{cmred})
with $a_{000}(0)$ replacing $A = a_{000}(\Lambda_0)$
(cf. (\ref{coeffs-ae})).
Note that $a_{000}$ has
a removable singularity at zero,
so we write
$a_{000}(0)
=
\lim_{\Lambda_0 \to 0}a_{000}(\Lambda_0)
=
(1 - \xs) \, \alpha$.
Using (\ref{defbarCa}),
it is plain to check that,
indeed,
the formula for $a_{000}(0)$
reported in (\ref{a000(0)})
equals $(1 - \xs) \, \alpha$.

\subsection{The bifurcating steady state
\label{ss-Lambda=O(1)}}
In this section,
we identify
the nontrivial fixed point
of the reduced system (\ref{redux-ODE-a=1}).
In particular,
we show that
this fixed point is given
to leading order
by the formulas
\begin{eqnarray}
 \Omega_0^*(\Lambda_0)
=
\frac{
 \Lambda_0
}{
 (1 - \xs) \, \alpha
}
\quad\mbox{and}\quad
 \Psi^*_k(\Lambda_0)
=
\frac{
 2 \, \Lambda_0^2 \, \cos(\sqrt{N_k} \, \xs)
}{
 (1 - \xs) \,  B \, N_k \, (N_k + \Lambda_0)
} ,
\Label{fp-Labda0=O(1)}
\end{eqnarray}
where $k \ge 0$
and
the parameter $\alpha$ was introduced
in (\ref{defbarCa}).
Plainly,
$\Omega_0^*$ remains positive,
and hence also ecologically relevant,
for all positive values of $\Lambda_0$
and all values of $0 \le \xs < 1$
(equivalently,
all positive values of $\V$
up to the co-dimension two point).
Further,
the leading order expression (\ref{fp-Labda0=O(1)})
for $\Omega_0^*$
exactly matches
\be
 \Omega_0^*
=
 \frac{\Lambda_0}{a_{000}(0)} ,
\quad\mbox{for}\
 \Lambda_0
\to
 0 ,
\Label{Omega0-fp-Lambda0->0}
\ee
cf.~our discussion
in the Introduction
and in Remark~4.1 above.
It will also be elucidated
in Section~\ref{sss-eco} below
that this fixed point
corresponds to a DCM
with an $\Or(\eps)$ biomass
and
an associated $\Or(\eps)$ nutrient depletion.

Note that the denominators
in the formulas
for $\Omega^*_0$ and $\Psi^*_k$
vanish for $\xs = 1$.
As explained in the Introduction,
this value is attained by $\xs$
at the co-dimension two point
where DCMs and BLs bifurcate concurrently.
This is another indication
that the nature of
the co-dimension two bifurcation
is of independent analytical interest.

\subsubsection{Derivation of (\ref{fp-Labda0=O(1)})
\label{sss-derive}}
First,
setting the left members
of (\ref{redux-ODE-a=1}) to zero,
we obtain an algebraic system
for the nontrivial steady states,
\begin{eqnarray}
 \Lambda_0
-
 A
\,
 \Omega_0
-
 B
 \sum_{m\ge 0}
 \Psi_m
=
 0 ,
\Label{fp-eq1}\\
 N_k
 \Psi_k
-
 A'_k
\,
 \Omega_0
\,
\left[
 A
\,
 \Omega_0
+
 B
 \sum_{m\ge 0}
 \Psi_m
\right]
=
 0 .
\hspace*{1cm}
\Label{fp-eq2}
\end{eqnarray}
Here,
$k \ge 0$
and
we have removed
a superfluous factor of $\Omega_0$
in (\ref{fp-eq1})
corresponding to the trivial steady state.
Substituting from this equation into (\ref{fp-eq2}),
we obtain the equivalent formulation
\begin{eqnarray}
 A \, \Omega_0
+
 B
 \sum_{m\ge 0}
 \Psi_m
=
 \Lambda_0
\quad\mbox{and}\quad
 N_k
 \Psi_k
-
 A'_k
\,
 \Lambda_0
\,
 \Omega_0
=
 0 .
\Label{fp-eq-sub}
\end{eqnarray}
This system is readily solved to yield
\be
 \Omega_0^*
=
 \frac{\Lambda_0}{\alpha' \, s \, B \, \Lambda_0 + A}
\quad\mbox{and}\quad
 \Psi_k^*
=
 \frac{A'_k}{N_k}
\,
 \Lambda_0 \, \Omega_0^* ,
\Label{fp-Labda0=O(1)-raw}
\ee
where $s$ is defined by the series
\[
 s
=
 \frac{1}{\alpha'}
 \sum_{m \ge 0}
 \frac{A'_m}{N_m}
=
 \sum_{m \ge 0}
 \frac{\cos(\sqrt{N_m} \, \xs)}{N_m \, (N_m + \Lambda_0)} .
\]
To produce a closed formula for $s$,
we recast this formula as
\be
\fl
 s
=
 \sum_{m \ge 0}
 \frac{\cos(\sqrt{N_m} \, \xs)}{N_m \, (N_m + \Lambda_0)}
=
 \frac{1}{\Lambda_0}
\left(
 \sum_{m \ge 0}
 \frac{\cos(\sqrt{N_m} \, \xs)}{N_m}
-
 \sum_{m \ge 0}
 \frac{\cos(\sqrt{N_m} \, \xs)}{N_m + \Lambda_0}
\right) ,
\Label{s2-rewrite}
\ee
with both series
in the right member
converging absolutely
and
uniformly with $\xs$.
The second series
appearing in the right member
of this last equation
is a Mittag-Leffler expansion;
analytic formulas for such expansions
can often be obtained
by means of the Fourier transform.
In particular,
\cite[Eq.~(1.63)]{O-1973}
(with
$a = \pi$, $b = \im \sqrt{\Lambda_0}$, and $l = 1$)
yields the explicit formula
\be
\fl
 \sum_{m \ge 0}
\frac{
 \cos\left(\sqrt{N_m} \, \xs\right)
}{
 N_m + \Lambda_0
}
=
\frac{
 \sin\left(\im \sqrt{\Lambda_0} \, (1-\xs)\right)
}{
 2 \im \sqrt{\Lambda_0}
 \cos(\im \sqrt{\Lambda_0})
}
=
\frac{
 \sinh\left(\sqrt{\Lambda_0} \, (1-\xs)\right)
}{
 2 \sqrt{\Lambda_0} \cosh\sqrt{\Lambda_0}
}
=
 \frac{a(\Lambda_0)}{2} ,
\Label{ML}
\ee
whence also
\be
 \sum_{m \ge 0}
\frac{
 \cos\left(\sqrt{N_m} \, \xs\right)
}{
 N_m
}
=
 \frac{a(0)}{2}
=
 \frac{1-\xs}{2} .
\Label{cos-sum}
\ee
Substituting into (\ref{s2-rewrite}),
we obtain
\be
 s
=
 \frac{1 - \xs}{2 \Lambda_0}
-
 \frac{a(\Lambda_0)}{2 \Lambda_0} ,
\Label{s2}
\ee
and therefore
(\ref{fp-Labda0=O(1)-raw})
for $\Omega_0^*$
becomes
\[
 \Omega_0^*
=
\frac{
 \Lambda_0
}{
 (\alpha - \alpha' \, B/2) \, a(\Lambda_0)
+
 (1-\xs) \, \alpha' \, B/2} .
\]
The final formulas
collected in (\ref{fp-Labda0=O(1)})
now follow by identity~(\ref{2a=a'B})
and
(\ref{fp-Labda0=O(1)-raw}) for $\Psi_k^*$.

\subsubsection{Ecological interpretation
\label{sss-eco}}
We next proceed to show
that the steady state (\emph{stationary pattern})
we identified above
corresponds to
an $\Or(\eps)$ biomass
with a corresponding
$\Or(\eps)$ depletion
of the nutrient.
Indeed,
(\ref{e-decomps}) yields
the leading order expression
\be
 \int_0^1
 \wpu(x)
\,
 dx
=
 \eps^{5/6} \, \delta \, \Omega_0^*
 \int_0^1
\,
 \wpu_0(x)
\,
 dx
\Label{biomass-aux}
\ee
for the biomass.
Here,
we have also recalled that $c = 1$
and
that $\Omega_1^* , \Omega_2^* , \ldots$
are higher order,
cf. (\ref{equil-a=1}).
Recalling the definition of $\delta$ in (\ref{delta})
and using the explicit
leading order formula (\ref{wo+}) for $\wpu_0$,
we obtain
\[
 \delta
 \int_0^1
 \wpu_0(x)
\,
 dx
=
 \eps^{-1/12}
\,
 \frac{C_1 C_2 \sL^{1/3}}{2 \sqrt{\pi}}
 \int_0^1
 F^{-1/4}(x)
 \exp
\left(
 \frac{J_-(x)-J_-(\xs)}{\sqrt{\eps}}
\right)
\,
 dx .
\]
As mentioned in Section~\ref{s-a000},
$J_-(\cdot)$ has a sole,
locally quadratic maximum
at $\xs$,
and hence
the integrand above
is exponentially small
except in an asymptotically small neighborhood
of that point.
Hence,
the integral
is of the type
considered in \ref{s-Laplace},
and Theorem~\ref{t-Laplace} yields,
to leading order,
\begin{eqnarray*}
\fl
 \delta
 \int_0^1
 \wpu_0(x)
\,
 dx
=
 \eps^{-1/12}
\,
 \frac{C_1 \, C_2 \, \sL^{1/3}}{2 \sqrt{\pi}}
\left(
 \eps^{1/4}
\,
 \frac{\sqrt{2\pi}}{F^{1/4}(\xs) \, \sqrt{-J''_-(\xs)}}
\right)
=
 \eps^{1/6}
\,
 C_1 \, C_2 \, \sL^{1/3} \, \sst^{-1/2} ,
\end{eqnarray*}
where we have also recalled that
$J''_- = -2^{-1} F^{-1/2} F'$.
Substituting back into (\ref{biomass-aux}),
together with
the formula for $\Omega_0^*$
given in (\ref{fp-Labda0=O(1)}),
we finally recover the first expression (\ref{biomass})
for the total biomass
given in the Introduction.
The second expression
may be derived by noting that
(\ref{defalphaLamb}) implies
the leading order result
$\nu/(1+j_H) = \ell + \V$,
as well as that
$\eps \, \Lambda_0 = \nu (1+j_H)^{-1} - \ell - \V$.

Similarly,
(\ref{e-decomps}) yields
the leading order formula
\be
 \int_0^1
 \n(x)
\,
 dx
=
 \eps^{5/6} \delta \, \Omega^*_0
\,
 \int_0^1
 \n_0(x)
\,
 dx
+
 \eps
 \sum_{k\ge 0}
 \Psi^*_k
 \int_0^1
 \s_k(x)
\,
 dx .
\Label{Sn}
\ee
Now,
$\int_0^1 \s_k(x) \, dx = (-1)^k/N_k$
by (\ref{vs}).
Further, the integral $\int_0^1 \n_0(x) \, dx$
can be calculated using (\ref{ODE-eta0}):
integrating both members over $[0,x]$
and
using the boundary condition at zero,
we find
\be
 \ell \, \Lambda_0
 \int_0^1
 \n_0(x)
\,
 dx
=
 \ell
\,
 \partial_x\eta_0(1)
+
 \int_0^1
 f(x)
\,
 \wpu_0(x)
\,
 dx .
\Label{Sno}
\ee
The derivative $\partial_x\eta_0(1)$
can be estimated at leading order
by (\ref{eta0}).
Differentiating both members of that formula,
we find
\begin{eqnarray*}
\fl
 \ell
\,
 \partial_x\n_0(1)
=
 \int_0^1
\left[
 \tanh\sqrt{\Lambda_0}
 \sinh\left(\sqrt{\Lambda_0} (1 - y)\right)
-
 \cosh\left(\sqrt{\Lambda_0} (1 - y)\right)
\right]
 f(y)
\,
 \wpu_0(y)
\,
 dy .
\end{eqnarray*}
It follows from (\ref{Sno}), then, that
\begin{eqnarray*}
\fl
\lefteqn{
 \ell \, \Lambda_0
 \int_0^1
 \n_0(x)
\,
 dx
}
\\
\fl
\quad
=
 \int_0^1
\left[
 1
+
 \tanh\sqrt{\Lambda_0}
 \sinh\left(\sqrt{\Lambda_0} (1 - y)\right)
-
 \cosh\left(\sqrt{\Lambda_0} (1 - y)\right)
\right]
\,
 f(y)
\,
 \wpu_0(y)
\,
 dy .
\end{eqnarray*}
Applying Theorem~\ref{t-Laplace},
we obtain
\begin{eqnarray}
 \int_0^1
 \n_0(x)
\,
 dx
=
 \eps^{1/6}
 \delta^{-1}
\,
 \frac{C_1 C_2 \, \sL^{1/3} \, \sst^{-1/2}}{\Lambda_0}
\left(
 1
-
\frac{
 \cosh\left(\sqrt{\Lambda_0} \, \xs\right)
}{
 \cosh\sqrt{\Lambda_0}
}
\right) ,
\Label{Sno-ae}
\end{eqnarray}
which is
the desired formula
for $\int_0^1 \n_0(x) \, dx$.
Recalling also (\ref{fp-Labda0=O(1)}) for $\Psi_k^*$,
we obtain from (\ref{Sn})
the leading order result
\begin{eqnarray}
\fl
 \int_0^1
 \n(x)
\,
 dx
=
 \eps
\,
 \Omega_0^*(\Lambda_0)
\left[
 \frac{C_1 C_2 \, \sL^{1/3} \, \sst^{-1/2}}{\Lambda_0}
\left(
 1
-
\frac{
 \cosh\left(\sqrt{\Lambda_0} \, \xs\right)
}{
 \cosh\sqrt{\Lambda_0}
}
\right)
+
 \alpha'
\,
 \bar{s}
\,
 \Lambda_0
\right] ,
\Label{Sn-ae}
\end{eqnarray}
where
\be
\fl
 \bar{s}
=
\frac{1}{\alpha'}
 \sum_{m \ge 0}
 (-1)^m
 \frac{A'_m(\Lambda_0)}{N_m^2}
=
 \sum_{m \ge 0}
 (-1)^m
\frac{
 \cos\left(\sqrt{N_m} \, \xs\right)
}{
 N_m^2 \, (N_m + \Lambda_0)
}
=
 \sum_{m \ge 0}
\frac{
 \sin\left(\sqrt{N_m} \, (1-\xs)\right)
}{
 N_m^2 \, (N_m + \Lambda_0)
} .
\Label{s3-expl}
\ee
This equation,
together with (\ref{fp-Labda0=O(1)}) for $\Omega_0^*$,
yields the total nutrient depletion level to leading order.

\subsection{Stability of the small pattern
\label{ss-stab}}
In this section,
we examine the stability
of the DCM-like fixed point
$(\Omega_0^* , \Psi^*)
=
(\Omega_0^* , \Psi_0^* , \Psi_1^* , \ldots)$
which we identified
in the previous section.
In particular,
we show that this fixed point
is stabilized through a transcritical bifurcation
at $\Lambda_0 = 0$
and that it subsequently undergoes
a destabilizing Hopf bifurcation.

\subsubsection{The eigenvalue equation
\label{sss-eig}}
We start by linearizing
the ODE system
\begin{eqnarray*}
 \dot\Omega_0
&=
 \Lambda_0
\,
 \Omega_0
-
 A
\,
 \Omega_0^2
-
 B
\,
 \Omega_0
 \sum_{m\ge 0}
 \Psi_m ,
\\
 \dot{\Psi}_k
&=
 -N_k
 \Psi_k
+
 A'_k
\left[
 A
\,
 \Omega_0^2
+
 B
\,
 \Omega_0
 \sum_{m\ge 0}
 \Psi_m
\right] ,
\quad\mbox{for all}\
 k \ge 0 ,
\end{eqnarray*}
around $(\Omega_0^* , \Psi^*)$.
Letting
$\Omega_0 = \Omega_0^* + \dWeq_0$
and
$\Psi_k = \Psi^*_k + \dPsi_k$
and recalling (\ref{fp-eq-sub}),
we find that
the corresponding linearized problem
reads
\begin{eqnarray}
\fl
 \dot{\dWeq}_0
&=
-
 A \, \Omega_0^*
\,
 \dWeq_0
-
 B \, \Omega_0^*
 \sum_{m\ge 0}
 \dPsi_m ,
\Label{dOmega0-ODE}\\
\fl
 \dot{\dPsi}_k
&=
 A'_k
\,
\left[
 \Lambda_0
+
 A
\,
 \Omega_0^*
\right]
\,
 \dWeq_0
+
\left[
 A'_k \, B \, \Omega_0^*
-
 N_k
\right]
\,
 \dPsi_k
+
 A'_k \, B \, \Omega_0^*
\,
 \sum_{m \ne k}
 \dPsi_m ,
\Label{dPsik-ODE}
\end{eqnarray}
where
we have only retained
the leading order component
from each term.

Truncating at the arbitrary value $k = K \in \N$,
we obtain the system
$\dot{\delta \Phi}
=
\LK \, \delta\Phi$,
where
$\delta\Phi
=
(\dWeq_0 , \dPsi_0 , \dPsi_1 , \ldots , \dPsi_K)^{\rm T}$
and
\[
\fl
 \LK
=
\left(
\begin{array}{cccccc}
 -A \, \Omega_0^* & -B \, \Omega_0^* & -B \, \Omega_0^* & \ldots & -B \, \Omega_0^* \vspace*{0.25cm} \\
 A'_0 (A \, \Omega_0^* + \Lambda_0) & A'_0 \, B \, \Omega_0^* - N_0 & A'_0 \, B \, \Omega_0^* & \ldots & A'_0 \, B \, \Omega_0^* \vspace*{0.25cm} \\
 A'_1 (A \, \Omega_0^* + \Lambda_0) & A'_1 \, B \, \Omega_0^* & A'_1 \, B \, \Omega_0^* - N_1 & \ldots & A'_1 \, B \, \Omega_0^* \vspace*{0.25cm}\\
 \vdots & \vdots & \vdots & \ddots & \vdots \vspace*{0.25cm}\\
 A'_K (A \, \Omega_0^* + \Lambda_0) & A'_K \, B \, \Omega_0^* & A'_K \, B \, \Omega_0^* & \ldots & A'_K \, B \, \Omega_0^* - N_K \vspace*{0.25cm}
\end{array}
\right) .
\]
To characterize the spectrum of this matrix,
we derive a formula
for its characteristic polynomial $\det(\Lo - \lambda I)$.
First,
we use the first row of $\Lo - \lambda I$
to eliminate the off-diagonal entries
of all other rows.
In this way,
we find that
the equation
$\det(\Lo - \lambda I) = 0$
is equivalent to
setting to zero the determinant
\[
\left\vert
\begin{array}{ccccc}
 \lambda + A \, \Omega_0^*
&
 B \, \Omega_0^*
&
 B \, \Omega_0^*
&
 \ldots
&
 B \, \Omega_0^*
\\
 A'_0 (\lambda - \Lambda_0)
&
 \lambda + N_0
&
 0
&
 \ldots
&
 0 \vspace*{0.25cm}
\\
 A'_1 (\lambda - \Lambda_0)
&
 0
&
 \lambda + N_1
&
 \ldots
&
 0 \vspace*{0.25cm}
\\
 \vdots
&
 \vdots
&
 \vdots
&
 \ddots
&
 \vdots \vspace*{0.25cm}
\\
 A'_K (\lambda - \Lambda_0)
&
 0
&
 0
&
 \ldots
&
 \lambda + N_K
\end{array}
\right\vert .
\]
Next,
we can use the $(k+2)-$nd column
to eliminate the $(k+2)-$nd entry
of the first column,
for $0 \le k \le K$,
as long as $\lambda \ne -N_k$.
Since $\lambda = -N_k$
if and only if
$A'_k = 0$
(as can be shown
by expanding the determinant
along the $(k+2)-$nd row),
we can eliminate all entries
of the first column.
(Note that
$A'_k$ may indeed be zero:
indeed,
$A'_k$ is proportional to
$\cos((k + 1/2) \pi \, \xs)$,
which may or may not be zero
depending on the values
of $k$ and $\xs$.)
Defining
${\cal K}
=
\{ k : A'_k \ne 0\}
\subset
\{0 , \ldots , K\}$,
${\cal K}_k
=
{\cal K} - \{k\}$,
and
eliminating the entries
of the first column
as detailed above,
we obtain
\be
\left\vert
\begin{array}{ccccc}
 Q(\lambda)
&
 B \, \Omega_0^*
&
 B \, \Omega_0^*
&
 \ldots
&
 B \, \Omega_0^* \vspace*{0.25cm}
\\
 0
&
 \lambda + N_0
&
 0
&
 \ldots
&
 0 \vspace*{0.25cm}
\\
 0
&
 0
&
 \lambda + N_1
&
 \ldots
&
 0 \vspace*{0.25cm}
\\
 \vdots
&
 \vdots
&
 \vdots
&
 \ddots
&
 \vdots \vspace*{0.25cm}
\\
 0
&
 0
&
 0
&
 \ldots
&
 \lambda + N_K
\end{array}
\right\vert
=
 0 .
\Label{Lo-det-Lambda0=O(1)}
\ee
Here,
\begin{eqnarray*}
 Q(\lambda)
&=
 (\lambda + A \, \Omega_0^*)
 \prod_{k \in {\cal K}}
 (\lambda + N_k)
+
 B \, \Omega_0^* \, (\Lambda_0 - \lambda)
 \sum_{k \in {\cal K}}
 A'_k
 \prod_{m \in {\cal K}_k}
 (\lambda + N_m) .
\end{eqnarray*}
As detailed above,
$\lambda = -N_k$ solves (\ref{Lo-det-Lambda0=O(1)})
if and only if
$A'_k = 0$
(equivalently,
if and only if
$k \not\in {\cal M}$).
Further,
$\Lambda_0 > 0$ cannot be an eigenvalue,
since $Q(\Lambda_0) > 0$
and
$\Lambda_0 + \N_k > 0$,
for all $k \in \{0 , \ldots , K\}$---note that
$A , N_0 , N_1 , \ldots , N_K$
are all positive constants.
Hence,
we can extend the set
over which we sum
in the formula above
to the entire set $\{0 , \ldots , K\}$
and
rewrite the equation for $Q(\lambda)$ in the form
\[
 Q(\lambda) =
\left[
 B \, \Omega_0^*
 \sum_{k = 0}^K
 \frac{A'_k}{N_k + \lambda}
-
\frac{
 \lambda + A \, \Omega_0^*
}{
 \lambda - \Lambda_0
}
\right]
 (\Lambda_0 - \lambda)
 \prod_{k \in {\cal K}}
 (N_k + \lambda) .
\]

As we just noted,
the elements of the set
$\{-N_k\}_{k \in \cal K}$
are not eigenvalues of $\Lo$.
Hence,
the eigenvalues of $\Lo$ are
$\{-N_k\}_{k \not\in \cal K}$
together with all solutions to
\[
 B \, \Omega_0^*
 \sum_{k = 0}^K
 \frac{A'_k}{N_k + \lambda}
=
\frac{
 \lambda + A \, \Omega_0^*
}{
 \lambda - \Lambda_0
} .
\]
Substituting for $A'_k$ from (\ref{5const})
and
for $\Omega_0^*$ from (\ref{fp-Labda0=O(1)}),
recalling the identity (\ref{2a=a'B}),
and letting $K \to \infty$,
we rewrite this equation in the form
\[
 \frac{2 \Lambda_0}{1-\xs}
\,
 \sum_{k \ge 0}
\frac{
 \cos\left(\sqrt{N_k} \, \xs\right)
}{
 (N_k + \lambda) \, (N_k + \Lambda_0)
}
=
\frac{
 \lambda + \Lambda_0 \, a(\Lambda_0)/(1-\xs)
}{
 \lambda - \Lambda_0
} .
\]
Here again,
we may write
\begin{eqnarray*}
\fl
 \sum_{k \ge 0}
\frac{
 \cos\left(\sqrt{N_k} \, \xs\right)
}{
 (N_k + \lambda) \, (N_k + \Lambda_0)
}
&=&
 \frac{1}{\lambda - \Lambda_0}
\left(
 \sum_{k \ge 0}
\frac{
 \cos\left(\sqrt{N_k} \, \xs\right)
}{
 N_k + \Lambda_0
}
-
 \sum_{k \ge 0}
\frac{
 \cos\left(\sqrt{N_k} \, \xs\right)
}{
 N_k + \lambda
}
\right)
\\
&=&
 \frac{1}{2}
 \frac{a(\Lambda_0) - a(\lambda)}{\lambda - \Lambda_0} ,
\end{eqnarray*}
so that the eigenvalue problem becomes
$(1-\xs) \, \lambda + \Lambda_0 \, a(\lambda) = 0$.
Recalling that $a(0) = 1-\xs$,
we recast this equation as
\be
 \lambda
\,
 \frac{a(0)}{a(\lambda)}
=
 -\Lambda_0 ,
\quad\mbox{where we recall that}\
 a(\lambda)
=
\frac{
 \sinh\left(\sqrt{\lambda}(1-\xs)\right)
}{
 \sqrt{\lambda}
\,
 \cosh\sqrt{\lambda}
} .
\Label{eig-eq-sym}
\ee
This equation is satisfied
by some $\lambda$
if and only if
it is also satisfied
by its complex conjugate $\lambda^*$,
as the right member is real
and
$(\lambda^{-1} a(\lambda))^*
=
(\lambda^*)^{-1} a(\lambda^*)$.
Hence,
we may restrict $\arg(\lambda)$
to lie in $[0 , \pi]$.
Further writing
$\mu := \sqrt{\lambda} = \mu_R + \im \, \mu_I$,
we rewrite the eigenvalue equation
in its final form,
\be
 p(\mu)
:=
-
\frac{
 (1-\xs)
\,
 \mu^3
\,
 \cosh\mu
}{
 \sinh\left((1-\xs) \, \mu\right)
}
=
 \Lambda_0 ,
\quad\mbox{with}\
 \arg(\mu)
\in
 [0,\pi/2] .
\Label{eig-eq}
\ee
We note here for later use that
\begin{eqnarray*}
\fl
 \frac{{\rm Re}(p(\mu))}{1-\xs}
&=&
 \mu_R (3 \mu_I^2 - \mu_R^2)
\frac{
 \sinh[(2-\xs) \mu_R]
 \cos(\xs \mu_I)
-
 \sinh(\xs \mu_R)
 \cos[(2-\xs) \mu_I]
}{
 \cosh[2 (1-\xs) \mu_R]
-
 \cos[2 (1-\xs) \mu_I]
}
\\
\fl
&{}&
+
 \mu_I (3 \mu_R^2 - \mu_I^2)
\frac{
 \cosh[(2-\xs) \mu_R]
 \sin(\xs \mu_I)
-
 \cosh(\xs \mu_R)
 \sin[(2-\xs) \mu_I]
}{
 \cosh[2 (1-\xs) \mu_R]
-
 \cos[2 (1-\xs) \mu_I]
} ,
\\
\fl
 \frac{{\rm Im}(p(\mu))}{1-\xs}
&=&
 \mu_I (\mu_I^2 - 3 \mu_R^2)
\frac{
 \sinh[(2-\xs) \mu_R]
 \cos(\xs \mu_I)
-
 \sinh(\xs \mu_R)
 \cos[(2-\xs) \mu_I)]
}{
 \cosh[2 (1-\xs) \mu_R]
-
 \cos[2 (1-\xs) \mu_I]
}
\\
\fl
&{}&
+
 \mu_R (3 \mu_I^2 - \mu_R^2)
\frac{
 \cosh[(2-\xs) \mu_R]
 \sin(\xs \mu_I)
-
 \cosh(\xs \mu_R)
 \sin[(2-\xs) \mu_I]
}{
 \cosh[2 (1-\xs) \mu_R]
-
 \cos[2 (1-\xs) \mu_I]
} .
\end{eqnarray*}

\subsubsection{Analysis of (\ref{eig-eq})
for $\Lambda_0 \downarrow 0$}
We first establish that,
as $\Lambda_0 \downarrow 0$,
the eigenvalues $\{\lambda_n\}_{n \ge -1}$
remain each in a neighborhood
of the discrete values
$-\Lambda_0, -N_0 , -N_1 , \ldots $.

\begin{figure}[t]
\hspace*{-2cm}
\scalebox{0.445}[0.32]{
\includegraphics{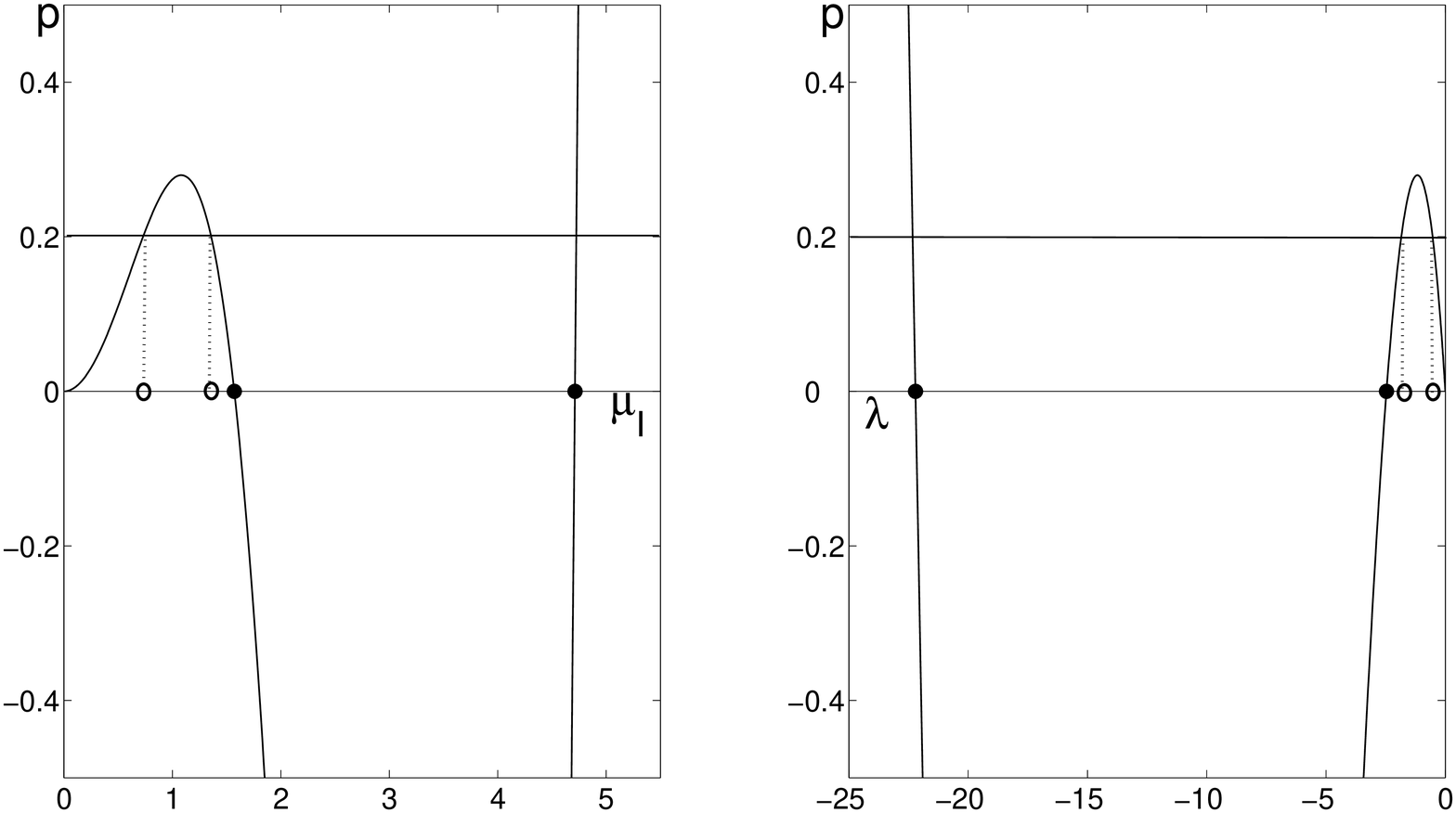}
}
\caption{
\label{f-p(imu)}
Plots of the function $p(\im \mu_I)$
(see (\ref{pim}))
versus $\mu_I$ (left panel)
and versus $\lambda = -\mu_I^2$ (right panel)
for $\xs = 0.7$.
Also plotted:
the level line at $p = \Lambda_0$,
here set at $0.2$;
the first two members
of the sequence
$\{\sqrt{N}_k\}_{k \ge 0}$ (left panel)
and
$\{-N_k\}_{k \ge 0}$ (right panel)
as solid dots;
and
the two smallest solutions
$\mu_I$ to (\ref{pim}) (left panel)
together with
the first two eigenvalues
$\lambda_{-1}$ and $\lambda_0$ (right panel)
they correspond to,
all as hollow dots.
}
\end{figure}

For $\Lambda_0 = 0$,
(\ref{eig-eq}) yields either $\mu = 0$
(equivalently, $\lambda = 0$)
or $\cosh\mu = 0$
(whence $\mu = \im \, \sqrt{N_m}$, $m \ge 0$ or,
equivalently,
$\lambda \in \{-N_m\}_{m \ge 0}$).
To investigate the possibility
of negative eigenvalues $\lambda$
for $\Lambda_0 > 0$,
we set $\mu_R = 0$ to find that
(\ref{eig-eq}) reduces to
\be
 p(\im \mu_I)
=
\frac{
 (1-\xs)
\,
 \mu_I^3
}{
 1 - \cos[2 (1-\xs) \mu_I]
}
 \sin[(1-\xs) \, \mu_I]
\,
 \cos\mu_I
=
 \Lambda_0 .
\Label{pim}
\ee
For $\Lambda_0 \downarrow 0$,
there is plainly
a small root of this equation,
$\mu_I = \sqrt{\Lambda_0} + \Or(\Lambda_0)$,
yielding the small eigenvalue
$\lambda = -\Lambda_0 + \Or(\Lambda_0^2)$.
Additionally,
all eigenvalues
of the set $\{-N_m\}_{m \ge 1}$
perturb and remain real
for small enough values of $\Lambda_0 > 0$.
Indeed,
$p(\im \cdot)$ intersects zero transversally
at $\{\sqrt{N_m}\}_{m \ge 0}$,
whence
the persistence of
any finite number of eigenvalues
from among this set
is automatically established.
That the remaining,
infinitely many eigenvalues also persist
can be established
by noting that,
if the maximum value of $p(\im \cdot)$
is positive between successive zeros,
then this value
grows unboundedly with $\mu_I$.
For the two first eigenvalues,
in particular,
we have the Taylor expansions
\[
 \lambda_{-1}
=
 -\Lambda_0
+
 \Or(\Lambda_0^2)
\quad\mbox{and}\quad
 \lambda_0
=
 -N_0
+
 4 
 \frac{\sin[(1-\xs) \, \pi/2]}{(1-\xs) \, \pi}
 \Lambda_0
+
 \Or(\Lambda_0) ,
\]
which demonstrate that both remain
in the interval $(-N_0,0)$
and approach each other
as $\Lambda_0$ increases,
see also Figure~\ref{f-p(imu)}.
These are precisely
the two first eigenvalues that collide
as $\Lambda_0$ is increased,
yielding a pair of
complex conjugate eigenvalues.

Next,
the possibility of \emph{positive} eigenvalues
$\lambda$---equivalently,
positive solutions of (\ref{eig-eq})---can be excluded
by noticing that $-\Lambda_0 < 0$
while $p(\mu) > 0$ for all $\mu > 0$.
In fact,
the possibility of eigenvalues
anywhere but in a neighborhood
of the negative axis
can be similarly excluded
by observing that
\[
\fl
 \abs{p(\mu)}
=
 (1 - \xs)
 \abs{\mu}^3
\left(
\frac{
 \cosh(2 \mu_R) + \cos(2 \mu_I)
}{
 \cosh[2 (1-\xs) \mu_R] - \cos[2 (1-\xs) \mu_I]
}
\right)^{1/2}
\to
 \infty ,
\quad\mbox{as}\quad
 \mu_R
\to
 \infty .
\]
Plainly,
for every value of $\Lambda_0$,
there exists a value $\mu_R^*(\Lambda_0) > 0$
which depends continuously on $\Lambda_0$,
satisfies
$\lim_{\Lambda_0 \to 0}\mu_R^*(\Lambda_0) = 0$,
and
is such that the equation
$\abs{p(\mu)} = \abs{\Lambda_0}$
\emph{cannot} be satisfied
for any $\mu_R > \mu_R^*(\Lambda_0)$.
It follows that
all solutions to (\ref{eig-eq})
must lie in the half plane
$\{ \mu \, \vert \, \mu_R \le \mu_R^*(\Lambda_0)\}$
which, in turn,
corresponds to a neighborhood
of the half axis
$\{\lambda \in \R
\, \vert \,
\lambda \le \abs{\mu_R^*(\Lambda_0)}^2\}$.
A local analysis around the origin
now establishes the absence of eigenvalues
with positive real parts,
for $\Lambda_0$ small enough,
and hence also the result.

%

\subsubsection{Complexification of eigenvalues
and the Hopf bifurcation}
As we briefly mentioned
in the last section
in conjunction with Figure~\ref{f-p(imu)},
the two principal eigenvalues
$\lambda_{-1}$ and $\lambda_0$
come closer together
as $\Lambda_0$ increases.
Eventually,
they collide
at a specific value $\mu_I' \in (0,\pi/2)$
and for
$\Lambda_0
=
\Lambda_0'
=
p(\im \mu_I')
=
\max_{\mu_I \in (0,\pi/2)} p(\im \mu_I)
>
0$.
For $\Lambda_0 > \Lambda_0'$,
this pair of eigenvalues becomes complex,
so it is natural to examine
whether it crosses into the right half-plane
through the imaginary axis.
(Note that no eigenvalues
can cross through zero,
as
(\ref{eig-eq-sym}) does not admit
a zero eigenvalue
for $\Lambda_0 > 0$.)

\begin{figure}[t]
\hspace*{-1.9cm}
\scalebox{0.44}[0.32]{
\includegraphics{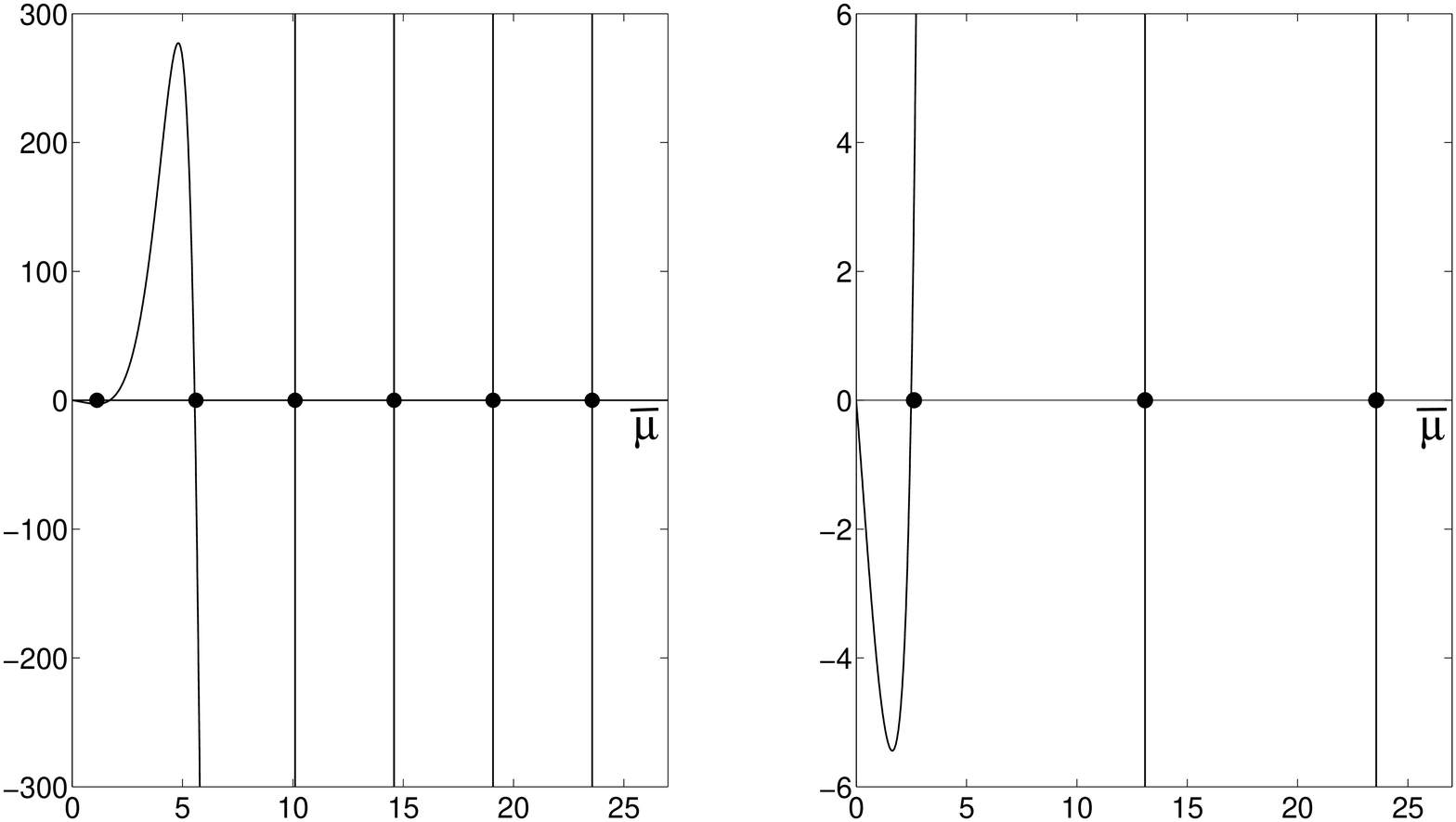}
}
\caption{
\label{f-im_cond}
Plots of the function
in the left member of (\ref{cond-Im-exp})
for two distinct values of $\xs$,
namely $\xs = 0.7$ (left panel)
and $\xs = 0.3$ (right panel).
The solid dots
mark the members
of the sequence
$\{(k + 1/4) \pi \, \xs^{-1}\}_{k \ge 0}$.
Note that the zeros
of the plotted function
approach this sequence
rather quickly,
with the quality
of the approximation
decreasing as $\xs \uparrow 1$.
Indeed,
in that regime,
all exponentials appearing
in (\ref{cond-Im-exp})
remain approximately equal
for a large range
of $\bar{\mu}-$values,
and hence
the first term becomes dominant
only in the far range.
}
\end{figure}

To locate imaginary eigenvalues
$\lambda = \im \lambda_I \in \im \R$,
we set $\mu_R = \mu_I = \bar{\mu} > 0$
and
rewrite the real and imaginary parts
of $p$ as
\begin{eqnarray*}
\fl
 {\rm Re}(p(\mu))
&=&
 2 (1-\xs) \bar{\mu}^3
\left[
\frac{
 \cosh[(2-\xs) \bar{\mu}]
 \sin(\xs \bar{\mu})
-
 \cosh(\xs \bar{\mu})
 \sin[(2-\xs) \bar{\mu}]
}{
 \cosh[2 (1-\xs) \bar{\mu}]
-
 \cos[2 (1-\xs) \bar{\mu}]
}
\right.
\\
\fl
&{}&
\left.
\hspace*{2.5cm}
+
\frac{
 \sinh[(2-\xs) \bar{\mu}]
 \cos(\xs \bar{\mu})
-
 \sinh(\xs \bar{\mu})
 \cos[(2-\xs) \bar{\mu}]
}{
 \cosh[2 (1-\xs) \bar{\mu}]
-
 \cos[2 (1-\xs) \bar{\mu}]
}
\right] ,
\\
\fl
 {\rm Im}(p(\mu))
&=&
 2 (1-\xs) \bar{\mu}^3
\left[
\frac{
 \cosh[(2-\xs) \bar{\mu}]
 \sin(\xs \bar{\mu})
-
 \cosh(\xs \bar{\mu})
 \sin[(2-\xs) \bar{\mu}]
}{
 \cosh[2 (1-\xs) \bar{\mu}]
-
 \cos[2 (1-\xs) \bar{\mu}]
}
\right.
\\
\fl
&{}&
\left.
\hspace*{2.5cm}
-
\frac{
 \sinh[(2-\xs) \bar{\mu}]
 \cos(\xs \bar{\mu})
-
 \sinh(\xs \bar{\mu})
 \cos[(2-\xs) \bar{\mu})]
}{
 \cosh[2 (1-\xs) \bar{\mu}]
-
 \cos[2 (1-\xs) \bar{\mu}]
}
\right] .
\end{eqnarray*}
The condition ${\rm Im}(p(\mu)) = 0$,
derived from (\ref{eig-eq}),
yields
\begin{eqnarray}
\fl
 \cosh[(2-\xs) \bar{\mu}]
 \sin(\xs \bar{\mu})
-
 \cosh(\xs \bar{\mu})
 \sin[(2-\xs) \bar{\mu}]
\nonumber\\
=
 \sinh[(2-\xs) \bar{\mu}]
 \cos(\xs \bar{\mu})
-
 \sinh(\xs \bar{\mu})
 \cos[(2-\xs) \bar{\mu})] .
\Label{cond-Im}
\end{eqnarray}
Therefore,
the equation ${\rm Re}(p(\mu)) = \Lambda_0$,
similarly derived from (\ref{eig-eq}),
becomes
\be
\fl
 4 (1-\xs) \bar{\mu}^3
\,
\frac{
 \cosh[(2-\xs) \bar{\mu}]
 \sin(\xs \bar{\mu})
-
 \cosh(\xs \bar{\mu})
 \sin[(2-\xs) \bar{\mu}]
}{
 \cosh[2 (1-\xs) \bar{\mu}]
-
 \cos[2 (1-\xs) \bar{\mu}]
}
=
 \Lambda_0 .
\Label{cond-Re}
\ee
Condition~(\ref{cond-Im}) determines
the values of $\bar{\mu}$
corresponding to imaginary eigenvalues
$\lambda = 2 \im \bar{\mu}^2$,
while (\ref{cond-Re}) yields
the corresponding values of $\Lambda_0$
for which these eigenvalues appear.
Since the former of these
can be recast as
\begin{eqnarray}
\fl
 \ex^{(2-\xs) \bar{\mu}}
 \sin\left(\xs \bar{\mu} - \frac{\pi}{4}\right)
&-&
 \ex^{\xs \bar{\mu}}
 \sin\left((2-\xs) \bar{\mu} - \frac{\pi}{4}\right)
\nonumber\\
&+&
 \ex^{-(2-\xs) \bar{\mu}}
 \sin\left(\xs \bar{\mu} + \frac{\pi}{4}\right)
-
 \ex^{-\xs \bar{\mu}}
 \sin\left((2-\xs) \bar{\mu} + \frac{\pi}{4}\right)
\!=\!
 0 ,
\hspace*{1cm}
\Label{cond-Im-exp}
\end{eqnarray}
we see that there exists
a whole, discrete sequence
$\{\bar{\mu}_k\}_{k \ge 0}$
of values $\bar{\mu}$,
see also Figure~\ref{f-im_cond}.
As $k \to \infty$,
$\{\bar{\mu}_k\}_{k \ge 0}$
limits to
$\{(k + 1/4) \pi \, \xs^{-1}\}_{k \ge 0}$,
the sequence of
the set of zeroes
of the first term
in (\ref{cond-Im-exp})
which becomes dominant
in the regime
$\bar{\mu} \to \infty$.
Equation~(\ref{cond-Re}) yields
the leading order result
\[
 \Lambda_0
=
 2 \sqrt{2} \, \pi^3
 (1-\xs) \, \xs^{-3}
 (-1)^k k^3 \, \ex^{(k+1/4)\pi} ,
\quad\mbox{as}\
 k \to \infty ,
\]
which establishes that
the values $\bar{\mu}_k$
corresponding to even values of $k$
yield a positive, increasing sequence
of values of $\Lambda_0$.
(Odd $k-$values yield
negative $\Lambda_0-$values.)
In particular,
the first Hopf bifurcation
occurs at an $\Or(1)$ value of $\Lambda_0$
when the complex conjugate pair
$(\lambda_{-1} , \lambda_0)$
crosses into the right half-plane
through $\bar{\mu}_0$.
Higher, even $k-$values
correspond to Hopf bifurcations
occurring at higher values of $\Lambda_0$,
presumably when
higher order eigenvalues
cross into the right half-plane.

These last remarks
conclude our discussion
of the DCM-like steady state
for $\Or(1)$ values of $\Lambda_0$.
In the next section,
we investigate
a logarithmic scaling for $\Lambda_0$
in which
the number of steady states
of the system
(\ref{fp-eq1})--(\ref{fp-eq2})
becomes two.

%

\subsection{A second DCM pattern
\label{ss-destroy}}
So far,
we have identified
a DCM pattern
corresponding to an $\Or(\eps)$ biomass
which is stabilized
through a transcritical bifurcation
at $\Lambda_0 = 0$
and
destabilized through a secondary, Hopf bifurcation
at an $\Or(1)$ value of $\Lambda_0$.
Here,
we show that,
the system (\ref{Omega0'})--(\ref{Omegak'})
admits a second,
asymptotically larger,
DCM-like steady state
corresponding to an $\Or(\eps^{1/2})$ biomass.
We refrain from establishing
the stability type,
origins and eventual fate
of that second steady state,
reserving those problems
to a later work.

We start by noting that
the inclusion of
the first higher order term
in the formula for $a'_{00k}$
reported in (\ref{coeffs-ae})
yields
\[
 a'_{00k}
=
-
 A'_k(\Lambda_0)
\,
 A(\Lambda_0)
+
 \eps^{1/2}
\,
 \alpha
\,
 \tilde{A}'_k(\Lambda_0) .
\]
This formula is derived
in Section~\ref{ss-a00k-Lambda0>>1},
see (\ref{a00k-ae-Lambda0>>1})
in particular.
Here,
the $\Lambda_0-$independent constants
$\alpha$ and $\alpha'$
were defined in
(\ref{defbarCa}) and (\ref{5const}),
respectively,
whereas the functions
$a$ and $a'$
are reported in (\ref{a000-ae}) and (\ref{5const}).
Also,
$\tilde{A}'_k(\Lambda_0)
=
\tilde{\alpha}' \, \tilde{a}'(\Lambda_0) \, \cos\left(\sqrt{N_k} \, \xs\right)$,
with
\begin{eqnarray}
\fl
 \tilde{\alpha}'
&=&
\frac{
 C_1 \, C_2 \, \sL^{1/3} \, \sst^{-1/2}
}{
 \sqrt{2} \, f(0)
} ,
\Label{ta'-def}
\\
\fl
 \tilde{a}'(\Lambda_0)
&=&
\frac{
 \sinh\left(\sqrt{\Lambda_0} \, (1-\xs)\right)
\,
 \int_0^\xs
 f(x) \, \cosh\left(\sqrt{\Lambda_0} \, x\right)
\,
 dx
}{
 \sqrt{\Lambda_0} \, \cosh\sqrt{\Lambda_0}
} .
\Label{ta'k-def}
\end{eqnarray}
This formula for $\tilde{a}'(\Lambda_0)$
is also valid
in a logarithmic regime for $\Lambda_0$,
see (\ref{Lambda0-log}) for details.
Since the first term
in the formula for $a'_{00k}$ above
decreases exponentially with $\Lambda_0$
(see (\ref{defbarCa}))
whereas
the second term
decreases only algebraically,
the two terms become
asymptotically comparable
for values of $\Lambda_0$
logarithmically large in $\eps$,
see Section~\ref{s-a00k} for details.

Replacing the formula for $a'_{00k}$
in (\ref{coeffs-ae})
by the formula above,
substituting into
(\ref{Omega0'})--(\ref{Omegak'}),
and working as
in Section~\ref{ss-Lambda=O(1)-slaving},
we obtain the system
\be
\begin{array}{rcl}
 \dot{\Omega}_0
&=&
 \Lambda_0
 \Omega_0
-
 A \, \Omega_0^2
-
 B \, \Omega_0
 \sum_{m\ge 0}
 \Psi_m
\ \ ,
\\
 \dot{\Psi}_k
&=&
 -N_k
 \Psi_k
+
\left[
 (A'_k \, A - \eps^{1/2} \, \alpha \, \tilde{A}'_k)
\,
 \Omega_0^2
+
 A'_k \, B \, \Omega_0
 \sum_{m\ge 0}
 \Psi_m
\right] .
\end{array}
\Label{redux-ODE-a=1-HOT}
\ee
This is the analogue of (\ref{redux-ODE-a=1})
with the inclusion of
higher order terms.
The fixed points $(\Omega_0^* , \Psi^*)$
of this system
are found,
here again,
by setting the left members to zero,
\be
\begin{array}{rcll}
 \Lambda_0
&=&
 A \, \Omega_0^*
+
 B
 \sum_{m\ge 0}
 \Psi_m^*
\ \ ,
&
 {}
\\
 0
&=&
 -N_k
 \Psi_k^*
+
 A'_k \, \Lambda_0 \, \Omega_0^*
-
 \eps^{1/2} \alpha \tilde{A}'_k
\,
 (\Omega_0^*)^2 .
\end{array}
\Label{fp-eq-HOT}
\ee
Solving the second equation for $\Psi_k^*$,
we find
an explicit expression for $\Psi_k^*$
in terms of $\Omega_0^*$,
\[
 \Psi_k^*
=
 \frac{A'_k}{N_k} \, \Lambda_0 \, \Omega_0^*
-
 \eps^{1/2}
 \alpha \frac{\tilde{A}'_k}{N_k}
\,
 (\Omega_0^*)^2 .
\]
Substituting this expression
into the first equation
in (\ref{fp-eq-HOT}),
we recover
a singularly perturbed
quadratic equation for $\Omega_0^*$,
\be
 \eps^{1/2} \alpha B
\left(
 \sum_{m\ge 0}
 \frac{\tilde{A}'_m}{N_m}
\right)
 (\Omega_0^*)^2
-
\left(
 A
+
 B \, \Lambda_0
\,
 \sum_{m\ge 0}
 \frac{A'_m}{N_m}
\right)
\,
 \Omega_0^*
+
 \Lambda_0
=
 0 .
\Label{Omega0-fp-quadr}
\ee
In Section~\ref{sss-derive},
we obtained the formula
\[
 A
+
 B \, \Lambda_0
\,
 \sum_{m\ge 0}
 \frac{A'_m}{N_m}
=
 \alpha \, a(0) ,
\]
while (\ref{ML}) yields
\[
 \sum_{m\ge 0}
 \frac{\tilde{A}'_m}{N_m}
=
 \tilde{\alpha}' \, \tilde{a}'(\Lambda_0)
 \sum_{m\ge 0}
 \frac{\cos(\sqrt{N_m} \, \xs)}{N_m}
=
 \frac{a(0)}{2}
 \tilde{\alpha}' \, \tilde{a}'(\Lambda_0) .
\]
It follows that
the quadratic equation
yielding $\Omega_0^*$
can be recast as
\[
 \eps^{1/2}
\,
 \frac{\tilde{\alpha}' \, B \, \tilde{a}'(\Lambda_0)}{2}
\,
 (\Omega_0^*)^2
-
 \Omega_0^*
+
 \frac{\Lambda_0}{\alpha \, a(0)}
=
 0 .
\]
The two solutions
of this equation are
\[
\fl
 \Omega_0^{*,\pm}
=
 \eps^{-1/2}
\,
\frac{
 1
\pm
\sqrt{
 1
-
 2 \eps^{1/2}
\,
 \tilde{\alpha}' \, B
\,
 \Lambda_0 \, \tilde{a}'(\Lambda_0)/(\alpha \, a(0)})
}{
 \tilde{\alpha}' \, B \, \tilde{a}'(\Lambda_0)
}
=
\Big\{
\begin{array}{l}
 2 \eps^{-1/2}/(\tilde{\alpha}' \, B \, \tilde{a}'(\Lambda_0)) ,
\\
 \Lambda_0/(\alpha \, a(0)) ,
\end{array}
\]
with the first one
corresponding to the asymptotically larger DCM-pattern
and
the second one
corresponding to the DCM-pattern
identified trough our earlier work.
We remark here that
this first steady-state
is, indeed,
within the reach
of our asymptotic methods,
as $\Omega_0^*$ and $\Psi_k^*$
safely remain asymptotically smaller
than the asymptotic bounds
$\eps^{-3/4}$ and $\eps^{-1}$
for which our work
in Section~\ref{sss-Lfpu-decomp}
remains valid.
Note also that
this steady state
is a \emph{nonlinear} function of $\Lambda_0$,
with the distinguished limits
\[
\fl
 \lim_{\Lambda_0 \to 0}
 \Omega_0^*(\Lambda_0)
=
\frac{
 2 \eps^{-1/2}
}{
 (1-\xs)
\,
 \tilde{\alpha}'
\,
 B
\,
 \int_0^\xs f(x) \, dx
}
\quad\mbox{and}\quad
 \Omega_0^*(\Lambda_0)
=
\frac{
 4 \eps^{-1/2}
}{
 \ell
\,
 \tilde{\alpha}'
\,
 B
}
\,
 \Lambda_0 ,
\ \mbox{as} \
 \Lambda
\to
 \infty .
\]
In particular,
this second pattern
approaches a nonzero value
as $\Lambda_0 \downarrow 0$
and
eventually grows linearly
for $\Lambda_0 \gg 1$.

\section{An asymptotic formula for $b_{m00}$
\label{s-bm00}}
\setcounter{equation}{0}
In this section,
we derive
the asymptotic formula
for $b_{m00}$ given in (\ref{coeffs-ae}),
where $m \in \N$
and
\be
 b_{m00}
=
 (1 - \nu)
 \int_0^1
\,
 f(x)
\,
 \s_m(x)
\,
 \w_0^2(x)
\,
 dx .
\Label{bm00-expl}
\ee
As detailed earlier,
the function $\w_0$,
appearing in (\ref{bm00-expl}),
decays exponentially outside
an $\Or(\eps^{1/3})-$neighborhood
of the origin
(cf. (\ref{wo})),
whereas the period
of the sinusoidal term $\s_m$
is equal to
$2 \pi/\sqrt{N_m} = 4/(2m + 1)$.
Below,
we analyze
the three different regimes---in which
the integrand is
predominantly localized,
concurrently localized and oscillatory,
and predominantly oscillatory---and
we derive
the leading order,
uniform asymptotic expansion
\be
\fl
 b_{m00}
=
\left\{
\begin{array}{l}
\displaystyle
 b ,
\quad\mbox{for}\
 m \ll \eps^{-1/3} ,
\vspace*{0.2cm}\\
\displaystyle
 b \,C_1^2
 \int_0^\infty
 \cos\left(
 \frac{\eps^{1/3} \, \sqrt{N_m} \, \tau}{\sL^{1/3}}
 \right)
 \Ai^2\left(\tau + A_1\right)
\,
 d\tau ,
\quad\mbox{for}\
 m = \Or(\eps^{-1/3}) ,
\vspace*{0.2cm}\\
\displaystyle
-
 b
\,
\frac{
 6 \, C_3 \, C_1^2 \, \sL^2
}{
 \eps \, N_m^2 \, f(0)
} ,
\quad\mbox{for}\
 m \gg \eps^{-1/3} .
\end{array}
\right.
\Label{bm00-ae}
\ee
Here,
$b = \sqrt{2} \, (1 - \nu) \, f(0)$,
cf. Section~\ref{ss-coeff-ae}.

\subsection{The case $m \ll \eps^{-1/3}$}
Here,
$2 \pi/\sqrt{N_m} \gg \eps^{1/3}$
and hence
the integrand is predominantly localized
around $x=0$.
Thus,
$\s_m$ may be approximated
to leading order by
$\s_m(0) = \sqrt{2}$
in that neighborhood.
Since $\| \w_0 \|_2 = 1$
(cf. our discussion
in Sections~\ref{ss-eigen}--\ref{ss-dual}),
we obtain the desired formula
\be
 b_{m00}
\sim
 b .
\Label{bm00-aeI}
\ee
%

\subsection{The case $m =\Or(\eps^{-1/3})$}
Here,
$2 \pi/\sqrt{N_m} =\Or(\eps^{1/3})$,
and hence
the neighborhood
of the origin
outside which
$\w_0$ decays exponentially
and
the period
of the sinusoidal term
are of the same asymptotic magnitude.
Defining the new variable
$\tau = \tau_1 \, x$
in (\ref{bm00-expl}),
with
$\tau_1 = \abs{A_1}/\xo$ (\ref{barx0}),
we obtain
\be
 b_{m00}
=
 \frac{\sqrt{2} \, (1 - \nu)}{\tau_1}
 \int_0^{\tau_1}
 f\left(\frac{\tau}{\tau_1}\right)
\,
 \cos\left(\sqrt{N_m} \, \frac{\tau}{\tau_1}\right)
\,
 \w_0^2\left(\frac{\tau}{\tau_1}\right)
\,
 d\tau .
\Label{bm00-resc}
\ee
Now,
(\ref{wo}) yields,
to leading order
and
for any $\tau_0 \ll \eps^{-1/3}$,
\[
\fl
 \w_0\left(\frac{\tau}{\tau_1}\right)
=
\left\{
\begin{array}{l}
 \eps^{-1/6}
\,
 C_1
\,
 \sL^{1/6}
 \Ai\left(\tau + A_1\right) ,
\hspace*{4.25cm}
\quad\mbox{for}\
 \tau \in [ 0 , -A_1 ) ,
\\
\frac{
 \eps^{-1/12}
\,
 C_1 \, C_2 \, \sL^{1/3}
}{
 2 \sqrt{\pi} \, F^{1/4}(\tau/\tau_1)
}
\,
\exp\left(
 -\frac{1}{\sqrt{\eps} \tau_1}
 \int_0^{\tau + A_1}
 \sqrt{F(\frac{t}{\tau_1} + \xo) - F(\xo)} \, dt
\right) ,
\\
\hspace*{9cm}
\mbox{for}\
 \tau \in ( -A_1 , \tau_0 ] ,
\end{array}
\right.
\]
where
we have also changed
the integration variable
by means of
$s = t/\tau_1 + \xo$.
These two formulas agree---as, indeed,
they should by construction---in the regime
$1 \ll \tau_0 \ll \eps^{-1/3}$.
Indeed,
recalling the asymptotic expansion
of $\Ai$ in a neighborhood of infinity
\cite{BO-1999},
we find that
the first branch of
the formula above
yields
\[
 \eps^{-1/6}
\,
\frac{
 C_1 \, \sL^{1/6}
}{
 2 \, \sqrt{\pi} \, \tau^{1/4}
}
\,
 \exp\left(-\frac{2}{3} (\tau + A_1)^{3/2}\right) .
\]
Similarly,
the formula
in the case
$\tau \in (\abs{A_1},\tau_0]$
becomes,
upon Taylor-expanding $F$,
\[
 \eps^{-1/12}
\,
\frac{
 C_1 \, C_2 \, \sL^{1/12} \, \tau_1^{1/4}
}{
 2 \sqrt{\pi} \, \tau^{1/4}
}
\,
\exp\left(
 -\frac{2}{3}
 \sqrt{\frac{\sL}{\eps}}
 \left(\frac{\tau + A_1}{\tau_1}\right)^{3/2}
\right) .
\]
That the two formulas agree
now follows from
the definition $\tau_1 = \abs{A_1}/\xo$
and
the formulas (\ref{barx0}) and (\ref{defC12sstar})
for $\xo$ and $C_2$.
Hence,
we may write
\[
 \w_0\left(\frac{\tau}{\tau_1}\right)
\sim
 \eps^{-1/6}
\,
 C_1
\,
 \sL^{1/6}
 \Ai\left(\tau + A_1\right) ,
\quad\mbox{for}\quad
 \tau \ll \eps^{-1/3} .
\]
Since the contribution
to the integral
in (\ref{bm00-resc})
of greater values of $\tau$
may be estimated
to be exponentially small,
we can write
\begin{eqnarray}
\fl
 b_{m00}
&=&
 \eps^{-1/3}
\,
\frac{
 \sqrt{2} \, (1 - \nu) \, C_1^2 \, \sL^{1/3}
}{
 \tau_1
}
 \int_0^\infty
\,
 f\left(\frac{\tau}{\tau_1}\right)
\,
 \cos\left(\sqrt{N_m} \, \frac{\tau}{\tau_1}\right)
\,
 \Ai^2\left(\tau + A_1\right)
\,
 d\tau
\nonumber\\
\fl
&=&
 b \, C_1^2
 \int_0^\infty
\,
 \cos\left(
 \frac{\eps^{1/3} \, \sqrt{N_m} \, \tau}{\sL^{1/3}}\right)
\,
 \Ai^2\left(\tau + A_1\right)
\,
 d\tau ,
\Label{bm00-aeII}
\end{eqnarray}
to leading order,
as desired.
Note that
this formula reduces
to (\ref{bm00-aeI}),
for $m \ll \eps^{-1/3}$.

\subsection{The case $m \gg \eps^{-1/3}$}
Here,
$2 \pi/\sqrt{N_m} \ll \eps^{1/3}$.
Similarly to our work
in the previous section,
we define
the new variable
$\tau = \eps^{-1/3} x$.
We find, then,
\[
 b_{m00}
=
 \sqrt{2}
\,
 \eps^{1/3}
\,
 (1 - \nu)
 \int_0^{\eps^{-1/3}}
\,
 g(\tau)
\,
 \cos\left(\eps^{1/3} \, \sqrt{N_m} \, \tau\right)
\,
 d\tau ,
\]
where
$g(\tau)
=
f\left(\eps^{1/3} \tau\right)
\,
\w_0^2\left(\eps^{1/3}\tau\right)$.
Using Theorem~\ref{t-Fourier1}
(with
$\lambda = \eps^{1/3} \, \sqrt{N_m}$,
$\Phi(t) = t = \tau$,
and
$h(\tau) = g(\tau)$)
and
the fact that
the right-boundary term
is exponentially smaller
than the left one,
as $\w_0(1)$
is exponentially smaller
than $\w_0(0)$
(cf. \ref{s-w0-ae}),
we obtain
\begin{eqnarray}
 b_{m00}
&=&
 \sqrt{2}
\,
 \eps^{1/3}
\,
 (1 - \nu)
\,
 {\rm Re}
\left(
 \sum_{k=0}^\infty
 g^{(k)}(0)
 \left(\frac{\im}{\eps^{1/3} \, \sqrt{N_m}}\right)^{k+1}
\right)
\nonumber\\
&=&
 \sqrt{2}
\,
 \frac{1 - \nu}{\eps^{1/3} \, N_m}
\,
 \sum_{k=0}^\infty
 (-1)^{k+1}
 g^{(2k+1)}(0)
 \left(\frac{1}{\eps^{1/3} \, \sqrt{N_m}}\right)^{2k} .
\Label{bm00-aeIII}
\end{eqnarray}
Recalling the definition of $g$,
and employing (\ref{wo})
and that
$\Ai(A_1) = \Ai''(A_1) = 0$,
we calculate
\[
 g'(0)
=
 0
\quad\mbox{and}\quad
 g'''(0)
=
-
 6
\,
 [\Ai'(A_1)]^2
\,
 C_1^2
\,
 \sL^2 .
\]
The desired result now follows,
while
(\ref{bm00-aeIII}) also reduces to (\ref{bm00-aeII})
for $m = \Or(\eps^{-1/3})$.

\section{An asymptotic formula for $a'_{00k}$
\label{s-a00k}}
\setcounter{equation}{0}
In this section,
we derive
the asymptotic formula for $a'_{00k}$
for $\Or(1)$ values of $\Lambda_0$
collected in (\ref{coeffs-ae}),
\be
 a'_{00k}
=
-
 A'_k(\Lambda_0)
\,
 A(\Lambda_0) ,
\quad\mbox{for}\
 0 \ne k \ll \eps^{-1/3} .
\Label{a00k-ae}
\ee
Further, we extend this result to
\be
\fl
 a'_{00k}
=
-
\left(
 A'_k(\Lambda_0)
\,
 A(\Lambda_0)
-
 \eps^{1/2}
\,
 \alpha
\,
 \tilde{A}'_k(\Lambda_0)
\right) ,
\quad\mbox{for}\
 0 \ne k \ll \eps^{-1/3} ,
\Label{a00k-ae-Lambda0>>1}
\ee
which remains valid
at least in the regime
\be
 \Lambda_0
=
 \frac{1}{4 \xs^2}
 \log^2\eps
+
 \frac{1}{\xs^2}
 \log\eps
 \log(-\log\eps)
+
 \frac{1}{\xs^2}
 \log^2\log\eps
+
 \mu
 \log\eps ,
\Label{Lambda0-log}
\ee
for all $\mu \in (-\infty , \mu_0]$
and $\mu_0 > 0$ any $\Or(1)$ value.
Here,
$A(\Lambda_0) = \alpha \, a(\Lambda_0)$,
$A'_k(\Lambda_0) = \alpha' \, a'_k(\Lambda_0)$,
and
$\tilde{A}'_k(\Lambda_0)
=
\tilde{\alpha}' \, \tilde{a}'(\Lambda_0)$.
The $\Lambda_0-$independent constants
$\alpha$, $\alpha'$, and $\tilde{\alpha}'$
were defined in
(\ref{defbarCa}), (\ref{5const}), and (\ref{ta'-def}),
respectively,
whereas the functions
$a$, $a'$, and $\tilde{a}'$
are reported in
(\ref{a000-ae}), (\ref{5const}), and (\ref{ta'k-def}).
We remark, here,
that these results
are only valid
for those values of $k$
for which $\s_k(\xs) \ne 0$.
For the remaining values of $k$,
Theorem~\ref{t-Laplace} yields
(algebraically) higher order results.
Also,
we note that
asymptotic formulas
for higher values of $k$
can be derived
as in the previous section,
albeit at considerable
extra computational cost.

\begin{figure}[t]
\hspace*{-1.7cm}
\scalebox{1.05}[0.5]{
\includegraphics{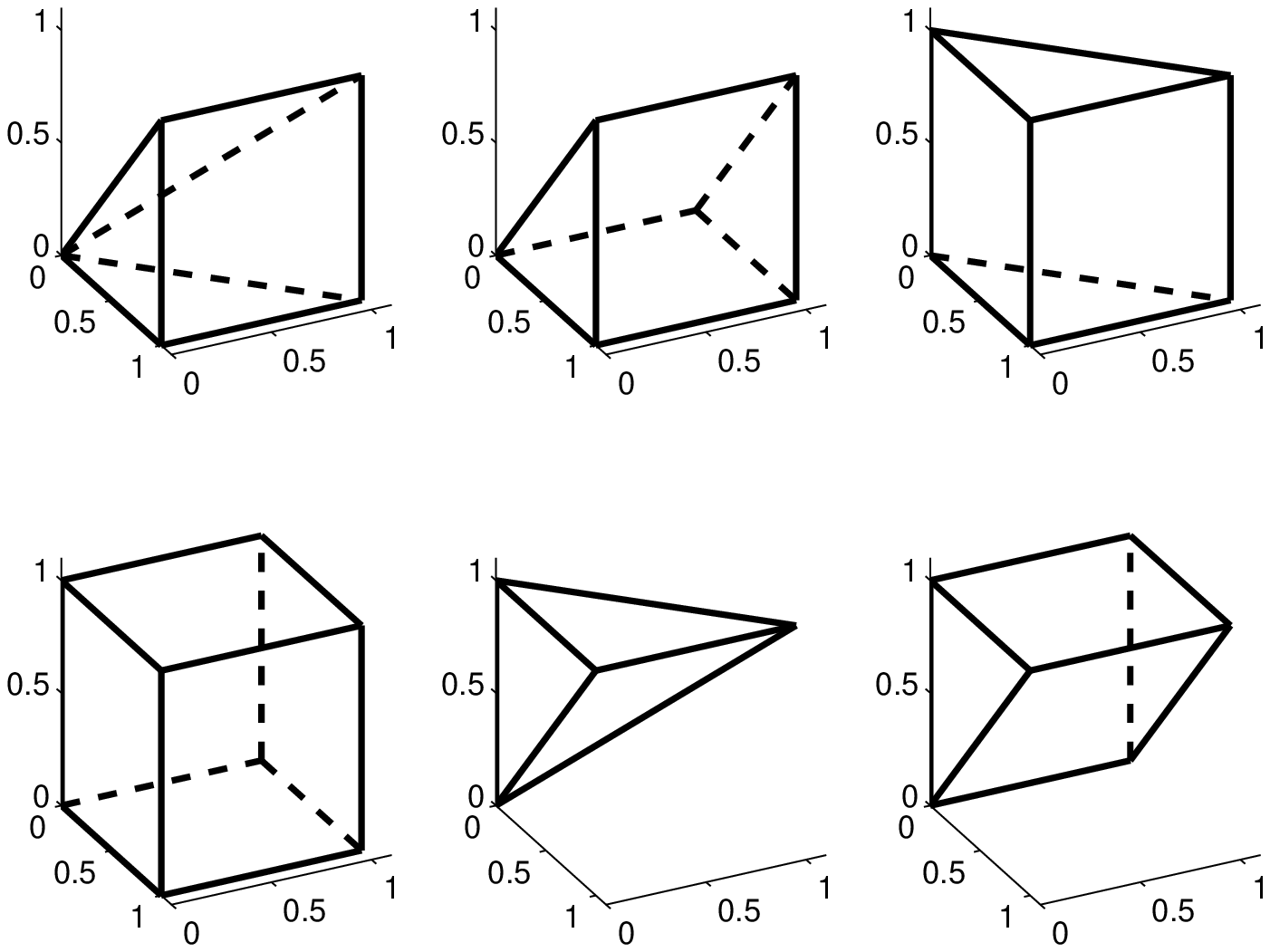}
}
\caption{
\label{f-a'00k-domains}
The domains of integration
for the integrals
$\I_1 , \ldots , \I_6$
in (\ref{a00k-expl}).
}
\end{figure}

We first write out explicitly
the expression for $a'_{00k}$
yielded by (\ref{ab-coeffs}),
\[
 a'_{00k}
=
 \eps^{-1/3}
 \delta
 \int_0^1
 a_0(x)
\,
 \w_0(x)
\,
 \y_k(x)
\,
 dx
+
 \eps^{2/3}
 \delta \ell^{-1}
 \int_0^1
 a_0(x)
\,
 \wpu_0(x)
\,
 \s_k(x)
\,
 dx .
\]
Recalling the definition of $a_0$
from (\ref{am-vs-bm})
and
working as in Section~\ref{s-a000},
we obtain further
\begin{eqnarray*}
 a'_{00k}
=&
 \eps^{-1/3}
 \delta^2
 \int_0^1
 \int_0^x
 h_1(x,y)
\,
 \wpu_0(y)
\,
 \w_0(x)
\,
 \y_k(x)
\,
 dy dx
\nonumber\\
&
+
 \eps^{-1/3}
 \delta^2
 \int_0^1
 \int_0^1
 h_2(x,y)
\,
 \wpu_0(y)
\,
 \w_0(x)
\,
 \y_k(x)
\,
 dy dx
\nonumber\\
&
+
 \eps^{2/3}
 \delta^2
 \ell^{-1}
 \int_0^1
 \int_0^x
 h_1(x,y)
\,
 \s_k(x)
\,
 \wpu_0(y)
\,
 \wpu_0(x)
\,
 dy dx
\nonumber\\
&
+
 \eps^{2/3}
 \delta^2
 \ell^{-1}
 \int_0^1
 \int_0^1
 h_2(x,y)
\,
 \s_k(x)
\,
 \wpu_0(y)
\,
 \wpu_0(x)
\,
 dy dx .
\end{eqnarray*}
Substituting, finally,
from (\ref{psit-n}),
we obtain
an integral formula
for $a'_{00k}$
which is amenable
to the sort of asymptotic analysis
employed in Sections~\ref{s-a000} and \ref{s-bm00},
\begin{eqnarray}
\fl
 a'_{00k}
=&
 \frac{\eps^{-1/3} \delta^2}{\ell}
\left[
 (W_\y)^{-1}
 \int_0^1
 \int_0^x
 \int_0^x
 h_{1,k}(x,y,z)
\,
 \w_0(x)
\,
 \y_{k,-}(x)
\,
 \wpu_0(y)
\,
 \ypu_{k,+}(z)
\,
 dz dy dx
\right.
\nonumber\\
\fl
&
+
 (W_\y)^{-1}
 \int_0^1
 \int_0^1
 \int_0^x
 h_{2,k}(x,y,z)
\,
 \w_0(x)
\,
 \y_{k,-}(x)
\,
 \wpu_0(y)
\,
 \ypu_{k,+}(z)
\,
 dz dy dx
\nonumber\\
\fl
&
-
 (W_\y)^{-1}
 D_k(0)
 \int_0^1
 \int_0^x
 \int_0^1
 h_{1,k}(x,y,z)
\,
 \w_0(x)
\,
 \y_{k,-}(x)
\,
 \wpu_0(y)
\,
 \ypu_{k,-}(z)
\,
 dz dy dx
\nonumber\\
\fl
&
-
 (W_\y)^{-1}
 D_k(0)
 \int_0^1
 \int_0^1
 \int_0^1
 h_{2,k}(x,y,z)
\,
 \w_0(x)
\,
 \y_{k,-}(x)
\,
 \wpu_0(y)
\,
 \ypu_{k,-}(z)
\,
 dz dy dx
\nonumber\\
\fl
&
+
 (W_\y)^{-1}
 \int_0^1
 \int_0^x
 \int_x^1
 h_{1,k}(x,y,z)
\,
 \w_0(x)
\,
 \y_{k,+}(x)
\,
 \wpu_0(y)
\,
 \ypu_{k,-}(z)
\,
 dz dy dx
\nonumber\\
\fl
&
+
 (W_\y)^{-1}
 \int_0^1
 \int_0^1
 \int_x^1
 h_{2,k}(x,y,z)
\,
 \w_0(x)
\,
 \y_{k,+}(x)
\,
 \wpu_0(y)
\,
 \ypu_{k,-}(z)
\,
 dz dy dx
\nonumber\\
\fl
&
+
 \eps
 \int_0^1
 \int_0^x
 h_1(x,y)
\,
 \s_k(x)
\,
 \wpu_0(x)
\,
 \wpu_0(y)
\,
 dy dx
\nonumber\\
\fl
&
\left.
+
 \eps
 \int_0^1
 \int_0^1
 h_2(x,y)
\,
 \s_k(x)
\,
 \wpu_0(x)
\,
 \wpu_0(y)
\,
 dy dx
\right] .
\Label{a00k-expl}
\end{eqnarray}
Here,
$h_{i,k}(x,y,z) = h_i(x,y) \, f(z) \, \s_k(z)$,
for $i = 1 , 2$,
and
the constants $D_k(0)$
are reported in (\ref{Do-ae})--(\ref{dn-def}).
Let $\I_1 , \ldots , \I_8$
denote the integrals
in the right member of (\ref{a00k-expl})
in the order that they appear
(the three-dimensional
domains of integration
for $\I_1 , \ldots , \I_6$
are sketched
in Figure~\ref{f-a'00k-domains}).
In what follows,
we omit
the term
$\theta \, \w^2_{0,-}(1;\xo) \, \w_{0,+}(x;\xo)$
in the expression (\ref{wo})
for $\w_0$,
as one can show
that its contribution
is exponentially small
compared to
the leading order terms
(see also
Sections~\ref{s-a000} and \ref{s-bm00}).
%

\subsection{A rewrite of (\ref{a00k-expl})}
In this section,
we group together integrals
appearing in
the right member
of (\ref{a00k-expl})
in order to achieve
a first reduction
in the numbers of terms
of that member.
We start with
rewriting the term
$-(W_\y)^{-1} \, D_k(0) \, \I_4 + (W_\y)^{-1} \, \I_6 + \eps \, \I_8$.
First,
\[
 \I_4
=
 \int\int_{\pi_x D_4}
\left(
 \int_0^1
 h_{2,k}(x,y,z)
\,
 \w_0(x)
\,
 \y_{k,-}(x)
\,
 dx
\right)
 \wpu_0(y)
\,
 \ypu_{k,-}(z)
\,
 dA_{yz} ,
\]
where
$\pi_x$ is
the orthogonal projection
on the $yz-$plane---and hence
$\pi_xD_4 = [0,1]^2$---and
$dA_{yz}$ is the area element
on that plane.
Since
$\y_{k,-} = \eps^{1/3} C_1^{-1} \sL^{-1/3} \w_0$
in a neighborhood
of the origin
(cf. (\ref{wo}) and (\ref{psin-})),
$\w_0$ is exponentially small
outside this neighborhood,
and
$\|\w_0\|_2 = 1$,
we write
\begin{eqnarray*}
 \I_4
&=&
 \eps^{1/3}
\,
 C_1^{-1}
\,
 \sL^{-1/3}
\,
 \int\int_{\pi_x D_4}
 h_{2,k}(0,y,z)
\,
 \wpu_0(y)
\,
 \ypu_{k,-}(z)
\,
 dA_{yz} .
\end{eqnarray*}
Recalling that $\ypu_{k,-} = E \, \y_{k,-}$,
according to our convention
in Section~\ref{s-WNLA},
and
substituting into the formula above
from (\ref{wo+}) and (\ref{psin-}),
we obtain
\begin{eqnarray*}
 \I_4
&=&
 \eps^{1/2}
\,
 \frac{C_2^2}{4 \pi}
\,
 \int\int_{\pi_x D_4}
\frac{
 h_{2,k}(0,y,z)
}{
 F^{1/4}(y) \, F^{1/4}(z)
}
\,
 \exp
\left(
 \frac{J_-(y) + J_-(z)}{\sqrt{\eps}}
\right)
\,
 dA_{yz} ,
\end{eqnarray*}
whence,
employing also (\ref{Do-ae}),
we find
\be
 (W_\y)^{-1} \, D_k(0) \, \I_4
=
 \eps^{-1/6}
\,
 \int\int_{\pi_x D_4}
 \Xi_4(y,z)
\,
 \exp
\left(
 \frac{\Pi_4(y,z)}{\sqrt{\eps}}
\right)
\,
 dA_{yz} .
\Label{a00k-I4-aux1}
\ee
Here,
\be
\fl
 \Xi_4(y,z)
=
 \frac{C_2^2 \, d_k}{4 \pi \, W_\y}
\,
\frac{
 h_2(0,y) \, f(z) \, \s_k(z)
}{
 F^{1/4}(y) \, F^{1/4}(z)
}
\quad\mbox{and}\quad
 \Pi_4(y,z)
=
 J_-(y) + J_-(z) .
\Label{a00k-I4-aux2}
\ee
Next,
we rewrite $\I_6$,
\[
\fl
 (W_\y)^{-1} \, \I_6
=
 (W_\y)^{-1}
\,
 \int\int_{\pi_x D_6}
\left(
 \int_0^z
 h_{2,k}(x,y,z)
\,
 \w_0(x)
\,
 \y_{k,+}(x)
\,
 dx
\right)
\,
 \wpu_0(y)
\,
 \ypu_{k,-}(z)
\,
 dA_{yz} .
\]
Employing (\ref{wo}) and (\ref{psin+}), now,
we obtain
\[
\fl
 (W_\y)^{-1} \, \I_6
=
 \eps^{1/6}
\,
 \frac{C_1 \, \sL^{1/3}}{2 \pi \, W_\y}
\,
 \int\int_{\pi_x D_6}
\left(
\int_0^z
 \frac{h_{2,k}(x,y,z)}{\sqrt{F(x)}}
dx
\right)
 \wpu_0(y)
\,
 \ypu_{k,-}(z)
\,
 dA_{yz} .
\]
Further using (\ref{wo+}) and, once again, (\ref{psin+}),
we find
\be
 (W_\y)^{-1} \, \I_6
=
 \eps^{1/3}
\,
 \int\int_{\pi_x D_6}
 \Xi_6(y,z)
\,
 \exp
\left(
 \frac{\Pi_6(y,z)}{\sqrt{\eps}}
\right)
 dA_{yz} .
\Label{a00k-I6-aux1}
\ee
Here,
${\pi_x D_6} = {\pi_x D_4}$,
$\Pi_6(y,z) = \Pi_4(y,z)$,
and
\be
 \Xi_6(y,z)
=
\frac{
 C_1^2 \, C_2^2 \, \sL^{2/3}
}{
 8 \pi^2 \, W_\y
}
\,
 \frac{f(z) \, \s_k(z)}{F^{1/4}(y) \, F^{1/4}(z)}
\,
\int_0^z
 \frac{h_2(x,y)}{\sqrt{F(x)}}
\,
dx .
\Label{a00k-I6-aux2}
\ee
Similarly,
renaming $(x,y)$
as $(y,z)$ in $\I_8$,
we derive the formula
\be
 \eps \, \I_8
=
 \eps^{5/6}
\,
 \int\int_{D_8}
 \Xi_8(y,z)
\,
 \exp
\left(
 \frac{\Pi_8(y,z)}{\sqrt{\eps}}
\right)
\,
 dA_{yz} ,
\Label{a00k-I8-aux1}
\ee
where
$D_8 = \pi_x D_6 = \pi_x D_4$,
$\Pi_8(y,z) = \Pi_6(y,z) = \Pi_4(y,z)$,
and
\be
 \Xi_8(y,z)
=
 \frac{C_1^2 \, C_2^2 \, \sL^{2/3}}{4\pi}
\,
\frac{
 h_2(y,z) \, \s_k(y)
}{
 F^{1/4}(y) \, F^{1/4}(z)
} .
\Label{a00k-I8-aux2}
\ee
Combining (\ref{a00k-I4-aux1})--(\ref{a00k-I8-aux2}),
we obtain
\begin{eqnarray}
\fl
 -(W_\y)^{-1} \, D_k(0) \, \I_4
+
 (W_\y)^{-1} \, \I_6
+
 \I_8
\nonumber\\
=
-
 \eps^{-1/6}
\,
 \int\int_{\pi_x D_4}
 \tilde{\Xi}_4(y,z)
\,
 \exp
\left(
 \frac{\Pi_4(y,z)}{\sqrt{\eps}}
\right)
\,
 dA_{yz} ,
\Label{a00k-I468-aux1}
\end{eqnarray}
where,
to leading order,
uniformly over ${\pi_x D_4}$,
and for all $\Or(1)$ values of $\Lambda_0$,
\begin{eqnarray}
 \tilde{\Xi}_4(y,z)
=
 \Xi_4(y,z)
=
 \frac{C_2^2 \, d_k}{4 \pi \, W_\y}
 \frac{h_2(0,y) \, f(z) \, \s_k(z)}{F^{1/4}(y) \, F^{1/4}(z)} .
\Label{a00k-I468-aux2}
\end{eqnarray}

Next,
we rewrite the term
$(W_\y)^{-1} \I_5 + \eps \, \I_7$.
We write first
\begin{eqnarray*}
 \I_5
&=&
 \int\int_{\pi_x D_5}
\left(
 \int_y^z
 h_{1,k}(x,y,z)
\,
 \w_0(x)
\,
 \y_{k,+}(x)
\,
 dx
\right)
\,
 \wpu_0(y)
\,
 \ypu_{k,-}(z)
\,
 dA_{yz} ,
\end{eqnarray*}
where
$\pi_x D_5
=
\{(y,z) \vert 0 \le y \le z , \, 0 \le z \le 1\}$.
Now,
(\ref{wo})--(\ref{wo+}) and (\ref{psin-})
yield further
\begin{eqnarray}
\fl
 (W_\y)^{-1} \, \I_5
&=&
 \eps^{1/6}
\,
 \frac{C_1 \, C_2 \, \sL^{1/3}}{2 \pi \, W_\y}
\,
 \int\int_{\pi_x D_5}
\left(
\int_y^z
 \frac{h_{1,k}(x,y,z)}{\sqrt{F(x)}}
\,
dx
\right)
 \wpu_0(y)
\,
 \ypu_{k,-}(z)
\,
 dA_{yz}
\nonumber\\
\fl
&=&
 \eps^{1/3}
\,
 \int\int_{\pi_x D_5}
 \Xi_5(y,z)
\,
 \exp
\left(
 \frac{\Pi_5(y,z)}{\sqrt{\eps}}
\right)
 dA_{yz} ,
\Label{a00k-I5-aux1}
\end{eqnarray}
where
we have defined the functions
\be
 \Xi_5(y,z)
=
 \frac{C_1^2 \, C_2^3 \, \sL^{2/3}}{8 \pi^2 \, W_\y}
\,
 \frac{f(z) \, \s_k(z)}{F^{1/4}(y) \, F^{1/4}(z)}
\,
\int_y^z
 \frac{h_1(x,y)}{\sqrt{F(x)}}
\,
 dx
\Label{a00k-I5-aux2}
\ee
and
$\Pi_5(y,z) = \Pi_4(y,z)$.
Next,
renaming $x$ into $z$
in $\I_7$,
we find
\be
 \eps \, \I_7
=
 \eps^{5/6}
\,
 \int\int_{D_7}
 \Xi_7(y,z)
\,
 \exp
\left(
 \frac{\Pi_7(y,z)}{\sqrt{\eps}}
\right)
\,
 dA_{yz} ,
\Label{a00k-I7-aux1}
\ee
where
\be
\fl
 D_7
=
 \pi_x D_5 ,
\quad
 \Xi_7(y,z)
=
 \frac{C_1^2 \, C_2^2 \, \sL^{2/3}}{4\pi}
\,
\frac{
 h_1(z,y) \, \s_k(z)
}{
 F^{1/4}(y) \, F^{1/4}(z)
} ,
\quad\mbox{and}\quad
 \Pi_7(y,z)
=
 \Pi_4(y,z) .
\Label{a00k-I7-aux2}
\ee
Combining (\ref{a00k-I5-aux1})--(\ref{a00k-I7-aux2}),
we find,
to leading order
and
uniformly over $D_8$,
\be
 (W_\y)^{-1} \, \I_5 + \eps \, \I_7
=
 \eps^{1/3}
\,
 \int\int_{D_7}
 \tilde{\Xi}_5(y,z)
\,
 \exp
\left(
 \frac{\Pi_4(y,z)}{\sqrt{\eps}}
\right)
\,
 dA_{yz} ,
\Label{a00k-I57-aux}
\ee
where
$\tilde{\Xi}_5(y,z)
=
\Xi_5(y,z) + \eps^{1/2} \, \Xi_7(y,z)$.

We now rewrite
$(W_\y)^{-1} \I_2$.
First,
\[
 \I_2
=
 \int\int_{\pi_y D_2}
 \tilde{H}_2(x)
\,
 f(z)
\,
 \s_k(z)
\,
 \w_0(x)
\,
 \y_{k,-}(x)
\,
 \ypu_{k,+}(z)
\,
 dA_{xz} ,
\]
where
$\tilde{H}_2(x)
=
\int_0^1
h_{2}(x,y) \, \wpu_0(y) \, dy$.
Substituting for $\wpu_0(y)$ from (\ref{wo+}),
we find further
\[
 \I_2
=
 \eps^{-1/12}
\,
 \frac{C_1 C_2 \sL^{1/3}}{2 \sqrt{\pi}}
\,
 \int\int_{\pi_y D_2}
 H_2(x)
\,
 \w_0(x)
\,
 \y_{k,-}(x)
\,
 f(z)
\,
 \s_k(z)
\,
 \ypu_{k,+}(z)
\,
 dA_{xz} ,
\]
where
$H_2(x)
=
\int_0^1
h_{2}(x,y)
\,
F^{-1/4}(y)
\,
\exp(J_-(y)/\sqrt{\eps}) \, dy$.
Using Theorem~\ref{t-Laplace},
now,
we obtain
\begin{eqnarray}
\fl
 (W_\y)^{-1}
\,
 \I_2
&=
 \eps^{1/6}
 \delta^{-1}
\,
 C''_2
 \int\int_{\pi_y D_2}
 h_{2}(x,\xs)
\,
 \w_0(x)
\,
 \y_{k,-}(x)
\,
 f(z)
\,
 \s_k(z)
\,
 \ypu_{k,+}(z)
\,
 dA_{xz}
\nonumber\\
&=
 \eps^{7/12}
\,
 \int\int_{\pi_y D_2}
 \Xi_2(x,z)
\,
 \exp
\left(
 \frac{\Pi_2(x,z)}{\sqrt{\eps}}
\right)
\,
 dA_{xz} ,
\Label{a00k-I2-aux1}
\end{eqnarray}
where
$C''_2$ is an $\Or(1)$ constant,
$\pi_y D_2
=
\{(x,z) \vert 0 \le z \le x , \, 0 \le x \le 1 \}$,
$\Pi_2(x,z)
=
J_-(\xs) + J_+(z) - 2I(x)$,
\be
 \Xi_2(x,z)
=
 C'_2
\,
\frac{
 h_{2}(x,\xs)
\,
 f(z)
\,
 \s_k(z)
}{
 \sqrt{F(x)} \, F^{1/4}(z)
} ,
\quad\mbox{with $C'_2$ an $\Or(1)$ constant.}\
\Label{a00k-I2-aux2}
\ee

Finally, we rewrite
$(W_\y)^{-1} D_k(0) \, \I_3$.
First,
\begin{eqnarray*}
 \I_3
&=&
\left(
 \int_0^1
 f(z)
\,
 \s_k(z)
\,
 \ypu_{k,-}(z)
\,
 dz
\right)
 \int\int_{\pi_z D_3}
 h_1(x,y)
\,
 \w_0(x)
\,
 \y_{k,-}(x)
\,
 \wpu_0(y)
\,
 dA_{xy} .
\end{eqnarray*}
Substituting from
(\ref{wo})--(\ref{wo+})
and
(\ref{psin-})
into this formula
and
interchanging the roles
of $y$ and $z$
in the single and double integrals,
we find
\be
 (W_\y)^{-1}
\,
 D_k(0)
\,
 \I_3
=
 \eps^{-1/3}
\,
 \tilde{I}_3
\,
 \int\int_{\pi_y D_2}
 \tilde{\Xi}_3(x,z)
\,
 \exp
\left(
 \frac{\tilde{\Pi}_3(x,z)}{\sqrt{\eps}}
\right)
\,
 dA_{xz} ,
\Label{a00k-I3-aux1}
\ee
where
$\tilde{I}_3
=
\int_0^1
F^{-1/4}(y) \, f(y) \, \s_k(y)
\,
\exp
\left(J_-(y)/\sqrt{\eps}\right)
\,
dy$
and
\be
 \tilde{\Xi}_3(x,z)
=
 \tilde{C}_3
\,
\frac{
 h_1(x,z)
}{
 \sqrt{F(x)} \, F^{1/4}(z)
}
\quad\mbox{and}\quad
 \tilde{\Pi}_3(x,z)
=
 J_-(z) - 2I(x) ,
\Label{a00k-I3-aux2}
\ee
for some $\Or(1)$ constant $\tilde{C}_3$.

\subsection{An asymptotic estimate for $a'_{00k}$
in the regime $\Lambda_0 = \Or(1)$}
In this section,
we estimate
the various terms derived above,
starting from
$-(W_\y)^{-1} \, D_k(0) \, \I_4 + (W_\y)^{-1} \, \I_6 + \eps \, \I_8$
(cf. (\ref{a00k-I468-aux1})--(\ref{a00k-I468-aux2})).
The exponent $\Pi_4$
becomes maximum at
the interior
critical point $(\xs,\xs)$,
and thus
Theorem~\ref{t-Laplace} yields
\begin{eqnarray*}
\fl
 -(W_\y)^{-1} \, D_k(0) \, \I_4
+
 (W_\y)^{-1} \, \I_6
+
 \I_8
&=
-
 \eps^{1/2}
\frac{
 2 \pi
}{
 \abs{J''_-(\xs)}
}
\,
\left(
 \eps^{-1/6}
\,
 \tilde{\Xi}_4(\xs,\xs)
\,
 \exp
\left(
 \frac{\Pi_4(\xs,\xs)}{\sqrt{\eps}}
\right)
\right)
\\
&=
 -\eps^{1/3} \, \delta^{-2} \, \tilde{C}_4 ,
\end{eqnarray*}
where
\begin{eqnarray*}
 \tilde{C}_4
&=&
 C_2^2 \, (\sst W_\y)^{-1} \, d_k
\,
 \s_k(\xs) \, f(\xs) \, \, h_2(0,\xs) .
\end{eqnarray*}

Next,
we estimate
$(W_\y)^{-1} \, \I_5 + \eps \, \I_7$,
cf. (\ref{a00k-I57-aux}).
The sole (quadratic) maximum
of $\Pi_4$
in $D_7$
lies at the critical point
$(\xs,\xs) \in \partial D_7$,
where
$\tilde{\Xi}_5(\xs,\xs) = 0$
and
$\tilde{\Xi}_7(\xs,\xs) \ne 0$.
Recalling the definition of $\tilde{\Xi}_5$
and
employing Theorem~\ref{t-Laplace}, then,
we obtain
\begin{eqnarray*}
 (W_\y)^{-1} \, \I_5 + \I_7
=
 \eps
\left(
 \eps^{1/3}
\,
 C_0
\,
 \exp
\left(
 \frac{\Pi_4(\xs,\xs)}{\sqrt{\eps}}
\right)
\right)
=
 \eps^{4/3} \delta^{-2} \, \tilde{C}_7 ,
\end{eqnarray*}
for some $\Or(1)$ constants $C_0$ and $\tilde{C}_7$.

We now estimate
the remaining three integrals
starting with $(W_\y)^{-1} \, \I_2$,
cf. (\ref{a00k-I2-aux1})--(\ref{a00k-I2-aux2}).
The exponent $\Pi_2$
has a sole maximum
at the point
$(\xs,\xs) \in \partial(\pi_y D_2)$
which is not
a critical point
(compare to
the maximization of $\Pi_4$
in Section~\ref{s-a000}).
Hence,
Theorem~\ref{t-Laplace} yields
\[
 (W_\y)^{-1}
\,
 \I_2
=
 \eps^{3/4}
\,
 C'''_2
\left(
 \eps^{7/12}
\,
 \Xi_2(\xs,\xs)
\,
 \exp
\left(
 \frac{\Pi_2(\xs,\xs)}{\sqrt{\eps}}
\right)
\right)
=
 \eps^{4/3} \, \delta^{-2} \, C_2 ,
\]
for some $\Or(1)$ constants
$C'''_2$ and $C_2$.
Next,
since
$D_1 \subset D_2$
and
the integrands
of $\I_1$ and $\I_2$
differ only
by an $\Or(1)$ multiple,
the above analysis
also yields that
$(W_\y)^{-1} \I_1$ is at most
of the same order
as $(W_\y)^{-1} \I_2$.
Finally,
we estimate
$(W_\y)^{-1} \, D_k(0) \, \I_3$,
cf. (\ref{a00k-I3-aux1})--(\ref{a00k-I3-aux2}).
First,
we estimate
\[
 \tilde{I}_3
=
 \int_0^1
\frac{
 f(y)
\,
 \s_k(y)
}{
 F^{1/4}(y)
}
\,
 \exp
\left(
 \frac{J_-(y)}{\sqrt{\eps}}
\right)
\,
 dy
=
 \eps^{1/4} \, \delta^{-1} \, C''_3 ,
\]
for some $\Or(1)$ constant $C''_3$.
Substituting into (\ref{a00k-I3-aux1}),
then,
we obtain
\begin{eqnarray*}
 (W_\y)^{-1}
\,
 D_k(0)
\,
 \I_3
&=&
 \eps^{-1/12}
\,
 \int\int_{\pi_y D_2}
 \Xi_3(x,z)
\,
 \exp
\left(
 \frac{\Pi_3(x,z)}{\sqrt{\eps}}
\right)
\,
 dA_{xz} ,
\end{eqnarray*}
where
\[
 \Xi_3(x,z)
=
 \tilde{C}'_3
\,
\frac{
 h_1(x,z)
}{
 \sqrt{F(x)} \, F^{1/4}(z)
}
\quad\mbox{and}\quad
 \Pi_3(x,z)
=
 J_-(\xs) + J_-(z) - 2I(x) ,
\]
for some $\Or(1)$ constant $\tilde{C}'_3$.
The exponent $\Pi_3$
has a sole maximum
at the point
$(\xss,\xss) \in \partial(\pi_y D_2)$
which is also not
a critical point
(compare to
the maximization of $\Pi_1$
in Section~\ref{s-a000}).
Hence,
Theorem~\ref{t-Laplace} yields
\begin{eqnarray*}
 (W_\y)^{-1} \, D_k(0) \, \I_3
&=&
 \eps^{3/4}
 C'_3
\left(
 \eps^{-1/12}
 \Xi_3(\xs,\xs)
 \exp
\left(
 \frac{\Pi_3(\xs,\xs)}{\sqrt{\eps}}
\right)
\right)
\\
&=&
 \eps^{2/3}
 C_3
 \exp
\left(
 \frac{\Pi_3(\xss,\xss)}{\sqrt{\eps}}
\right) ,
\end{eqnarray*}
for some $\Or(1)$ constants
$C_3$ and $\tilde{C}'_3$
and where
$\Pi_3(\xss,\xss) < 2 J_-(\xs)$.

In total, then,
and to leading order,
we obtain
the leading order formula
\be
 a'_{00k}
=
-
\frac{
 C_2^2 \, d_k \, \s_k(\xs) \, h_2(0,\xs)
}{
 \sst \, W_\y
} ,
\quad\mbox{for}\
 k \ll \eps^{-1/3} .
\Label{a00k-ae-aux}
\ee
Here,
we have used that
$f(\xs) = \ell$ to leading order,
while
$h_2$ is given in (\ref{h2-def})
and
(cf.~(\ref{Wpsi}) and (\ref{dn-def}))
\[
\fl
 W_\y
=
 \Ai'(A_1) \abs{\Bi(A_1)}
=
 \frac{1}{\pi}
\quad\mbox{and}\quad
 d_k
=
\frac{
 \sL^{2/3}
}{
 \pi \, C_3 \, (N_k + \Lambda_0)
} .
\]
To derive
the desired formula (\ref{a00k-ae})
from (\ref{a00k-ae-aux}),
we note that
(cf. (\ref{h1-def})--(\ref{h2-def}))
\be
 h_2(0,\xs)
=
 (1 - \nu) \, f(0)
\,
\frac{
 \sinh\left(\sqrt{\Lambda_0} (1-\xs)\right)
}{
 \sqrt{\Lambda_0} \, \cosh\sqrt{\Lambda_0}
} .
\Label{h-calc}
\ee
Hence,
(\ref{a00k-ae-aux}) becomes
\begin{eqnarray*}
\fl
 a'_{00k}
=
-
\frac{
 (1 - \nu) \, C_2^2 \, \sL^{2/3} \, f(0) \, \s_k(\xs)
}{
 \sst \, C_3
}
\,
\frac{
 \sinh\left(\sqrt{\Lambda_0}(1-\xs)\right)
}{
 \sqrt{\Lambda_0}
\,
 (N_k + \Lambda_0)
\,
 \cosh\sqrt{\Lambda_0}
} .
\end{eqnarray*}
The desired formula (\ref{a00k-ae})
may now be derived
from this equation
by recalling (\ref{vs})
and the definitions collected in (\ref{5const}).

\subsection{Higher order terms
in the asymptotic estimate for $a'_{00k}$
\label{ss-a00k-Lambda0>>1}}
As became evident
in the material presented above,
certain terms among those we estimated
are $\Lambda_0-$dependent,
and hence
they do not necessarily remain higher order
for asymptotically large values of $\Lambda_0$.
As we will see in this section,
certain terms
which are higher order
for $\Lambda_0 = \Or(1)$
become leading order
for $\Lambda_0 \gg 1$.
Apart from that,
these higher order terms
have an important effect
even for $\Lambda_0 = \Or(1)$,
as they lead to
the singularly perturbed problem (\ref{Omega0-fp-quadr})
for the steady states
of the reduced system
(\ref{Omega0'})--(\ref{Omegak'}).

To quantify these terms,
we recall from the last section that
\be
\fl
 -(W_\y)^{-1} \, D_k(0) \, \I_4
+
 (W_\y)^{-1} \, \I_6
+
 \I_8
=
-
 \eps^{1/3}
\frac{
 2 \pi
}{
 \abs{J''_-(\xs)}
}
\,
 \delta^{-2}
\,
 \tilde{\Xi}_4(\xs,\xs) .
\Label{I146-aux}
\ee
By definition of $\tilde{\Xi}_4$,
\begin{eqnarray}
 \tilde{\Xi}_4(\xs,\xs)
=
 \Xi_4(\xs,\xs)
-
 \eps^{1/2}
\,
 \Xi_6(\xs,\xs)
-
 \eps
\,
 \Xi_8(\xs,\xs) ,
\end{eqnarray}
where $\Xi_4$, $\Xi_6$, and $\Xi_8$
are expressed in terms of
the function $h_2$
defined in (\ref{h2-def})---see
(\ref{a00k-I4-aux2}),
(\ref{a00k-I6-aux2}),
and
(\ref{a00k-I8-aux2}),
respectively.
As we saw in the last section,
\begin{eqnarray*}
\fl
-
 \frac{\eps^{-1/3} \delta^2 \, D_k(0)}{\ell \, W_\y}
 I_4
=
-
\frac{
 (1 - \nu) \, C_2^2 \, \sL^{2/3} \, f(0) \, \s_k(\xs)
}{
 \sst \, C_3
}
\,
\frac{
 \sinh\left(\sqrt{\Lambda_0}(1-\xs)\right)
}{
 \sqrt{\Lambda_0}
\,
 (N_k + \Lambda_0)
\,
 \cosh\sqrt{\Lambda_0}
} .
\end{eqnarray*}
At the same time,
we calculate
\begin{eqnarray*}
\fl
-
 \frac{\eps^{-1/3} \delta^2}{\ell \, W_\y}
 I_6
&=&
 \eps^{1/2}
\,
\frac{
 (1-\nu) \, C_1^2 \, C_2^2 \, \sL^{2/3} \s_k(\xs)
}{
 2 \sst
}
\frac{
 \sinh\left(\sqrt{\Lambda_0} \, (1-\xs)\right)
\,
 \int_0^\xs
 f(x)
\,
 \cosh\left(\sqrt{\Lambda_0} \, x\right)
\,
 dx
}{
 \sqrt{\Lambda_0} \, \cosh\sqrt{\Lambda_0}
} ,
\\
\fl
 \eps \, \frac{\eps^{-1/3} \delta^2}{\ell}
 I_8
&=&
 \eps
\,
\frac{
 (1-\nu) \, C_1^2 \, C_2^2 \, \sL^{2/3} \s_k(\xs)
}{
 \sst
}
\frac{
 \cosh\left(\sqrt{\Lambda_0} \, \xs\right)
\,
 \sinh\left(\sqrt{\Lambda_0} \, (1-\xs)\right)
}{
 \sqrt{\Lambda_0} \, \cosh\sqrt{\Lambda_0}
} .
\end{eqnarray*}
There are two distinguished limits
for these expressions,
namely,
\be
\begin{array}{rcl}
\fl
-
 \frac{\eps^{-1/3} \delta^2 \, D_k(0)}{\ell \, W_\y}
 I_4
&=&
-
\frac{
 (1 - \nu) \, (1-\xs) \, C_2^2 \, \sL^{2/3} \, f(0) \, \s_k(\xs)
}{
 \sst \, C_3
}
\,
\frac{
 1
}{
 N_k + \Lambda_0
} ,
\\
\fl
 \frac{\eps^{-1/3} \delta^2}{\ell \, W_\y}
 I_6
&=&
 \eps^{1/2}
\,
\frac{
 (1-\nu) \, (1-\xs) \, C_1^2 \, C_2^2 \, \sL^{2/3} \s_k(\xs)
\,
 \int_0^\xs
 f(x)
\,
 dx
}{
 2 \sst
} ,
\\
\fl
 \frac{\eps^{-1/3} \delta^2}{\ell}
 \eps \, I_8
&=&
 \eps
\,
\frac{
 (1-\nu) \, (1-\xs) \, C_1^2 \, C_2^2 \, \sL^{2/3} \s_k(\xs)
}{
 \sst
} ,
\end{array}
\quad\mbox{for}\quad
 \Lambda_0
\ll
 1 ,
\ee
and
\be
\begin{array}{rcl}
\fl
-
 \frac{\eps^{-1/3} \delta^2 \, D_k(0)}{\ell \, W_\y}
 I_4
&=&
-
\frac{
 (1 - \nu) \, C_2^2 \, \sL^{2/3} \, f(0) \, \s_k(\xs)
}{
 \sst \, C_3
}
\,
\frac{
 \ex^{-\sqrt{\Lambda_0} \, \xs}
}{
 \sqrt{\Lambda_0} \, (N_k + \Lambda_0)
} ,
\\
\fl
 \frac{\eps^{-1/3} \delta^2}{\ell \, W_\y}
 I_6
&=&
 \eps^{1/2}
\,
\frac{
 (1-\nu) \, \ell \, C_1^2 \, C_2^2 \, \sL^{2/3} \s_k(\xs)
}{
 4 \sst
}
\,
\frac{1}{\Lambda_0} ,
\\
\fl
 \frac{\eps^{-1/3} \delta^2}{\ell}
 \eps \, I_8
&=&
 \eps
\,
\frac{
 (1-\nu) \, C_1^2 \, C_2^2 \, \sL^{2/3} \s_k(\xs)
}{
 2 \sst
}
\,
\frac{1}{\sqrt{\Lambda_0}} ,
\end{array}
\quad\mbox{for}\quad
 \Lambda_0
\gg
 1 ,
\ee
where we have used
Theorem~\ref{t-Laplace1}
to estimate the integral
appearing in the definition~(\ref{a00k-I6-aux2})
of $\Xi_6$.
It immediately follows that
$\eps \, \Xi_8(\xs,\xs) \ll \eps^{1/2} \, \Xi_6(\xs,\xs)$
for all $\Lambda_0 \ll \eps^{-1/2}$.

Next,
we estimate
$(W_\y)^{-1} \, \I_5 + \eps \, \I_7$
in the regime $\Lambda_0 \gg 1$.
First,
we recall (\ref{a00k-I57-aux}),
\[
 (W_\y)^{-1} \, \I_5 + \eps \, \I_7
=
 \eps^{1/3}
 \int\int_{D_7}
 \tilde{\Xi}_5(y,z)
\,
 \exp
\left(
 \frac{\Pi_4(y,z)}{\sqrt{\eps}}
\right)
\,
 dA_{yz} ,
\]
where
$\tilde{\Xi}_5(y,z)
=
\Xi_5(y,z) + \eps^{1/2} \, \Xi_7(y,z)$.
The functions
$\Xi_5$ and $\Xi_7$
are expressible in terms of
the function $h_1$
defined in (\ref{h1-def}),
see (\ref{a00k-I5-aux2}) and (\ref{a00k-I7-aux2}),
respectively.
Working as for $h_2$ above,
we procure
the leading order asymptotic relation
\[
 h_1(x,y)
=
 r
\,
 f(x)
\,
 \left(1 - \frac{f(x)}{\nu}\right)
-
 \theta\left(\sqrt{\Lambda_0} \, (x-y)\right)
\,
 h_2(x,y) .
\]
Here,
$\theta(s) = (1 - \ex^{-2s})/2$---and hence
$\theta(0) = 0$---while
the first term
in the right member
is $\Lambda_0-$independent
and hence remains bounded
in this regime also.
Using this expression,
we can establish that
$(W_\y)^{-1} \, \I_5 + \I_7$ is at most
of order $\eps^{4/3} \, \Lambda_0^{-1}$
and hence higher order.

Similarly,
(\ref{a00k-I2-aux1}) yields
to leading order
\[
 (W_\y)^{-1}
\,
 \I_2
=
 \eps^{4/3}
\,
 \delta^{-2}
\,
\frac{
 (1 - \nu)
\,
 \ell^2
\,
 C'_2
\, 
 C'''_2
\,
 \s_k(\xs)
}{
 2 \, F^{3/4}(\xs)
}
\,
 \frac{1}{\sqrt{\Lambda_0}} ,
\quad\mbox{for}\quad
 \Lambda_0
\gg
 1 ,
\]
where $C'_2$ and $C'''_2$ are
$\Or(1)$ constants.
Hence,
this term is also higher order.
The term
$(W_\y)^{-1} \I_1$
can be bounded in a similar way,
whereas
$(W_\y)^{-1} \, D_k(0) \, \I_3$ is,
here also,
exponentially smaller
than all other terms.

In total, then,
and to leading order,
we obtain the formula
\begin{eqnarray*}
\fl
 a'_{00k}
&=&
  (1 - \nu) \, C_2^2 \, \sL^{2/3} \, \sst^{-1} \, \s_k(\xs)
\left(
-
\frac{
 f(0)
}{
 C_3
}
\,
\frac{
 \sinh\left(\sqrt{\Lambda_0} \, (1-\xs)\right)
}{
 \sqrt{\Lambda_0} \, (N_k + \Lambda_0) \, \sinh\sqrt{\Lambda_0}
}
\right.
\\
\fl
&{}&
\hspace*{2cm}
\left.
+
 \eps^{1/2}
\,
\frac{
 C_1^2
}{
 2
}
\,
\frac{
 \sinh\left(\sqrt{\Lambda_0} \, (1-\xs)\right)
\,
 \int_0^\xs
 f(x)
\,
 \cosh\left(\sqrt{\Lambda_0} \, x\right)
\,
 dx
}{
 \sqrt{\Lambda_0} \, \cosh\sqrt{\Lambda_0}
}
\right) .
\end{eqnarray*}
This formula precisely matches
(\ref{a00k-ae-Lambda0>>1}).
The two associated distinguished limits are
\[
\fl
 a'_{00k}
=
\frac{
 (1 - \nu) \, (1-\xs) \, C_2^2 \, \sL^{2/3} \, \s_k(\xs)
}{
 \sst
}
\left(
-
\frac{
 f(0)
}{
 (N_k + \Lambda_0) C_3
}
+
 \frac{C_1^2 \, \int_0^\xs f(x) \, dx}{2}
\right) ,
\quad\mbox{for}\
 \Lambda_0
\ll
 1 ,
\]
and
\[
\fl
 a'_{00k}
=
\frac{
 (1 - \nu) \, C_2^2 \, \sL^{2/3} \, \s_k(\xs)
}{
 \sst
}
\left(
-
\frac{
 f(0)
}{
 C_3
}
\,
\frac{
 \ex^{-\sqrt{\Lambda_0} \, \xs}
}{
 \sqrt{\Lambda_0} \, (N_k + \Lambda_0)
}
+
\frac{
 \ell \, C_1^2
}{
 4
}
\,
 \frac{\eps^{1/2}}{\Lambda_0}
\right) ,
\quad\mbox{for}\
 \Lambda_0
\gg
 1 .
\]
Note that,
in this last formula,
the first term in the parentheses
dominates the second one
for all $o(1)$ values of $\mu$
(cf.~(\ref{Lambda0-log}));
the two terms
only become commensurate
for $\Or(1)$ values of $\mu$.

\section{An asymptotic formula for $b'_{m0k}$
\label{s-bm0k}}
\setcounter{equation}{0}
Finally,
we derive 
the asymptotic formula
for $b'_{m0k}$
%
\be
 b'_{m0k}
=
-
 A'_k(\Lambda_0)
\,
 B ,
\quad\mbox{for}\
 0 \ne k , m \ll \eps^{-1/3} ,
\Label{bm0k-ae}
\ee
which has already been reported in (\ref{coeffs-ae}).
We also remark that,
here also,
this result is valid
for those values of $k$
for which $\s_k(\xs) \ne 0$.
Theorem~\ref{t-Laplace} yields
an (algebraically) higher order result
for the remaining values of $k$.

Definition~(\ref{ab-coeffs})
and
(\ref{psit-n})
yield the expression
\begin{eqnarray}
\fl
 b'_{m0k}
&=&
 \eps^{-1/6} \delta \, (1 - \nu)
 \int_0^1
 f(x)
\,
 \s_m(x)
\,
 \w_0(x)
\,
 \y_k(x)
\,
 dx
\nonumber\\
\fl
&{}&
+
 \eps^{5/6} \, \delta \, \ell^{-1}
 (1 - \nu)
 \int_0^1
\,
 f(x)
\,
 \s_m(x)
\,
 \wpu_0(x)
\,
 \s_k(x)
\,
 dx
\nonumber\\
\fl
&=&
 \frac{\eps^{-1/6} \delta \, (1-\nu)}{\ell}
\left[
 \frac{1}{W_\y}
 \int_0^1
 \int_0^x
 f(x)
\,
 f(y)
\,
 \s_m(x)
\,
 \s_k(y)
\,
 \w_0(x)
\,
 \y_{k,-}(x)
 \ypu_{k,+}(y)
\,
 dy dx
\right.
\nonumber\\
\fl
&{}&
\hspace*{2.25cm}
-
 \frac{D_k}{W_\y}
 \int_0^1
 \int_0^1
 f(x)
\,
 f(y)
\,
 \s_m(x)
\,
 \s_k(y)
\,
 \w_0(x)
\,
 \y_{k,-}(x)
 \ypu_{k,-}(y)
\,
 dy dx
\nonumber\\
\fl
&{}&
\hspace*{2.25cm}
+
 \frac{1}{W_\y}
 \int_0^1
 \int_x^1
 f(x)
\,
 f(y)
\,
 \s_m(x)
\,
 \s_k(y)
\,
 \w_0(x)
\,
 \y_{k,+}(x)
\,
 \ypu_{k,-}(y)
\,
 dy dx
\nonumber\\
\fl
&{}&
\hspace*{2.25cm}
\left.
+
 \eps
 \int_0^1
 f(x)
\,
 \s_m(x)
\,
 \s_k(x)
\,
 \wpu_0(x)
\,
 dx
\right] .
\Label{bm0k-expl}
\end{eqnarray}
Let $\I_1, \ldots, \I_4$
denote the integrals
in the right member
of this formula
in the order
that they appear in it.
We will derive
the leading order terms
in the asymptotic expansions
of these integrals
using Theorem~\ref{t-Laplace},
as in the previous section
and also for $k,m \ll \eps^{-1/3}$.
In what follows,
we omit the terms
$\theta \w^2_{0,-}(1;\xo) \w_{0,+}(1;\xo)$
and
$\theta \w^2_{0,-}(1;\xo) \wpu_{0,+}(1;\xo)$
in (\ref{wo}) and (\ref{wo+}),
respectively,
as one can show
that their contribution
is exponentially small
compared to
the leading order terms
(see also
Section~\ref{s-a000} and \ref{s-bm00}--\ref{s-a00k}).

First,
we derive a formula for
$-(W_\y)^{-1} D_k \, \I_2 + (W_\y)^{-1} \I_3 + \eps \, \I_4$.
We write
\begin{eqnarray*}
 \I_2
&=&
 \int_0^1
\left(
 \int_0^1
 f(x)
\,
 \s_m(x)
\,
 \w_0(x)
\,
 \y_{k,-}(x)
\,
 dx
\right)
 f(y)
\,
 \s_k(y)
\,
 \ypu_{k,-}(y)
\,
 dy
\\
&=&
 \eps^{1/3}
\,
 \sqrt{2}
\,
 f(0)
\,
 C_1^{-1}
\,
 \sL^{-1/3}
\,
 \int_0^1
 f(y)
\,
 \s_k(y)
\,
 \ypu_{k,-}(y)
\,
 dy ,
\end{eqnarray*}
where
we have used that
$\y_{k,-} = \eps^{1/3} \, C_1^{-1} \sL^{-1/3} \w_0$
in a neighborhood
of the origin,
that $\w_0$ is
exponentially small
outside this neighborhood,
the identity
$\|\w_0\|_2 = 1$,
and
(\ref{vs}).
Employing (\ref{psin-}),
next,
we obtain
\[
 \I_2
=
 \eps^{7/12}
\,
\frac{
 f(0) \, C_2
}{
 \sqrt{2 \pi} \, C_1 \, \sL^{1/3}
}
\,
 \int_0^1
\frac{
 f(y)
\,
 \s_k(y)
}{
F^{1/4}(y)
}
\,
 \exp\left( \frac{J_-(y)}{\sqrt{\eps}} \right)
\,
 dy .
\]
Substituting for $D_k$ from (\ref{Do-ae}),
we obtain
\be
 (W_\y)^{-1} \, D_k \, \I_2
=
 \eps^{-1/12}
\,
 \int_0^1
 \Xi_2(y)
\,
 \exp\left( \frac{\Pi_2(y)}{\sqrt{\eps}} \right)
\,
 dy ,
\Label{bm0k-I2-aux1}
\ee
where
we have defined
the functions
\be
 \Xi_2(y)
=
\frac{
 C_2 \, f(0) \, d_k
}{
 \sqrt{2 \pi} \, C_1 \, W_\y \, \sL^{1/3}
}
\,
\frac{
 f(y)
\,
 \s_k(y)
}{
F^{1/4}(y)
}
\quad\mbox{and}\quad
 \Pi_2(y)
=
 J_-(y) .
\Label{bm0k-I2-aux2}
\ee
Next,
we change the order
in which integration
is carried out
in $\I_3$
and
use (\ref{wo})
and
(\ref{psin-})--(\ref{psin+})
to rewrite this integral
as
\begin{eqnarray}
\fl
 (W_\y)^{-1} \, \I_3
&=&
 (W_\y)^{-1}
\,
 \int_0^1
\left(
 \int_0^y
 f(x)
\,
 \s_m(x)
\,
 \w_0(x)
\,
 \y_{k,+}(x)
\,
 dx
\right)
 f(y)
\,
 \s_k(y)
\,
 \ypu_{k,-}(y)
\,
 dy
\nonumber\\
\fl
&=&
 \eps^{5/12}
\,
 \int_0^1
 \Xi_3(y)
\,
 \exp\left( \frac{\Pi_3(y)}{\sqrt{\eps}} \right)
\,
 dy ,
\Label{bm0k-I3-aux1}
\end{eqnarray}
where
$\Pi_3(y) = \Pi_2(y)$
and
\be
 \Xi_3(y)
=
\frac{
 C_1 \, C_2 \, \sL^{1/3}
}{
 4 \pi^{3/2} \, W_\y
}
\left(
 \int_0^y
\frac{
 f(x)
\,
 \s_m(x)
}{
 \sqrt{F(x)}
}
\,
 dx
\right)
\frac{
 f(y)
\,
 \s_k(y)
}{
 F^{1/4}(y)
} .
\Label{bm0k-I3-aux2}
\ee
Finally,
using (\ref{wo+})
and
renaming the integration variable $x$
into $y$,
we obtain
\be
 \eps \, \I_4
=
 \eps^{-13/12}
\,
 \int_0^1
 \Xi_4(y)
\,
 \exp\left( \frac{\Pi_4(y)}{\sqrt{\eps}} \right)
\,
 dy ,
\Label{bm0k-I4-aux1}
\ee
where
\be
 \Xi_4(y)
=
 \frac{C_1 \, C_2 \, \sL^{1/3}}{2 \sqrt{\pi}}
\,
\frac{
 f(y)
\,
 \s_m(y)
\,
 \s_k(y)
}{
 F^{1/4}(y)
}
\quad\mbox{and}\quad
 \Pi_4(y)
=
 \Pi_3(y)
=
 \Pi_2(y) .
\Label{bm0k-I4-aux2}
\ee
Combining
(\ref{bm0k-I2-aux1})--(\ref{bm0k-I4-aux2}),
we obtain
\be
 -(W_\y)^{-1} D_k \, \I_2 + (W_\y)^{-1} \I_3 + \I_4
=
-
 \eps^{-1/12}
\,
 \int_0^1
 \tilde{\Xi}_2(y)
\,
 \exp\left( \frac{\Pi_2(y)}{\sqrt{\eps}} \right)
\,
 dy ,
\Label{bm0k-I234-aux1}
\ee
where,
to leading order
and uniformly over $[0,1]$,
\be
 \tilde{\Xi}_2(y)
=
 \Xi_2(y)
=
\frac{
 C_2 \, d_k \, \s_k(y) \, f(0) \, f(y)
}{
 \sqrt{2 \pi} \, C_1 \, \sL^{1/3} \, W_\y \, F^{1/4}(y)
} .
\Label{bm0k-I234-aux2}
\ee

Regarding $\I_1$,
we use (\ref{wo+})
and
(\ref{psin-})--(\ref{psin+}),
to write it
in the form
\be
 (W_\y)^{-1}
\,
 \I_1
=
 \eps^{5/12}
\,
 \int\int_D
 \Xi_1(x,y)
\,
 \exp\left(\frac{\Pi_1(x,y)}{\sqrt{\eps}}\right)
\,
 dA ,
\Label{bm0k-I1-aux1}
\ee
where
$D
=
\{ (x,y) \vert 0\le x \le 1 \,\mbox{and}\, 0 \le y \le x \}$,
$\Pi_1(x,y) = J_+(y) - 2 I(x)$,
and
\be
 \Xi_1(x,y)
=
\frac{
 C_1 \, C_2 \, \sL^{1/3}
}{
 4 \pi^{3/2} \, W_\y
}
\,
\frac{
 f(x) \, f(y) \, \s_m(x) \, \s_k(y)
}{
 \sqrt{F(x)} \, {F^{1/4}(y)}
} .
\Label{bm0k-I1-aux2}
\ee

First,
we estimate
$-(W_\y)^{-1} \, D_k \, \I_2 + (W_\y)^{-1} \, \I_3 + \eps \, \I_4$,
cf. (\ref{bm0k-I234-aux1})--(\ref{bm0k-I234-aux2}).
The exponent $\Pi_2$
assumes its maximum
at the interior critical point
$\xs \in (0,1)$,
and hence
Theorem~\ref{t-Laplace} yields
\[
\fl
 -(W_\y)^{-1} \, D_k \, \I_2 + (W_\y)^{-1} \, \I_3 + \eps \, \I_4
=
-
 \eps^{1/4}
\,
\frac{
 \sqrt{2 \pi}
}{
 \sqrt{-J''_-(\xs)}
}
\,
\left(
 \eps^{-1/12} \, \delta^{-1} \, \tilde{\Xi}_2(\xs)
\right)
=
 -\eps^{1/6} \, \delta^{-1} \, \tilde{C}_2 .
\]
Here,
\[
 \tilde{C}_2
=
\frac{
 \sqrt{2} \, C_2 \, \sL^{1/3} \, f(0) \, f(\xs) \, \s_k(\xs)
}{
 C_1 \, C_3 \, \sst^{1/2} \, (N_k + \Lambda_0)
} .
\]
Next,
we estimate $\I_1$,
cf. (\ref{bm0k-I1-aux1})--(\ref{bm0k-I1-aux2}).
The exponent $\Pi_1$
assumes its maximum
at the point
$(\xs,\xs) \in \partial D$
which is not
a critical point
of $\Pi_1$
(compare to
the maximization of $\Pi_4$
in Section~\ref{s-a000}).
As a result,
Theorem~\ref{t-Laplace} yields
\[
 (W_\y)^{-1}
\,
 \I_1
=
 \eps^{3/4}
\,
 C'_1
\,
\left(
 \eps^{5/12} \, \delta^{-1} \, \Xi_1(\xs,\xs)
\right)
=
 \eps^{7/6} \, \delta^{-1} \, C''_1
\]
to leading order,
and
with $C'_1$ and $C''_1$
being $\Or(1)$ constants.

In total, then,
and
to leading order,
we obtain
\[
 b'_{m0k}
=
-
\frac{
 \sqrt{2} \, (1-\nu) \, C_2 \, \sL^{1/3} \, f(0) \, \s_k(\xs)
}{
 C_1 \, C_3 \, \sst^{-1/2} \, (N_k + \Lambda_0)
} ,
\quad\mbox{for}\
 m , k \ll \eps^{-1/3} .
\]
Formula~(\ref{bm0k-ae})
now immediately follows.

\section{Discussion
\label{s-disc}}
\setcounter{equation}{0}
As argued in the Introduction,
there are two contextual themes
central to this article.
The first one
relates to understanding
the \emph{nonlinear}, \emph{long-term} dynamics
of small patterns of DCM type
generated through the linear destabilization mechanism
identified in \cite{ZDPS-2009}.
The second theme
concerns the development
of a concrete approach
to studying the dynamics
generated by the (rescaled) PDE model (\ref{PDE})
\emph{near} a linear destabilization
but
\emph{beyond} the region of applicability
of the center manifold reduction.
In this article,
we have reported significant results
(outlined in the Introduction)
touching on both themes.
These results,
in turn,
inspire further investigation
within this dual context.

Regarding our first focal point,
and
in view of our discovery that
the bifurcating, small-amplitude, DCM pattern
undergoes a Hopf bifurcation,
the central question is naturally
what happens \emph{beyond}
this secondary bifurcation.
This question can be answered
by the methods developed here,
as it is in principle possible
to deduce analytically
the sub- or supercriticality
of the Hopf bifurcation
undergone by (\ref{redux-ODE-a=1}).
The numerical simulations of \cite{HPKS-2006}
indicate that this bifurcation
may be only the first
of a cascade of subsequent
period-doubling bifurcations
leading to a region
of spatio-temporal chaotic dynamics
and
throughout which
the phytoplankton profile
maintains a DCM-like structure.
There is, of course,
no \emph{a priori} reason
for this cascade
to occur entirely
within the regime
$\lambda - \lambda_* = {\cal O}(\eps)$
covered by our analysis here.
In fact,
the simulations of \cite{HPKS-2006} suggest that,
for the parameter combinations considered there,
this is indeed not the case.
On the other hand,
our analysis is able
to determine regions
in parameter space
where this cascade
can or cannot occur
(for instance,
in the event that
the Hopf bifurcation turns out
to be subcritical).
Moreover,
the possibility
that there exist regions
in parameter space
where the entire cascade
is within the reach
of our asymptotic methods
cannot be excluded.
A similar question concerns,
naturally,
the origins and fate
of the second DCM pattern
identified in Section~\ref{ss-destroy}.

These last remarks
bring us to
the second theme.
The approach we developed here
will be used---and
if necessary extended---in forthcoming work
investigating the remaining issues
pertaining to our linear destabilization results
in \cite{ZDPS-2009}---namely,
determining the nonlinear behavior
associated with the destabilization of BL-type.
Our analysis in \cite{ZDPS-2009}
strongly suggests that,
for realistic choices
of the parameters
pertinent to shallower water columns
(\emph{e.g.}, estuaries and lakes),
patterns of benthic layer (BL) type
are equally relevant
to the dynamics generated by (\ref{PDE-orig})
as the DCM patterns considered here.
In fact,
preliminary numerical simulations
strongly suggest that co-dimension two-type patterns
combining DCM and BL characteristics
play an important role
in the region where
the trivial state is unstable.
From a mathematical point of view,
the co-dimension two point
may also be seen
as an `\emph{organizing center}'
for the more complex behavior
exhibited by the system
studied numerically in \cite{HPKS-2006}.
That is,
the cascade of period doubling bifurcations
reported in \cite{HPKS-2006}
may be based on the presence
of that co-dimension two point.
In view of that,
the derivation and analysis
of an extended reduced system
for parameter values
valid within an ${\cal O}(\eps)$ neighborhood
of that point
may prove highly engaging.

The same methodology
can also be applied to \emph{extended} models.
A natural extension of (\ref{PDE-orig})
is a multi-species model,
\emph{i.e.},
a model similar to (\ref{PDE-orig})
in which \emph{several} phytoplankton species
compete for the same nutrient.
At the linear level,
the species evolution decouples \cite{ZDPS-2009}.
Nonlinear coupling, however, is present
through shadowing (light limitation)
and
nutrient uptake (nutrient limitation),
and hence
the presence of every extra species
affects the life cycle
of each species.
Reaction--diffusion models of this sort
for eutrophic environments---\emph{i.e.},
in the presence of
an ample nutrient supply---have been
developed and investigated
both numerically \cite{HOW-1999b}
and theoretically \cite{DH-2010}.
The \emph{oligotrophic} case,
on the other hand---where
these multi-species models
are coupled to
a PDE for the nutrient---has so far
only been investigated numerically \cite{HPKS-2006}.

Another natural,
if not outright necessary,
extension
is the inclusion
of horizontal spatial directions.
Plainly,
the dynamics generated by (\ref{PDE-orig})
will be strongly influenced
by the flow in directions
perpendicular to the one-dimensional water column
considered here:
oceanic currents
are bound to mix
neighboring water columns
and thus also
enrich the collection
of emerging planktonic patterns.
Finally,
and as already described
in the Introduction,
we are currently studying
the simplified model problem (\ref{model})
through which we hope to understand
the applicability and limitations
of the general method developed here.
This approach may also serve
as a first step
towards obtaining a rigorous validation
of our method.

\ack
The authors wish to thank an anonymous referee
for pointing out references \cite{L-1981a,LK-1985} to us.
They also acknowledge
generous hosting by
the Mathematical Biosciences Institute (MBI)
at Ohio State University
for two weeks in June 2011,
which led to a complete reworking
of Section~\ref{s-emerge}.
The authors also extend their thanks
to Marty Golubitsky
for insightful conversations
while at MBI.

\appendix
\section{An asymptotic formula for $\w_0$
\label{s-w0-ae}}
\setcounter{equation}{0}
\setcounter{equation}{0}
The formula for
the principal part
in the asymptotic expansion of $\w_0$
reads
\be
\fl
 \w_0(x)
\sim
\left\{
\begin{array}{lr}
 \eps^{-1/6}
\,
 \sL^{1/6}
\,
 C_1
 \Ai\left( A_1 (1 - \xo^{-1} x) \right) ,
&
\mbox{for} \
 x \in [ 0 , \xo ) ,
\\
\frac{
 \eps^{-1/12} \, C_1 \, C_2 \, \sL^{1/3}
}{
 2 \sqrt{\pi} \, F^{1/4}(x)
}
\,
\left[
 \w_{0,-}(x;\xo)
+
 \theta
\,
 \w^2_{0,-}(1;\xo)
\,
 \w_{0,+}(x;\xo)
\right] ,
&
\mbox{for} \
 x \in ( \xo , 1 ] ,
\end{array}
\right.
\Label{wo}
\ee
cf. \cite{ZDPS-2009},
where $\xo$, $C_1$, $C_2$, $F$, $\sL$ and $\theta$
have been defined
in (\ref{defsigF}),
(\ref{barx0}),
(\ref{defC12sstar}),
and (\ref{defgammas1}).
We remark that
$C_1$ is a normalizing constant
ensuring that $\norm{\w_0}_2 = 1$.
(This factor does not appear
in the formula for $\w_0$
we give in \cite{ZDPS-2009},
since $\w_0$
was not normalized there.) Also,
\[
 \w_{0,\pm}(x;\xo)
=
 \exp\left(\pm\frac{I(x)}{\sqrt{\eps}}\right) ,
 \]
where $I$ has been defined in (\ref{defIJpm}).
An asymptotic formula
for $\wpu_0 = E \, \w_0$
is readily derived
using (\ref{wo}) above,
\be
\fl
 \wpu_0(x)
\sim
\left\{
\begin{array}{lr}
 \eps^{-1/6}
\,
 \sL^{1/6}
\,
 C_1
 \ex^{ \sqrt{\V/\eps} \, x }
 \Ai\left( A_1 (1 - \xo^{-1} x) \right) ,
&
\mbox{for} \
 x \in [ 0 , \xo ) ,
\\
\frac{
 \eps^{-1/12} \, C_1 \, C_2 \, \sL^{1/3}
}{
 2 \sqrt{\pi} \, F^{1/4}(x)
}
\,
\left[
 \wpu_{0,-}(x;\xo)
+
 \theta
\,
 \w^2_{0,-}(1;\xo)
\,
 \wpu_{0,+}(x;\xo)
\right] ,
&
\mbox{for} \
 x \in ( \xo , 1 ] ,
\end{array}
\right.
\Label{wo+}
\ee
where
we have defined
the functions
\[
 \wpu_{0,\pm}(x;\xo)
=
 E(x)
\,
 \w_{0,\pm}(x;\xo)
=
 \exp\left(\frac{J_\pm(x)}{\sqrt{\eps}}\right) ,
\]
with
$J_\pm$ as in (\ref{defIJpm}).
We remark that
$J_-$ becomes maximum
at the well-defined point
$\xs \in (0,1)$---the location
of the DCM, see (\ref{xstar})---whereas
$J_+$ increases monotonically.
Also,
the terms involving
$\w_{0,+}$
in (\ref{wo})
and
$\wpu_{0,+}$
in (\ref{wo+})
are exponentially smaller
than the terms
$\w_{0,-}(x)$
and
$\wpu_{0,-}(x)$,
respectively,
everywhere except for
an $\Or(\sqrt{\eps})-$region
of $x = 1$.
Indeed,
for all $x < 1$,
\be
 J_+(x) - 2 I(1)
=
 J_-(x) - 2 (I(1) - I(x))
<
 J_-(x) .
\Label{J-ineq}
\ee
In particular,
$\|\wpu_0\|_\infty$ can be bounded by
an $\Or(\eps^{-1/12} \delta^{-1})$ constant,
where
$\delta = \exp(-J_-(\xs))/\sqrt{\eps}$
is an exponentially small parameter
(cf. (\ref{delta})).

\section{An asymptotic formula for $\n_0$
\label{s-n0-ae}}
\setcounter{equation}{0}
We recall that
$\n_0$ is the solution
to the boundary value problem (\ref{ODE-eta}),
\[
 \eps \, \dxx\n_0
-
 \lambda_0 \, \n_0
=
 -\eps \ell^{-1} f \, \wpu_0 ,
\quad\mbox{where} \
 \dx \n_0(0)
=
 \n_0(1)
=
 0 .
\]
Recalling that
$\lambda_0 = \eps \Lambda_0$
in our bifurcation analysis,
we find that
\be
 \dxx\n_0 - \Lambda_0 \n_0
=
 -\ell^{-1} f \wpu_0 ,
\quad\mbox{where} \
 \dx \n_0(0)
=
 \n_0(1)
=
 0 .
\Label{ODE-eta0}
\ee
The functions
$\n_{0,\pm}(x) = \ex^{\pm \sqrt{\Lambda_0} x}$
form a pair
of fundamental solutions to
the homogeneous problem.
Using variation of constants,
then,
we obtain a special solution
to the inhomogeneous ODE,
\begin{eqnarray*}
 \n_{0,{\mathrm sp}}(x)
&=
 (2 \ell \sqrt{\Lambda_0})^{-1}
\left[
 \bmp{\n_{0,+} f}{x}{0}
 \n_{0,-}(x)
-
 \bmp{\n_{0,-} f}{x}{0}
 \n_{0,+}(x)
\right] .
\Label{eta0-sp}
\end{eqnarray*}
Here,
we have defined
the family of functionals
\be
\fl
 \bmp{\cdot}{x}{n}
=
 \int_0^x \cdot(s) \, \wpu_n(s) \ ds ,
\quad\mbox{parameterized by} \
 x \in [0,1]
\ \mbox{and} \
 n \ge 0 .
\Label{bump-fnal}
\ee
The solution
to (\ref{ODE-eta0}) is,
then,
\be
\left\{
\begin{array}{lcl}
 \n_0(x)
&=&
\left[
 C_\n^+
-
 (2 \ell \sqrt{\Lambda_0})^{-1}
 \bmp{\n_{0,-} f}{x}{0}
\right]
 \n_{0,+}(x)
\\
&{}&\ \
+
\left[
 C_\n^-
+
 (2 \ell \sqrt{\Lambda_0})^{-1}
 \bmp{\n_{0,+} f}{x}{0}
\right]
 \n_{0,-}(x) .
\end{array}
\right.
\Label{eta0-def}
\ee
Imposing the boundary conditions
for $\n_0$
and
using the identity
$\bmp{\cdot}{0}{0} = 0$,
we find that
the constants
$C_\n^-$ and $C_\n^+$
satisfy the linear system
\begin{eqnarray*}
 \sqrt{\Lambda_0}
\
 C_\n^+
 -
 \sqrt{\Lambda_0}
\
 C_\n^-
=
 0 ,
\\
\left[
 2 \ell \sqrt{\Lambda_0}
 C_\n^+
-
 \bmp{\n_{0,-} f}{1}{0}
\right]
 \ex^{\sqrt{\Lambda_0}}
+
\left[
 2 \ell \sqrt{\Lambda_0}
 C_\n^-
+
 \bmp{\n_{0,+} f}{1}{0}
\right]
 \ex^{-\sqrt{\Lambda_0}}
=
 0 ,
\end{eqnarray*}
the solution to which is
$C_\n^+
=
C_\n^-
=
C_\n/(2 \ell \sqrt{\Lambda_0})$,
with
\[
 C_\n
=
\frac{
 \bmp{\n_{0,-} f}{1}{0}
 \n_{0,+}(1)
-
 \bmp{\n_{0,+} f}{1}{0}
 \n_{0,-}(1)
}{
 2 \cosh\sqrt{\Lambda_0}
} .
\]
Thus,
(\ref{eta0}) becomes
\be
\fl
 \n_0(x)
=
 (2 \ell \sqrt{\Lambda_0})^{-1}
\left[
 2 C_\n \cosh\left(\sqrt{\Lambda_0} \, x\right)
+
 \bmp{\n_{0,+} f}{x}{0}
 \n_{0,-}(x)
-
 \bmp{\n_{0,-} f}{x}{0}
 \n_{0,+}(x)
\right] .
\Label{eta0-aux}
\ee
Further employing
the definition (\ref{bump-fnal}),
we calculate
\begin{eqnarray*}
\fl
\lefteqn{
 \bmp{\n_{0,+} f}{x}{0}
 \n_{0,-}(x)
-
 \bmp{\n_{0,-} f}{x}{0}
 \n_{0,+}(x)
}
\\
&=&
 \int_0^x
\left[
 \n_{0,-}(x) \n_{0,+}(y)
-
 \n_{0,+}(x) \n_{0,-}(y)
\right]
\,
 f(y)
\,
 \wpu_0(y)
\,
 dy
\\
&=&
 -2
 \int_0^x
 \sinh\left(\sqrt{\Lambda_0} (x-y)\right)
\,
 f(y)
\,
 \wpu_0(y)
\,
 dy
\\
&=&
 -2
\,
 \bmp{\sinh\left(\sqrt{\Lambda_0} (x - \cdot)\right) f}{x}{0} .
\end{eqnarray*}
Additionally,
\begin{eqnarray*}
 C_\n
&=&
\frac{
 \bmp{\n_{0,-} f}{1}{0}
 \n_{0,+}(1)
-
 \bmp{\n_{0,+} f}{1}{0}
 \n_{0,-}(1)
}{
 2 \cosh\sqrt{\Lambda_0}
}
\\
&=&
\frac{1}{2 \cosh\sqrt{\Lambda_0}}
 \int_0^1
\left[
 \n_{0,-}(y)
\,
 \n_{0,+}(1)
-
 \n_{0,+}(y)
\,
 \n_{0,-}(1)
\right]
\,
 f(y)
\,
 \wpu_0(y)
\,
 dy
\\
&=&
 \frac{1}{\cosh\sqrt{\Lambda_0}}
 \int_0^1
 \sinh\left(\sqrt{\Lambda_0}(1-y)\right)
\,
 f(y)
\,
 \wpu_0(y)
\,
 dy
\\
&=&
 \frac{1}{\cosh\sqrt{\Lambda_0}}
\,
 \bmp{\sinh\left(\sqrt{\Lambda_0} (1 - \cdot)\right) f}{1}{0} ,
\end{eqnarray*}
and hence
(\ref{eta0-aux}) becomes
\begin{eqnarray}
 \n_0(x)
&=&
 \frac{1}{\ell \sqrt{\Lambda_0}}
\left[
\frac{
 \cosh\left(\sqrt{\Lambda_0} \, x\right)
}{
 \cosh\sqrt{\Lambda_0}
}
\,
 \bmp{\sinh\left(\sqrt{\Lambda_0} (1 - \cdot)\right) f}{1}{0}
\right.
\nonumber\\
&{}&
\hspace*{1.5cm}
\left.
-
 \bmp{\sinh\left(\sqrt{\Lambda_0} (x - \cdot)\right) f}{x}{0}
\right]
\nonumber\\
&=&
 \frac{1}{\ell \sqrt{\Lambda_0}}
\left[
\frac{
 \cosh\left(\sqrt{\Lambda_0} \, x\right)
}{
 \cosh\sqrt{\Lambda_0}
}
\,
 \int_0^1
 \sinh\left(\sqrt{\Lambda_0} (1 - y)\right)
\,
 f(y)
\,
 \wpu_0(y)
\,
 dy
\right.
\nonumber\\
&{}&
\hspace*{1.5cm}
\left.
-
 \int_0^x
 \sinh\left(\sqrt{\Lambda_0} (x - y)\right)
\,
 f(y)
\,
 \wpu_0(y)
\,
 dy
 \right] .
\Label{eta0}
\end{eqnarray}

To estimate $\| \n_0 \|_\infty$ over $[0,1]$,
we first show
that $\n_0$ is positive
and
that it assumes its maximum
in an $\Or(\eps^{1/4})$ neighborhood of $\xs$.
First,
an estimate based on (\ref{eta0})
establishes readily that
$\n_0(x) > 0$ for all $x \in (0,1)$,
\begin{eqnarray*}
\fl
 \n_0(x)
&\ge
 \int_0^1
\left[
\frac{
 \cosh\left(\sqrt{\Lambda_0} \, x\right)
}{
 \cosh\sqrt{\Lambda_0}
}
 \sinh\left(\sqrt{\Lambda_0} (1 - y)\right)
-
 \sinh\left(\sqrt{\Lambda_0} (x - y)\right)
\right]
\,
\frac{
 f(y)
\,
 \wpu_0(y)
}{
 \ell \sqrt{\Lambda_0}
}
\,
 dy
\\
\fl
&=
\frac{
 \sinh\left(\sqrt{\Lambda_0} (1 - x)\right)
}{
 \ell \sqrt{\Lambda_0} \, \cosh\sqrt{\Lambda_0}
}
\,
 \int_0^1
 \cosh\left(\sqrt{\Lambda_0} \, y\right)
\,
 f(y)
\,
 \wpu_0(y)
\,
 dy
>
 0 ,
\end{eqnarray*}
for $x \in (0,1)$.
To locate the maximum,
we differentiate both members of (\ref{eta0})
and obtain
\begin{eqnarray}
\fl
 \ell
\,
 \partial_x\n_0(x)
=&
\frac{
 \sinh\left(\sqrt{\Lambda_0} \, x\right)
}{
 \cosh\sqrt{\Lambda_0}
}
\,
 \int_0^1
 \sinh\left(\sqrt{\Lambda_0} (1 - y)\right)
\,
 f(y)
\,
 \wpu_0(y)
\,
 dy
\nonumber\\
\fl
&{}
-
 \int_0^x
 \cosh\left(\sqrt{\Lambda_0} (x - y)\right)
\,
 f(y)
\,
 \wpu_0(y)
\,
 dy
\nonumber\\
\fl
&{}
-
\frac{
 \sinh\left(\sqrt{\Lambda_0} (x - y)\right)
}{
 \sqrt{\Lambda_0}
}
\,
 f(x)
\,
 \wpu_0(x) .
\Label{no'}
\end{eqnarray}
Theorem~\ref{t-Laplace} can be used
to yield the principal part
of the two integrals
in this formula,
whereas the term
proportional to $\wpu_0$
can be estimated via (\ref{wo+}).
For the values of $\Lambda_0$
we are interested in,
the localized term
in either integrand is $\wpu_0$,
while
the $\Lambda_0-$dependent terms
vary on an asymptotically larger length scale.
Thus,
\[
 \partial_x\n_0(x)
=
 \frac{\sinh\left(\sqrt{\Lambda_0} \, x\right)}{\ell \cosh\sqrt{\Lambda_0}}
\,
 \int_0^1
 \sinh\left(\sqrt{\Lambda_0} (1 - y)\right)
\,
 f(y)
\,
 \wpu_0(y)
\,
 dy
>
 0 ,
\]
to leading order
and
for $x < \xs$
and
$\abs{x-\xs} \gg \eps^{1/4}$,
since the second and third terms
in the right member of (\ref{no'})
are exponentially small
compared to the first one.
Similarly,
\begin{eqnarray*}
\fl
 \partial_x\n_0(x)
=&
 \frac{1}{\ell \, \cosh\sqrt{\Lambda_0}}
\left[
 \sinh\left(\sqrt{\Lambda_0} \, x\right)
\,
 \int_0^1
 \sinh\left(\sqrt{\Lambda_0} (1 - y)\right)
\,
 f(y)
\,
 \wpu_0(y)
\,
 dy
\right.
\nonumber\\
\fl
&{}
\hspace*{2.75cm}
\left.
-
 \cosh\sqrt{\Lambda_0}
 \int_0^x
 \cosh\left(\sqrt{\Lambda_0} (x - y)\right)
\,
 f(y)
\,
 \wpu_0(y)
\,
 dy
\right] ,
\end{eqnarray*}
for $x > \xs$
and
$\abs{x-\xs} \gg \eps^{1/4}$,
since the second and third terms
in the same formula
are of the same asymptotic order
and
the third one is exponentially smaller.
Changing the upper limit of the second integral to one
(and thus only introducing
an exponentially small error)
and
combining the two integrals,
we find
\[
 \partial_x\n_0(x)
=
-
 \frac{\cosh\left(\sqrt{\Lambda_0} \, (1-x)\right)}{\ell \, \cosh\sqrt{\Lambda_0}}
 \int_0^1
 \cosh\left(\sqrt{\Lambda_0} \, y\right)
\,
 f(y)
\,
 \wpu_0(y)
\,
 dy
<
 0 ,
\]
Since $\n_0 \in {\rm C}^2(0,1)$,
now,
it follows that $\n_0'(x_1) = 0$
at a point $x_1$ such that
$\abs{\xs - x_1} = \Or(\eps^{1/4})$,
as desired.
Hence,
we can now use (\ref{eta0})
to estimate further
\[
\fl
 \|\n_0\|_\infty
\le
 \n_0(x_1)
\le
\frac{
 \cosh\left(\sqrt{\Lambda_0} \, x_1\right)
}{
 \ell
\,
 \sqrt{\Lambda_0}
\,
 \cosh\sqrt{\Lambda_0}
}
 \int_0^1
 \sinh\left(\sqrt{\Lambda_0} (1 - y)\right)
\,
 f(y)
\,
 \wpu_0(y)
\,
 dy .
\]
Using our asymptotic estimate on $x_1$
and
Theorem~\ref{t-Laplace},
we find
\[
 \|\n_0\|_\infty
\le
 C
\,
 \eps^{1/6}
\,
 \delta^{-1}
\,
\frac{
 \cosh\left(\sqrt{\Lambda_0} \, \xs\right)
\,
 \sinh\left(\sqrt{\Lambda_0} (1 - \xs)\right)
}{
 \sqrt{\Lambda_0}
\,
 \cosh\sqrt{\Lambda_0}
} .
\]
for some $\Lambda_0-$independent,
$\Or(1)$ constant $C$.
Since the $\Lambda_0-$dependent quantity
in the bound above
remains bounded by an $\Or(1)$ constant
also for $\Lambda_0 \gg 1$,
we finally conclude that
$\|\n_0\|_\infty$ can be bounded by
an $\Or(\eps^{1/6} \delta^{-1})$ constant.

\section{Asymptotic formulas for $\y_n$, $n \ge 0$
\label{s-y-ae}}
\setcounter{equation}{0}
The function $\y_n$
is the solution
to the boundary value problem
\[
\fl
 \eps \, \dxx\y_n
+
 (f(x) - \ell - \V - \nu_n) \, \y_n
=
 -\eps \ell^{-1} f E \, \s_n ,
\quad\mbox{where}\
 \G{\y_n}{0}
=
 \G{\y_n}{1}
=
 0 ,
\]
cf. (\ref{ODEt-psi}).
Here,
$\G{\y_n}{x}
=
\y_n(x)
-
\sqrt{\eps/\V} \, \dx \y_n(x)$
and
we recall that
\be
 \s_n(x)
=
 \sqrt{2} \cos(\sqrt{N_n} \, x) ,
\Label{dl-def}
\ee
see (\ref{Nn-def}).
Recalling also the definitions
$F(x)
=
f(0) - f(x)$
and
$\lambda_*
=
f(0) - \ell - \V$,
as well as that
$\lambda_*
=
\lambda_0 + \eps^{1/3} \mu_0$
by (\ref{lambdan}),
we write
\[
 f(x) - \ell - \V
=
 \lambda_0 - F(x) + \eps^{1/3} \mu_0 ,
\]
with
$\mu_0
=
\sL^{2/3}
\abs{A_1}
+
\Or(\eps^{1/6})$.
Finally,
since $\lambda_0 = \eps \Lambda_0$
and
$\nu_n = -\eps N_n$,
we may rewrite (\ref{ODEt-psi})
in the final form
\be
\fl
 \eps \, \dxx\y_n
-
 \left[ F(x) - \eps^{1/3} \mu_0 - \eps \left(N_n + \Lambda_0\right) \right] \, \y_n
=
 -\frac{\eps \, E \, f \, \s_n}{\ell} ,
\Label{ODE-psit-n-aux}
\ee
together with
the boundary conditions
$\G{\y_n}{0}
=
\G{\y_n}{1}
=
0$.
In what follows,
we derive asymptotic formulas for $\y_n$
and
for values of $n$
satisfying $n \ll \eps^{-1/3}$.
In that case,
$\eps (N_n + \Lambda_0) \ll \eps^{1/3}$---recall our assumption
that $\Lambda_0 \ll \eps^{-2/3}$
in Section~\ref{sss-bifODEs}---and hence
this term is perturbative to
$\eps^{1/3} \mu_0$.
Hence,
we may write
\be
 \eps^{1/3} \mu_0
+
 \eps (N_ n + \Lambda_0)
=
 F(x_n) ,
\quad\mbox{where}\quad
 x_n
=
 \xo (1 + o(1))
\Label{xn-ae}
\ee
is a turning point
for (\ref{ODE-psit-n-aux}).
Then,
(\ref{ODE-psit-n-aux}) becomes
\be
 \eps \, \dxx\y_n
-
 \left[ F(x) - F(x_n) \right] \, \y_n
=
 -\eps \ell^{-1} f E \, \s_n ,
\Label{ODE-psit-n}
\ee
equipped with
the boundary conditions
(\ref{ODEt-psi}).
The solution
to this boundary-value problem
may be found
by variation of constants,
\be
\fl
 \y_n(x)
=
\left[
 C_\y^+
-
 (\ell W_\y)^{-1}
 G_-(x)
\right]
 \y_{n,+}(x)
+
\left[
 C_\y^-
+
 (\ell W_\y)^{-1}
 G_+(x)
\right]
 \y_{n,-}(x) .
\Label{psit-n-def}
\ee
Here,
$\y_{n,\pm}$ is any pair
of fundamental solutions to
$\eps \, \dxx\y_n
=
\left[ F(x) - F(x_n) \right] \y_n$
and
$W_\y
=
\y_{n,-}
\dx\y_{n,+}
-
\y_{n,+}
\dx\y_{n,-}$
is the associated Wronskian.
(To derive the result above,
one needs to show
that $W_\y$ is constant.
This is plain to show
by using the identity
$\partial_x W_\y(x) = 0$,
for all $x\in[0,1]$,
which follows from
the definition of $W_\y$
and
the ODE
that $\y_\pm$ satisfy.)
Further,
\be
 G_\pm(x)
=
 \int_0^x f(y) \s_n(y) \, \ypu_{n,\pm}(y) dy ,
\Label{Gpm}
\ee
where
$\ypu_{n,\pm} = E \, \y_{n,\pm}$.
Using (\ref{psit-n-def}),
we further obtain
\[
\fl
 \dx\y_n(x)
=
\left[
 C^+_\y
-
 (\ell W_\y)^{-1}
 G_-(x)
\right]
 \dx\y_{n,+}(x)
+
\left[
 C^-_\y
+
 (\ell W_\y)^{-1}
 G_+(x)
\right]
 \dx\y_{n,-}(x) ,
\]
and thus
the boundary conditions
yield the system
\begin{eqnarray*}
 C^+_\y
 \G{\y_{n,+}}{0}
+
 C^-_\y
 \G{\y_{n,-}}{0}
=
 0 ,
\\
\left[
 C^+_\y
-
 \frac{1}{\ell W_\y}
 G_-(1)
\right]
 \G{\y_{n,+}}{1}
+
\left[
 C^-_\y
+
 \frac{1}{\ell W_\y}
 G_+(1)
\right]
 \G{\y_{n,-}}{1}
=
 0 .
\end{eqnarray*}
The solution
to this system
is
\be
 C^+_\y
=
 -\frac{1}{\ell W_\y} \, D_\y \, \G{\y_{n,-}}{0}
\quad\mbox{and}\quad
 C^-_\y
=
 \frac{1}{\ell W_\y} \, D_\y \, \G{\y_{n,+}}{0} ,
\Label{Cpsit}
\ee
where
\be
 D_\y
=
\frac{
 G_-(1)
\,
 \G{\y_{n,+}}{1}
-
 G_+(1)
\,
 \G{\y_{n,-}}{1}
}{
 \G{\y_{n,+}}{0}
\,
 \G{\y_{n,-}}{1}
-
 \G{\y_{n,-}}{0}
\,
 \G{\y_{n,+}}{1}
} .
\Label{Dpsi}
\ee
Thus, also,
(\ref{psit-n-def}) becomes
\be
 \y_n(x)
=
 (\ell W_\y)^{-1}
\,
\left[
 \Gamma_-(x)
\,
 \y_{n,-}(x)
-
 \Gamma_+(x)
\,
 \y_{n,+}(x)
\right] ,
\Label{psit-n-aux}
\ee
where
\begin{eqnarray}
 \Gamma_-(x)
&=&
 G_+(x)
+
 D_\y \, \G{\y_{n,+}}{0}
\Label{Gamma-}
\\
 \Gamma_+(x)
&=&
 G_-(x)
+
 D_\y \, \G{\y_{n,-}}{0} .
\Label{Gamma+}
\end{eqnarray}

These formulas hold
for an arbitrary pair
$\y_{n,\pm}$
of fundamental solutions.
Working as in \cite{ZDPS-2009},
where
the problem
was considered
in detail
in the absence
of the perturbative term
$\eps(N_ n + \Lambda_0)$,
we can derive
the following leading order formulas
for a specific pair
of solutions $\y_{n,\pm}$:
\begin{eqnarray}
 \y_{n,-}(x)
&=&
\left\{
\begin{array}{lr}
 \eps^{1/6}
\,
 \sL^{-1/6}
\,
 \Ai\left( A_1 (1 - \xo^{-1} x) \right) ,
&
\mbox{for} \
 x \in [ 0 , \xo ) ,
\\
 \eps^{1/4}
\,
 \frac{C_2}{2 \sqrt{\pi} \, F^{1/4}(x)}
\,
 \w_{0,-}(x;\xo) ,
&
\mbox{for} \
 x \in ( \xo , 1 ] ,
\end{array}
\right.
\Label{psin-} \\
 \y_{n,+}(x)
&=&
\left\{
\begin{array}{lr}
 \eps^{1/6}
\,
 \sL^{-1/6}
\,
 \Bi\left( A_1 (1 - \xo^{-1} x) \right) ,
&
\mbox{for} \
 x \in [ 0 , \xo ) ,
\\
 \eps^{1/4}
\,
 \frac{1}{\sqrt{\pi} \, C_2 \, F^{1/4}(x)}
\,
 \w_{0,+}(x;\xo) ,
&
\mbox{for} \
 x \in ( \xo , 1 ] .
\end{array}
\right.
\Label{psin+}
\end{eqnarray}
Here,
we have used that
$x_n = \xo + o(\sqrt{\eps})$.
The identity
$\partial_x W_\y = 0$,
which was
reported earlier,
leads to
\be
 W_\y(x)
=
 W_\y(A_1)
=
 -\Ai'(A_1) \Bi(A_1)
=
 \lim_{\chi \to \infty}
 W_\y(\chi)
=
 1/\pi > 0 ,
\Label{Wpsi}
\ee
for all $x \in [0,1]$
and
for this particular pair.
(To calculate the limit,
we used the asymptotic expansions
of $\Ai(\chi)$ and $\Bi(\chi)$
as $\chi \to \infty$---see,
\emph{e.g.}, \cite{BO-1999}.)
Next,
we simplify the formula (\ref{Dpsi})
by investigating
the asymptotic magnitude
of the terms
in its right member.
By definition~(\ref{ODEt-psi}),
\[
 \G{\y_{n,\pm}}{0}
=
 \y_{n,\pm}(0)
-
 \sqrt{\eps/\V} \, (\dx \y_{n,\pm})(0) .
\]
Equations (\ref{xn-ae})
and
(\ref{psin-})--(\ref{psin+})
yield
\begin{eqnarray*}
 \G{\y_{n,-}}{0}
&=&
 -\eps^{5/6}
\,
 \sL^{-5/6}
\,
 \Ai'(A_1)
\, 
 (N_ n + \Lambda_0)
+
 \Or(\eps^{7/6}) ,
\\
 \G{\y_{n,+}}{0}
&=&
 \eps^{1/6}
\,
 \sL^{-1/6}
\,
 \Bi(A_1)
+
 \Or(\eps^{1/3}) .
\Label{G(0)-ae}
\end{eqnarray*}
(Here,
we have Taylor expanded
$\Ai( A_1 (1 - \xo^{-1} x) )$
around its zero $x = 0$.)
Next,
\begin{eqnarray}
 \G{\y_{n,\pm}}{1}
&\sim&
 \eps^{1/4}
 \frac{(1 \mp \sqrt{\sR/\V}) \, c_\pm}{\sR^{1/4}}
 \exp\left(\pm\frac{I(1)}{\sqrt{\eps}}\right) ,
\Label{G(1)-ae}
\end{eqnarray}
recall (\ref{defgammas1}).
These formulas imply that
$\G{\y_{n,+}}{0} \G{\y_{n,-}}{1}$
is exponentially smaller than
$\G{\y_{n,-}}{0} \G{\y_{n,+}}{1}$,
and thus
\be
 D_\y
=
\frac{
 D_n(1)
\,
 G_+(1)
-
 G_-(1)
}{
 \G{\y_{n,-}}{0}
} ,
\quad\mbox{where}\quad
 D_n(1)
=
\frac{
 \G{\y_{n,-}}{1}
}{
 \G{\y_{n,+}}{1}
}
\Label{Dpsi-simple-aux}
\ee
and
down to
exponentially small terms.
Next,
the relative asymptotic magnitudes
of the terms in
$G_-(1) - D_n(1) G_+(1)$
may be derived
using the definitions
(\ref{ODEt-psi})
and
(\ref{Gpm})
together with Laplace's approximation
(cf. Theorem~\ref{t-Laplace}).
One finds that
$G_-(1)$
is dominated by
$\exp(\eps^{-1/2} J_-(\xs))$,
whereas
$D_n(1) G_+(1)$
by
$\exp(\eps^{-1/2} J_-(1))$,
and hence
the latter is exponentially smaller
than the former.
Hence,
\be
 D_\y
=
-\frac{
 G_-(1)
}{
 \G{\y_{n,-}}{0}
} .
\Label{Dpsi-simple}
\ee
It follows, then, that
\be
 \Gamma_-(x)
=
 G_+(x)
-
 D_n(0) \, G_-(1)
\quad\mbox{and}\quad
 \Gamma_+(x)
=
 G_-(x)
-
 G_-(1) ,
\Label{G-simple}
\ee
and
down to
exponentially small terms.
Here,
\be
 D_n(0)
=
\frac{
 \G{\y_{n,+}}{0}
}{
 \G{\y_{n,-}}{0}
}
=
 \eps^{-2/3}
\,
 d_n(\Lambda_0),
\Label{Do-ae}
\ee
where (recall (\ref{Wpsi}))
\be
 d_n(\Lambda_0)
=
-
\frac{
 \sL^{2/3}
\,
 \Bi(A_1)
}{
 \Ai'(A_1)
\,
 (N_ n + \Lambda_0)
}
=
\frac{
 \sL^{2/3}
}{
 \pi
\,
 C_3
\,
 (N_ n + \Lambda_0)
}
 >
0 .
\Label{dn-def}
\ee
Combining this formula
with (\ref{psit-n-aux}),
we find
\begin{eqnarray}
\fl
 \y_n(x)
&=
 (\ell W_\y)^{-1}
\,
\left[
 G_+(x)
 \y_{n,-}(x)
-
 G_-(x)
 \y_{n,+}(x)
+
 G_-(1)
\left(
 \y_{n,+}(x)
-
 D_n(0)
 \y_{n,-}(x)
\right)
\right]
\nonumber\\
\fl
&=
 (\ell W_\y)^{-1}
\,
\left[
\left(
 G_+(x)
-
 D_n(0) G_-(1)
\right)
 \y_{n,-}(x)
+
\left(
 G_-(1)
-
 G_-(x)
\right)
 \y_{n,+}(x)
\right]
\nonumber\\
\fl
&=
 (\ell W_\y)^{-1}
\,
\left[
 \y_{n,-}(x)
\left(
 \int_0^x
 f(y) \s_n(y) \, \ypu_{n,+}(y) dy
-
 D_n(0)
 \int_0^1
 f(y) \s_n(y) \, \ypu_{n,-}(y) dy
\right)
\right.
\nonumber\\
\fl
&\hspace*{2.5cm}
\left.
+
 \y_{n,+}(x)
 \int_x^1
 f(y) \s_n(y) \, \ypu_{n,-}(y) dy
\right] .
\Label{psit-n}
\end{eqnarray}
%

\section{Asymptotic approximation of integrals
\label{s-Laplace}}
\setcounter{equation}{0}

\subsection{Localized integrals}
Our main tool
in this section
will be Laplace's method
and,
in particular,
the following three theorems
based on \cite[Ch.~II, VIII, IX]{W-2001}.
\noappendix
\begin{THEOREM}
\label{t-Laplace}
(\cite[Theorem~IX.3]{W-2001})
Let $n \in \N$,
$D \subset \R^n$
be a domain
with piecewise smooth boundary $\partial D$,
and
$u_0 \in \bar{D}$.
Let, also,
the functions
$\Pi \in C^2(\bar{D} , \R)$
and
$\Xi \in C(\bar{D} , \R)$
satisfy the conditions
\begin{eqnarray*}
&\mbox{(a)}& \
\inf_{\bar{D}-B(u_0;\delta)}\Pi(u) > \Pi(u_0) ,
\quad \mbox{for all}\
 \delta > 0 ,
\\
&\mbox{(b)}& \
 \sigma\left(D^2\Pi(u_0)\right)
\subset
 \mathring{\R}_+ ,
\\
&\mbox{(c)}& \
\mbox{the integral}\
 \I_D(\lambda)
:=
 \int\!\!\cdots\!\!\int_D \Xi(u) \, \ex^{-\lambda \Pi(u)} du
\ \mbox{converges absolutely}\
\\
&{}&
 \ \mbox{for all sufficiently large} \ \lambda .
\end{eqnarray*}
(Here,
$D^2\Pi$ denotes
the Hessian matrix
of $\Pi$.)
Then,
\[
 \I_D(\lambda)
\sim
 \ex^{-\lambda \Pi(u_0)}
\,
 \sum_{k = 0}^\infty
 c_k \, \lambda^{-(k+n/2)}
\quad
 (\lambda \to \infty) ,
\]
where
one may derive
explicit formulas
for the constants $\{c_k\}_k$.
In particular,
\begin{eqnarray*}
\fl
{\rm (I)} \quad
 \I_D(\lambda)
&\sim&
 \left(\frac{2 \pi}{\lambda}\right)^{n/2}
\,
\frac{
 \Xi(u_0) \, \ex^{-\lambda \Pi(u_0)}
}{
 \sqrt{\det D^2\Pi(u_0)}
} ,
\quad\mbox{if}\quad
 u_0 \in \mathring{D}
\ \mbox{and} \
 \Xi(u_0) \ne 0 ,
\\
\fl
{\rm (II)} \quad
 \I_D(\lambda)
&\sim&
 \left(\frac{2 \pi}{\lambda}\right)^{(n+2)/2}
\,
 C_0 \, \ex^{-\lambda \Pi(u_0)}
 ,
\quad\mbox{if}\quad
 u_0 \in \mathring{D}
\ \mbox{and} \
 \Xi(u_0) = 0 ,
\\
\fl
{\rm (III)} \quad
 \I_D(\lambda)
&\sim&
 \left(\frac{2 \pi}{\lambda}\right)^{n/2}
\,
\frac{
 \Xi(u_0) \, \ex^{-\lambda \Pi(u_0)}
}{
 2 \sqrt{\det D^2\Pi(u_0)}
} ,
\quad\mbox{if}\quad
 u_0 \in \partial D ,
\
 \Xi(u_0) \ne 0 ,
\ \mbox{and} \
 D\Pi(u_0) = 0 ,
\\
\fl
{\rm (IV)} \quad
 \I_D(\lambda)
&\sim&
 \left(\frac{2 \pi}{\lambda}\right)^{(n+1)/2}
\,
\frac{
 \Xi(u_0) \, \ex^{-\lambda \Pi(u_0)}
}{
 2 \pi \, \sqrt{\det J}
} ,
\quad\mbox{if}\quad
 u_0 \in \partial D ,
\
 \Xi(u_0) \ne 0 ,
\ \mbox{and} \
 D\Pi(u_0) \ne 0 ,
\end{eqnarray*}
as $\lambda \to \infty$,
for some constant $C_0$
which is at most $\Or(1)$
with respect to $\lambda$
and
under the assumption that
$\partial D$ is smooth around $u_0$
in the cases where
$u_0 \in \partial D$.
Here,
$J$ is a matrix
related to $D^2\Pi(u_0)$
and
to the local characteristics
of $\partial D$
around $u_0$.
\end{THEOREM}
\begin{THEOREM}
\label{t-Laplace1}
Let $a < b$
and
$u_0 \in [a,b]$.
Let, also,
the functions
$\Pi \in C^2([a,b] , \R)$
and
$\Xi \in C([a,b] , \R)$
satisfy the conditions
\begin{eqnarray*}
&\mbox{(a)}& \
 \inf_{[a,b]-B(u_0;\delta)}\Pi(u) > \Pi(u_0) ,
\quad \mbox{for all}\
 \delta > 0 ,
\\
&\mbox{(b)}& \
\mbox{the integral}\
 \I(\lambda)
:=
 \int_a^b \Xi(u) \, \ex^{-\lambda \Pi(u)} du
\ \mbox{converges absolutely}\
\\
&{}&
 \ \mbox{for all sufficiently large} \ \lambda .
\end{eqnarray*}
Then,
\[
 \I(\lambda)
\sim
 \ex^{-\lambda \Pi(u_0)}
\,
 \sum_{k = 1}^\infty
 c_k \, \lambda^{-k/2}
\quad
 (\lambda \to \infty) ,
\]
where
one may derive
explicit formulas
for the constants $\{c_k\}_k$.
In particular,
as $\lambda \to \infty$,
\begin{eqnarray*}
\fl
{\rm (I)} \quad
 \I(\lambda)
&\sim&
 \frac{\ex^{-\lambda \Pi(u_0)}}{\lambda^{1/2}}
\,
\frac{
 \sqrt{2 \pi} \, \Xi(u_0)
}{
 \sqrt{\Pi''(u_0)}
} ,
\quad\mbox{if}\quad
 u_0 \in (a,b)
\quad\mbox{and}\quad
 \Xi(u_0) \ne 0 ,
\\
\fl
{\rm (II)} \quad
 \I(\lambda)
&\sim&
 \frac{\ex^{-\lambda \Pi(u_0)}}{\lambda^{3/2}}
\,
\frac{
 \sqrt{\pi}
\,
\left(
 \Xi''(u_0)
-
 \frac{\Xi'(u_0) \, \Pi'''(u_0)}{\Pi''(u_0)}
\right)
}{
 \sqrt{2} \, [\Pi''(u_0)]^{3/2}
} ,
\quad\mbox{if}\quad
 u_0 \in (a,b)
\quad\mbox{and}\quad
 \Xi(u_0) = 0 ,
\\
\fl
{\rm (III)} \quad
 \I(\lambda)
&\sim&
 \frac{\ex^{-\lambda \Pi(u_0)}}{\lambda}
\,
\frac{
 \Xi(u_0)
}{
 \abs{\Pi'(u_0)}
} ,
\quad\mbox{if}\quad
 u_0 \in \{a,b\} ,
\
 \Xi(u_0) \ne 0 ,
\quad\mbox{and}\quad
 \Pi'(u_0) \ne 0 ,
\\
\fl
{\rm (IV)} \quad
 \I(\lambda)
&\sim&
 \frac{\ex^{-\lambda \Pi(u_0)}}{\lambda^{1/2}}
\,
\frac{
 \sqrt{\pi} \, \Xi(u_0)
}{
 \sqrt{2 \Pi''(u_0)}
} ,
\quad\mbox{if}\quad
 u_0 \in \{a,b\} ,
\
 \Xi(u_0) \ne 0 ,
\quad\mbox{and}\quad
 \Pi'(u_0) = 0 ,
\\
\fl
{\rm (V)} \quad
 \I(\lambda)
&\sim&
 \frac{\ex^{-\lambda \Pi(u_0)}}{\lambda^2}
\,
\frac{
 \pm \Xi'(u_0)
}{
 [\Pi'(u_0)]^2
} ,
\quad\mbox{if}\quad
 u_0
=
\left\{
\begin{array}{c}
 a \ (+) \\
 b \ (-)
\end{array}
\right. ,
\
 \Xi(u_0) = 0 ,
\quad\mbox{and}\quad
 \Pi'(u_0) \ne 0 ,
\\
\fl
{\rm (VI)} \quad
 \I(\lambda)
&\sim&
 \frac{\ex^{-\lambda \Pi(u_0)}}{\lambda}
\,
\frac{
 \pm \Xi'(u_0)
}{
 \Pi''(u_0)
} ,
\quad\mbox{if}\quad
 u_0
=
\left\{
\begin{array}{c}
 a \ (+) \\
 b \ (-)
\end{array}
\right. ,
\
 \Xi(u_0) = 0 ,
\quad\mbox{and}\quad
 \Pi'(u_0) = 0 .
\end{eqnarray*}
\end{THEOREM}
\begin{THEOREM}
\label{t-Laplace2}
Let $D \subset \R^2$
be a two-dimensional domain
with piecewise smooth boundary $\partial D$
and
$u_0 \in \partial D$.
Let, also,
the functions
$\Pi \in C^2(\bar{D} , \R)$
and
$\Xi \in C(\bar{D} , \R)$
satisfy the conditions
\begin{eqnarray*}
&\mbox{(a)}& \
\inf_{\bar{D}-B(u_0;\delta)}\Pi(u) > \Pi(u_0) ,
\quad \mbox{for all}\
 \delta > 0 ,
\\
&\mbox{(b)}& \
\mbox{the integral}\
 \I_D(\lambda)
:=
 \int\!\!\cdots\!\!\int_D \Xi(u) \, \ex^{-\lambda \Pi(u)} du
\ \mbox{converges absolutely}\
\\
&{}&
 \ \mbox{for all sufficiently large} \ \lambda .
\end{eqnarray*}
Assume, further, that
$\partial D$ has a corner at $u_0$
and, in particular, that
$\partial D$ is given
(locally around $u_0$)
by the curves
$k(x,y) = 0$ and $h(x,y) = 0$
with
$Dk(u_0) \times Dh(u_0) \ne 0$.
Let the vectors
$v_k$ and $v_h$
satisfy
\[
 v_k \perp Dk(u_0) ,
\quad
 v_h \perp Dh(u_0) ,
\quad\mbox{and}\quad
 \| v_k \times v_h \| = 1 .
\]
If $v_k$ and $v_h$
can be selected
to further satisfy
the conditions
\be
 \Pi_k
:=
 \left\langle v_k , D\Pi(u_0) \right\rangle
>
 0
\quad\mbox{and}\quad
 \Pi_h
:=
 \left\langle v_h , D\Pi(u_0) \right\rangle
>
 0 ,
\Label{Pik,Pih}
\ee
then
\[
 \I_D(\lambda)
\sim
 \ex^{-\lambda \Pi(u_0)}
\,
 \sum_{k = 0}^\infty
 c_k \, \lambda^{-(k+2)}
\quad
 (\lambda \to \infty) ,
\]
where
one may derive
explicit formulas
for the constants $\{c_k\}_k$.
In particular,
\begin{eqnarray*}
{\rm (I)} \quad
 \I_D(\lambda)
&\sim&
 \frac{1}{\lambda^2}
\,
\frac{
 \Xi(u_0) \, \ex^{-\lambda \Pi(u_0)}
}{
 2 \Pi_k \Pi_h \sqrt{\Pi_k^2 + \Pi_h^2}
} ,
\quad\mbox{if}\quad
 \Xi(u_0) \ne 0 ,
\\
{\rm (II)} \quad
 \I_D(\lambda)
&\sim&
 \frac{1}{\lambda^3}
\,
\frac{
\left(
 \Pi_k \Xi_h
+
 \Pi_h \Xi_k
\right)
\,
 \ex^{-\lambda \Pi(u_0)}
}{
 \Pi_k^2 \Pi_h^2 \sqrt{\Pi_k^2 + \Pi_h^2}
} ,
\quad\mbox{if}\quad
 \Xi(u_0) = 0 ,
\end{eqnarray*}
as $\lambda \to \infty$.
Here,
$\Xi_k := \langle v_k , D\Xi(u_0) \rangle$
and
$\Xi_h := \langle v_h , D\Xi(u_0) \rangle$,
compare to (\ref{Pik,Pih}).
\end{THEOREM}
%

\subsection{Oscillatory integrals}
\begin{THEOREM}
\label{t-Fourier1}
Let $a < b$,
$\Xi \in C([a,b] , \R)$,
and
$\Phi \in C^2([a,b] , \R)$.
Assume that
\[
\fl
 \Phi(t) = \Phi(a) + (t-a) \, \Phi_1(t)
\quad\mbox{and}\quad
 \Phi'(t) > 0 ,
\quad \mbox{for all}\
 t \in [a,b]
\ \mbox{and with} \
 \Phi_1(a) \ne 0 .
\]
Then,
the integral
$\I(\lambda)
:=
\int_a^b \Xi(t) \, \ex^{\im \lambda \Phi(t)} dt$
has the following asymptotic expansion:
\[
 \I(\lambda)
\sim
 \sum_{k = 0}^\infty
\left[
 h^{(k)}(0)
\,
 \ex^{\im \lambda \Phi(a)}
-
 h^{(k)}(\Phi(b) - \Phi(a))
\,
 \ex^{\im \lambda \Phi(b)}
\right]
 \left(\frac{\im}{\lambda}\right)^{k+1}
\quad
 (\lambda \to \infty) ,
\]
where
we have defined the function
\[
 h(\tau)
=
 \Xi(t(\tau)) \, t'(\tau) .
\]
Here,
$\tau(t) = \Phi(t) - \Phi(a)$
or, equivalently,
$t(\tau) = \Phi^{-1}(\Phi(a) + \tau)$.
\end{THEOREM}

\end{document}